\def\eiso{E_{\rm iso}}

\def\eop{E^{\rm obs}_{\rm peak}}
\def\esp{E^{\rm src}_{\rm peak}}

\def\ep{E_{\rm peak}}

\documentclass[12pt,preprint]{aastex}

\shorttitle{BAT 2 catalog}
\shortauthors{Sakamoto et al.}

\begin{document}


\title{The Second {\it Swift} BAT Gamma-Ray Burst Catalog}


\author{
T. Sakamoto\altaffilmark{1,2,3}, 
S. D. Barthelmy\altaffilmark{3}, 
W. H. Baumgartner\altaffilmark{1,2,3}, 
J. R. Cummings\altaffilmark{1,2,3}, 
E. E. Fenimore\altaffilmark{4},
N. Gehrels\altaffilmark{3}, 
H. A. Krimm\altaffilmark{1,5,3}, 
C. B. Markwardt\altaffilmark{3},
D. M. Palmer\altaffilmark{4},
A. M. Parsons\altaffilmark{3},
G. Sato\altaffilmark{6},
M. Stamatikos\altaffilmark{7}, 
J. Tueller\altaffilmark{3},
T. N. Ukwatta\altaffilmark{8},
B. Zhang\altaffilmark{9}
}

\altaffiltext{1}{Center for Research and Exploration in Space Science 
and Technology (CRESST), NASA Goddard Space Flight Center, Greenbelt, MD 
20771}
\altaffiltext{2}{Joint Center for Astrophysics, University of Maryland, 
	Baltimore County, 1000 Hilltop Circle, Baltimore, MD 21250}
\altaffiltext{3}{NASA Goddard Space Flight Center, Greenbelt, MD 20771}
\altaffiltext{4}{Los Alamos National Laboratory, P.O. Box 1663, Los
Alamos, NM, 87545}
\altaffiltext{5}{Universities Space Research Association, 10211 Wincopin 
	Circle, Suite 500, Columbia, MD 21044-3432} 
\altaffiltext{6}{Institute of Space and Astronautical Science, 
JAXA, Kanagawa 229-8510, Japan}
\altaffiltext{7}{Center for Cosmology and Astro-Particle Physics, 
	Department of Physics, The Ohio State University, 
	191 West Woodruff Avenue, Columbus, OH 43210}
\altaffiltext{8}{Michigan State University, 
                 3245  BPS Building, East Lansing, MI 48824}
\altaffiltext{9}{Department of Physics and Astronomy, University of Nevada, Las Vegas, 
NV 89154}

\begin{abstract}
We present the second {\it Swift} Burst Alert Telescope (BAT) catalog of gamma-ray 
bursts (GRBs), which contains 476 bursts detected by the BAT between 
2004 December 19 and 2009 December 21.  This catalog (hereafter the BAT2 catalog) 
presents burst trigger time, location, 90\% error radius, duration, fluence, 
peak flux, time-averaged spectral parameters and time-resolved spectral 
parameters measured by the BAT.  In the correlation study of various observed 
parameters extracted from the BAT prompt emission data, we distinguish among long-duration 
GRBs (L-GRBs), short-duration GRBs (S-GRBs), and 
short-duration GRBs with extended emission (S-GRBs with E.E.) to 
investigate differences in the prompt emission properties.  
The fraction of L-GRBs, S-GRBs and S-GRBs with E.E. in the catalog 
are 89\%, 8\% and 2\% respectively.  
We compare the BAT prompt emission properties with the BATSE, {\it BeppoSAX} 
and {\it HETE}-2 GRB samples.  
We also correlate the observed prompt emission properties with the redshifts for 
the GRBs with known redshift.  The BAT $T_{90}$ and $T_{50}$ durations peak at 70~s and 
30~s, respectively.  We confirm that the spectra of the BAT S-GRBs are generally 
harder than those of the L-GRBs.  The time-averaged spectra of the BAT S-GRBs with E.E. are similar 
to those of the L-GRBs.  Whereas, the spectra of the initial short spikes of the S-GRBs 
with E.E. are similar to those of the S-GRBs.  
We show that the BAT GRB samples are significantly softer than 
the BATSE bright GRBs, and that the time-averaged $\eop$ of 
the BAT GRBs peaks at 80 keV which is significantly lower energy than those of the BATSE 
sample which peak 
at 320 keV.  The time-averaged spectral properties of the BAT GRB sample are similar 
to those of the {\it HETE}-2 GRB samples.  By time-resolved spectral analysis, we find 
that only 10\% of the BAT observed photon indices are outside the allowed region of the 
synchrotron shock model.  
We see no obvious observed trend in the BAT $T_{90}$ and the observed 
spectra with redshifts. 
The $T_{90}$ and $T_{50}$ distributions measured at the 
140-220 keV band in the GRB 
rest frame from the BAT known redshift GRBs peak at 19~s and 8~s, respectively.  
We also provide an update on the status of the on-orbit BAT calibrations.  

\end{abstract}





\section{Introduction}

The {\it Swift} mission \citep{gehrels2004} has revolutionized our 
understanding of gamma-ray bursts (GRBs) and their usage to study the 
early universe with the sophisticated on-board localization capability 
of the {\it Swift} Burst Alert Telescope (BAT; \citet{barthelmy}).   
The autonomous spacecraft slews to point the X-Ray Telescope 
(XRT;\citet{burrows}) and the UV/Optical Telescope (UVOT; \citet{roming}) 
to the GRB location to start an immediate follow-up observation.  The 
precise GRB location provided by {\it Swift} permits coordinated 
multi-wavelength observations on the ground.  
Since the release of the first BAT GRB catalog \citep{sakamoto2008}, 
there have been more notable discoveries.  
On April 23, 2009, {\it Swift} BAT detected 
GRB 090423 at a redshift of 8.2.  This implies that massive stars were 
produced about 630 Myr after the Big Bang \citep{tanvir2009,salvaterra2009}.  
In GRB 080319B, extraordinary, bright, variable optical emission, 
which peaked at the visual 
magnitude of 5.3, has been observed while the prompt gamma-ray emission was 
still active.  
This observation suggests for the first time that there can 
exist a prompt optical emission component that tracks the gamma-ray light curve 
but belongs to a different spectral component \citep{racusin2008}.
{\it Swift} has been increasing 
the number of identifications of host galaxies for short duration bursts.  Based on 
the {\it Swift} sample, we are observing a wide variety of burst 
characteristics \citep[e.g.,][]{berger2009}.  
Furthermore, progress is being made in studying the properties 
of host galaxies of dark bursts, allowing us to understand the differences 
in the birth places of dark GRBs, as compared to the hosts of GRBs with bright 
optical afterglows \citep[e.g,.][]{perley2009}.  

The first BAT GRB catalog (the BAT1 catalog; \citet{sakamoto2008}) consisted of 237 
bursts from 2004 December 19 to 2007 June 16.  The BAT1 catalog contained burst 
trigger time, location, 90\% error radius, duration, fluence, peak flux and 
time-averaged spectral parameters.  
Here we present the second BAT GRB catalog (the BAT2 catalog), 
including 476 GRBs detected by BAT from 2004 December 19 to 2009 December 21.  
The GRB samples in the BAT1 catalog have been re-analyzed by the latest BAT 
software and calibration files.  In addition to the parameters in the BAT1 
catalog, the BAT2 catalog includes time-resolved spectral parameters.   
The catalog also presents correlations between the prompt 
emission properties of known-redshift GRBs.  

In \S 2, we summarize the updates 
to the in-orbit calibration of the BAT instrument.  
In \S 3, we describe the analysis methods used in compiling the catalog.  
In \S 4, we describe the content of the tables in the catalog and show the 
prompt emission properties of the BAT GRBs from the catalog.  Our 
conclusions are summarized in \S 5.  All quoted errors in this work 
are at the 90\% confidence level.  

\section{Updates to In-orbit Calibrations}
The BAT is a highly sensitive, large field of view (FOV) 
(2.2 sr for $>$ 10\% coded FOV), coded-aperture 
telescope that detects and localizes GRBs in real time.  The BAT 
detector plane is composed of 32,768 pieces of CdZnTe 
(CZT: $4 \times 4 \times 2$ mm), and the coded-aperture mask is 
composed of $\sim$ 52,000 lead tiles ($5 \times 5 \times 1$ mm) with 
a 1 m separation between mask and detector plane.  The energy range 
is 14--150 keV for imaging, which is a technique to subtract the 
background based on the modulation resulting from 
the coded mask, and 
spectra can be obtained with no position information up to 350 keV.  
Further detailed descriptions and references of the BAT instrument  
can be found in the BAT1 GRB catalog.

\begin{figure}
\centerline{
\includegraphics[width=12cm,angle=-90]{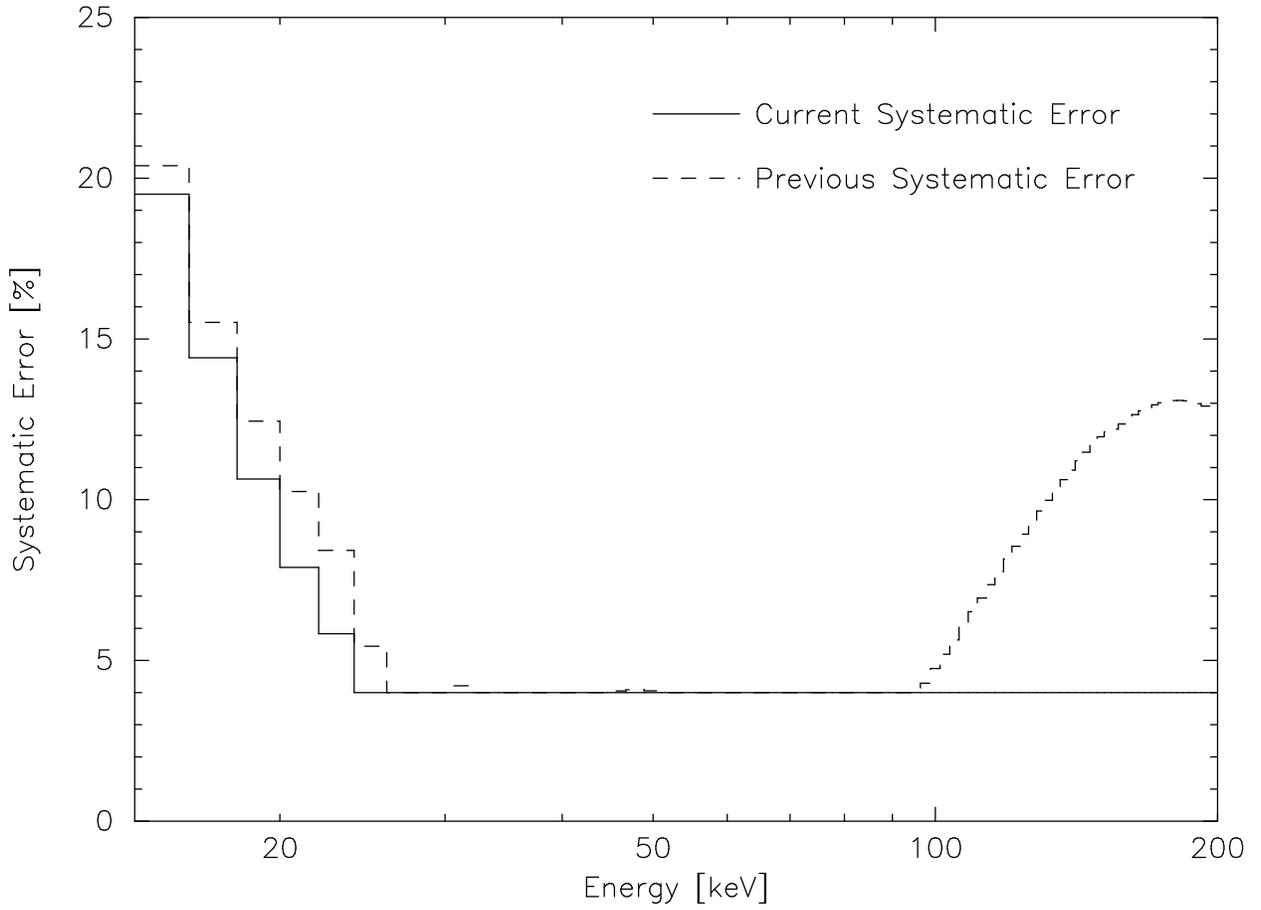}
}
\caption{Systematic error as a function of energy.  The current systematic 
error is shown in a solid line and the previous systematic error (used in the 
analysis of the BAT1 catalog) is shown in a dashed line. \label{bat_sys}}
\end{figure}

There have been two major updates to the energy calibration of the BAT 
since the publication of the BAT1 catalog.  The first update is the 
identification of a problem in the energy response above 100 keV.  
The mobility-lifetime products of electrons and holes ($\mu\tau$) 
which are used to model the energy response of an individual 
CZT detector \citep{sato2005,suzuki2005} 
are found to be 1.7 times larger than those originally determined 
in the ground calibration.  
This fix eliminates the ad-hoc correction formerly applied above 100 keV 
to reproduce the 
assumed photon index and flux values based on the Crab spectrum (see below) 
and also reduces the level of 
systematic errors (see Figure \ref{bat_sys}).  
These changes in the energy response and the systematic error 
are available to the public (BAT CALDB 20081026).  

The second update is a correction to the measured gain of the detectors.  
An analysis of four years of on-board $^{241}$Am spectra shows a shift 
of 2.5 keV in the position of the 59.5 keV line.  
The BAT team has developed new calibration files 
(coefficients to convert from PHA channel to energy) 
to correct this gain change as a function of time.  After applying the 
new gain correction, the scatter of the 59.5 keV line energy is $\sim$0.1 keV 
over the four-year period.  
The calibration file to correct this time-dependent gain 
change will be released to the public in summer of 2011.  

\begin{figure}
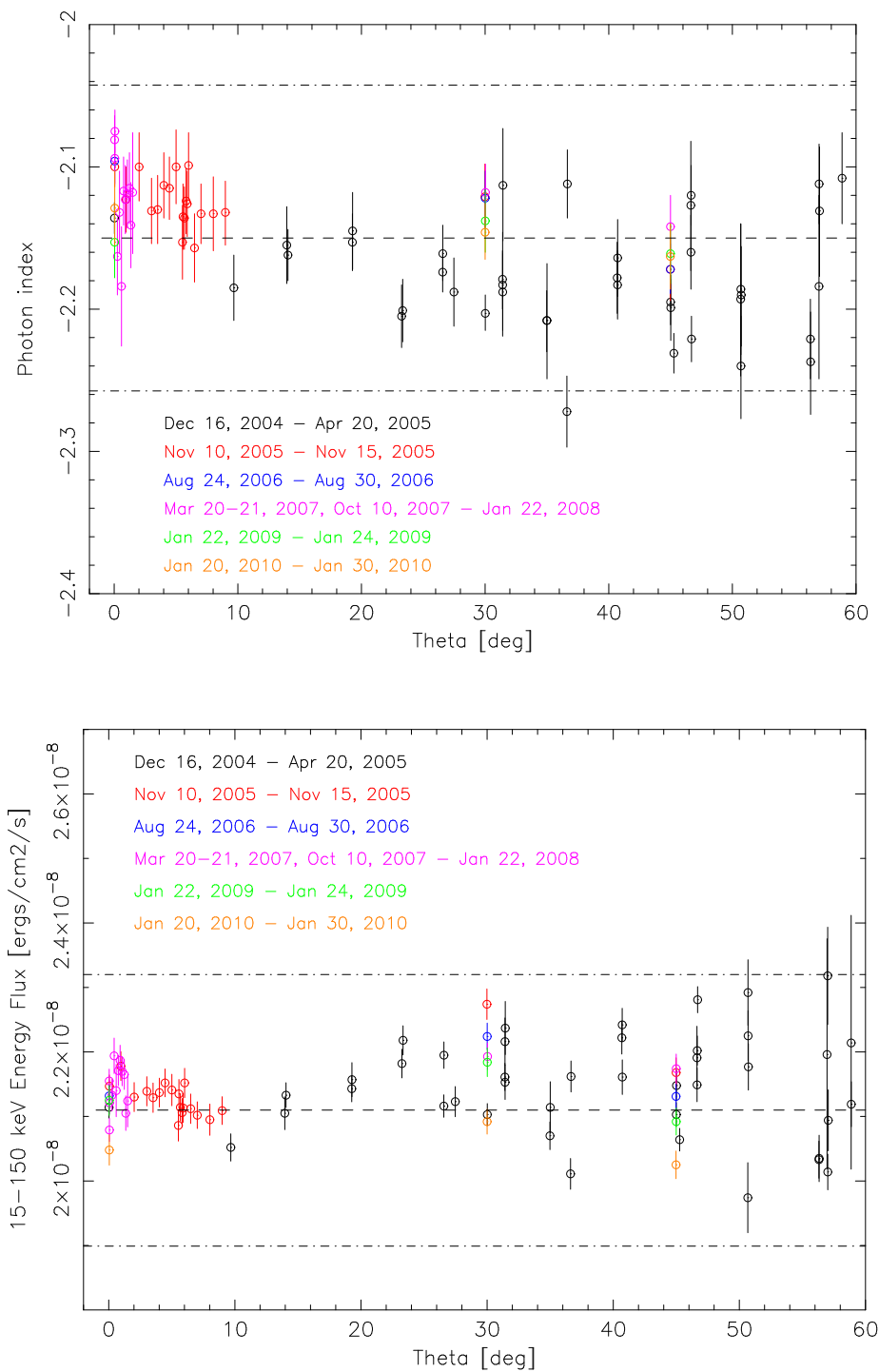

\centerline{
\includegraphics[width=9cm,angle=-90]{fig2a.eps}
}
\vspace{1cm}
\centerline{
\includegraphics[width=9cm,angle=-90]{fig2b.eps}
}
\caption{Power-law photon index (top) and the flux in the 15-150 keV band 
as a function of the incident angle of the Crab observed in different 
time periods.  The horizontal dashed lines are the Crab canonical 
values of $-2.15$ for the photon index and $2.11 \times 10^{-8}$ ergs 
cm$^{-2}$ s$^{-1}$ for the flux.  The dashed dotted lines are $\pm$5\% 
of the photon index and $\pm$10\% of the flux canonical values. \label{crab_phindex_flux}}
\end{figure}

Figure \ref{crab_phindex_flux} shows the Crab photon index and the flux in the 
15-150 keV band as a function of incident angle using the latest BAT software 
and calibration files.  
Both the photon index and the flux 
are within $\pm$5\% and $\pm$10\%, respectively, of the assumed Crab values\footnote{
The assumed Crab values in the BAT energy range are $-2.15$ for the photon index and 
$2.11 \times 10^{-8}$ ergs cm$^{-2}$ s$^{-1}$ for the flux in the 15-150 keV band.}
based on \citet{rothschild1998} and \citet{jung1989} over the BAT field of view.  
The deviation of the derived parameters from the 
Crab canonical values are larger toward the edge of the BAT field of view.  
Therefore, a larger systematic error in the spectral parameters could exist 
if the source is located at the edge of the field of BAT.  
We also notice that the photon index is systematically harder by $\sim$0.02 
than the Crab canonical value of $-2.15$ for observations with $\theta < 10^{\circ}$.  

\section{Analysis for the GRB catalog}

We used standard BAT software (HEASOFT 6.8) and the latest 
calibration database to process the BAT GRBs from December 2004 
(GRB 041217) to December 2009 (GRB 091221).  
The various data types of the BAT instrument are described in details in 
\citet{barthelmy}.  
The GRBs included 
in the BAT1 catalog have been reprocessed.  The BAT2 catalog sample 
also include bursts which did not trigger BAT in flight, but were 
found in ground processing.  The burst pipeline 
script, {\tt batgrbproduct}, was used to process the BAT event data.  
Since the burst emission is longer 
than the interval during which the instrument collected data in event mode 
for GRB 060124 and GRB 060218, we used both event data and survey 
data (Detector Plane Histogram data) to calculate the fluence and the 
time-averaged spectral parameters.  We used raw light curve data (quad-rate 
data) to measure the durations of most of the bursts found in ground processing 
because the event data does not cover the whole burst episode 
for these bursts.  For most bursts we use {\tt battblocks}, 
one of the tasks run in {\tt batgrbproduct}, 
to determine the burst duration using the Bayesian Block algorithm \citep{scargle1998}.\footnote{{\tt batgrbproduct} 
calls {\tt battblocks} using, in turn, mask-weighted 
light curves with bin widths 4 ms, 16ms, 1s and 16s to measure $T_{90}$ and $T_{50}$ durations.  
It then applies a set of criteria starting with the shortest bin size to decide whether or not 
to accept a duration estimate.  If a measured duration is greater than 2 times the bin size, and 30\% 
greater than the previous best estimate, then the current estimate becomes the new best estimate.  
}   
For some bursts {\tt battblocks} failed to find a valid 
burst interval.  In these cases, 
we fitted the light curve with background subtracted using the 
modulation resulting from the coded mask (a called mask-weighted light curve) 
over the full BAT energy range.  We applied a 
linear-rise, exponential decay model (``BURS'' model in ftools 
qdp\footnote{http://heasarc.gsfc.nasa.gov/docs/software/ftools/others/qdp/qdp.html})
to find the burst time intervals ($T_{100}$, $T_{90}$, $T_{50}$ and peak 
1-s intervals) and created the $T_{100}$ and peak 1-s PHA files (80 channels) 
based on these time intervals.  We include comments in Table \ref{tbl:bat_summary} for 
the bursts which have anomalies either in the data or the processing.  

For the time-averaged spectral analysis, we used the time interval from
the emission start time to the emission end time ($T_{100}$ interval) 
which is determined by Bayesian-block technique ({\tt battblocks}).  
For many bursts, the spacecraft slew for the XRT and UVOT pointing occurred while 
the prompt emission was ongoing.  This creates a complication for the derivation of 
the spectral response since 
the BAT energy response generator, {\tt batdrmgen}, can only 
produce a response at a single fixed incident angle of the source.  
To find a response during a slew, we created 
detector energy response matrices (DRM) for each five-second period 
during the time $T_{100}$ interval, taking into account the changing position 
of the GRB in detector coordinates 
especially during the spacecraft slew.  We then weighted these DRMs 
by the five-second count rates and created an averaged DRM using 
{\tt addrmf}.  The count-weighted BAT DRM have been tested on a subset of GRBs 
which were simultaneously detected by the Konus-Wind and the {\it Suzaku} 
Wide-band All-sky Monitor instruments.  The joint spectral analysis 
using the weighted BAT DRM for the jointly detected GRBs shows no systematic 
trend in the BAT-derived 
parameters compared to the spectral parameters derived by other GRB 
instruments \citep{grb_cc}.  

We extracted time-resolved spectra for the relevant intervals 
determined with {\tt battblocks}.  
Since the first and the last intervals 
identified by {\tt battblocks} are the pre- and post-burst backgrounds, 
the spectrum for these two intervals were not created.  
For the time-resolved spectral analysis, 
we created a DRM for each spectrum by taking into account the GRB position 
in detector coordinates and updating the keywords of the spectral files using 
{\tt batupdatephakw} before running {\tt batdrmgen} to generate the DRM.  
We also created individual DRMs for the peak spectra used to calculate 
the peak flux (see below).  

The spectra were fitted with a simple power-law (PL) model, 
\begin{eqnarray}
f(E) = K_{50}^{\rm PL}\left(\frac{E}{50  \: {\rm keV}}\right)^{\alpha^{\rm PL}} 
\label{eq:pl}
\end{eqnarray}
where $\alpha^{\rm PL}$ is the power-law photon index and $K_{50}^{\rm
PL}$ is the normalization at 50 keV in units of 
photons cm$^{-2}$ s$^{-1}$ keV$^{-1}$.  We also used a cutoff power-law (CPL) model, 
\begin{eqnarray}
f(E) = K_{50}^{\rm CPL}\left(\frac{E}{50 \: {\rm keV}}\right)^{\alpha^{\rm CPL}} 
\exp\left(\frac{-E\,(2+\alpha^{\rm CPL})}{\ep}\right)
\label{eq:cpl}
\end{eqnarray}
where $\alpha^{\rm CPL}$ is the power-law photon index, $\ep$ is the
peak energy in the $\nu$F$_{\nu}$ spectrum and $K_{50}^{\rm CPL}$ is the 
normalization at 50 keV in units of photons cm$^{-2}$ s$^{-1}$ keV$^{-1}$. 
All of the BAT spectra are acceptably fit by either a PL or a CPL model.  
The same criterion as in the BAT1 catalog, $\Delta \chi^{2}$ 
between a PL and a CPL fit greater than 6 ($\Delta \chi^{2}
\equiv \chi^{2}_{\rm PL} - \chi^{2}_{\rm CPL}$ $>$ 6), 
was used to determine if the CPL model is a better spectral model 
for the data.  Note that none of the BAT spectra show a significant improvement 
in $\chi^{2}$ with a Band function \citep{band1993} fit compared to that of a CPL model fit.  

The fluence, the 1-s and the 20-ms peak fluxes were derived from 
the spectral fits.  The fluences were calculated by fitting the time-averaged 
spectrum with the best fit spectral model.  The 1-s and 20-ms peak 
fluxes were calculated by fitting the spectrum of the 1-s and the 20 ms 
interval bracketing the highest peak in the light curve.  Those intervals were 
identified by {\tt battblocks}.  Similarly, we used the best fit spectral
model to calculate the peak fluxes.  Since the shortest 
burst duration observed by BAT is around 20 ms, we chose this window 
size to measure the peak flux on the shortest time scale. 
Note that since the total number of photons 
for a 20-ms spectrum is generally small, we created a spectrum in 10 logarithmically spaced channels 
from 14 keV to 150 keV to use for the fit.   
We are not always able to report a 20-ms peak flux.  For 29 GRBs, an unexplained systematic 
effect leads to an unacceptable reduced $\chi^{2}$ 
($\chi^{2}_{\nu}$ $>$ 2) in both the PL and CPL fit.  Furthermore, for 31 GRBs we could 
not create the 20-ms peak spectrum because {\tt battblocks} failed to find the interval.  

For GRBs with known redshift, we calculated the $T_{90}$ and $T_{50}$ durations 
in the 140-220 keV band in the GRB rest frame.  By  
fixing the energy range in the GRB rest frame, there is no need to apply 
a correction to the measured duration because of the difference in 
the GRB width as a function of the observed energy band \citep[e.g.,][]{fenimore1995}.  
We created the light curves 
in the energy range 140/(1+z) keV to 220/(1+z) keV with {\tt batbinevt}.  
This energy band was chosen so that the redshift of the nearest GRB z=0.1257 (GRB 060614), 
constrains the upper boundary: 220 keV / (1+0.1257) = 195 keV, 
and the most distant GRB z=8.26 (GRB 090423), constrains the lower boundary: 
140 keV / (1+8.26) = 15 keV, where 15 to 195 keV is the BAT observed energy range.  
We used the same algorithm in {\tt batgrbproduct} 
to find the best $T_{90}$ 
and $T_{50}$ durations in the observed 140/(1+z) - 220/(1+z) keV band.  
We then divided the duration by (1+z) to correct for the time dilation effect due to 
cosmic expansion.  

\section{The Catalog}

The BAT2 catalog includes GRBs detected by BAT in five years of operation 
between 2004 December 19 and 2009 December 31.  The 476 GRBs in the catalog 
include 25 GRBs 
found in ground processing (eight of the 25 were found by the BAT slew 
survey; \citet{batss}) 
and four possible GRBs.  The definition of a possible GRB is an event for which 
the BAT image significance 
based on the ground analysis is less than 6 $\sigma$ and there is no XRT X-ray 
counterpart.  
The first column of Table \ref{tbl:bat_summary} is the GRB name.  The next column 
is the BAT trigger number.  The next column specifies the BAT trigger time 
in UTC in the form of {\it YYYY-MM-DD hh:mm:ss.sss} where {\it YYYY} 
is year, {\it MM} month, {\it DD} day of month, {\it hh} hour, 
{\it mm} minute, and {\it ss.sss} second.  Note that the definition 
of the BAT trigger time is 
the start time of the image from which the GRB is detected on-board.    
The next six columns give, respectively,  
the location derived in ground processing for BAT\footnote{{\tt batgrbproduct} 
creates a BAT sky image in the 15-350 keV band using 
event data from before the autonomous spacecraft slew to a GRB location.  The foreground 
and background intervals used to create the image are determined by {\tt battblocks}.  
If {\tt battblocks} failed to find the intervals, {\tt batgrbproduct} uses the same 
intervals identified by the BAT on-board software.  The tool {\tt batcelldetect} is 
used to find the GRB location in the background subtracted image.
}
and XRT 
in equatorial (J2000) coordinates, 
the signal-to-noise ratio of the BAT image at that location, and the radius 
of the 90\% confidence region in arcminutes.  
The 90\% error radius is calculated 
based on the signal-to-noise ratio of the image using the following equation 
derived from the BAT hard X-ray survey 
process\footnote{http://heasarc.gsfc.nasa.gov/docs/swift/analysis/bat\_digest.html}
\begin{displaymath}
r_{90\%} = 10.92 \times {\rm SNR}^{-0.7}\;({\rm arcmin}), 
\end{displaymath}
where SNR is the signal-to-noise ratio of the GRB point source in BAT image.  However, due to 
limitations in our knowledge of the BAT point spread function, we quote the minimum 
allowed value of $r_{90\%}$ as 1$^{\prime}$ in the catalog.  
The next two columns specify the 
burst durations which contain from 5\% to 95\% ($T_{90}$) and from 25\% to 
75\% ($T_{50}$) of the total burst fluence.  These durations are calculated in 
the 15--350 keV band.\footnote{The coded mask is transparent to photons above 150 keV.
Thus, photons above 150 keV are treated as background in the mask-weighted
method.  The effective upper boundary is thus $\sim$150 keV.}  The next two columns are 
the start and stop times measured from the BAT trigger time of the available event data.  
The last column contains the comments.  

The energy fluences calculated in various energy bands are summarized in 
Table \ref{tbl:bat_fluence}.  The first column is the GRB name.  The next column specifies 
the spectral model which was used in deriving the fluences (PL: simple power-law model; 
Eq.(\ref{eq:pl}), CPL: cutoff power-law model; Eq.(\ref{eq:cpl})).  The next five columns 
are the fluences in the 15-25 keV, 25-50 keV, 50-100 keV, 100-150 keV, and 
15-150 keV bands.  The unit of fluence is 10$^{-8}$ ergs cm$^{-2}$.  
The next two columns specify the start and stop times relative to the 
BAT trigger time of the interval used to calculate the fluences.  The last column contains 
the comments.  Note that fluences are not reported for GRBs with incomplete data.  

Table \ref{tbl:bat_peak1sphflux} and \ref{tbl:bat_peak1seneflux} 
summarize the 1-s peak photon and energy fluxes, respectively, in various energy bands.  
The first column is the GRB name.  The next column specifies 
the spectral model used in deriving the 1-s peak flux.  The next five columns 
are the peak photon and energy fluxes in the 15-25 keV, 25-50 keV, 50-100 keV, 
100-150 keV, and 15-150 keV band.  The unit of the flux is 
photons cm$^{-2}$ s$^{-1}$ for the peak photon flux and 10$^{-8}$ ergs cm$^{-2}$ s$^{-1}$ 
for the peak energy flux.  The penultimate column in Table \ref{tbl:bat_peak1sphflux} specifies 
the start time relative to the BAT trigger time for the peak second.  
The last column contains the comments.  

Table \ref{tbl:bat_peak20msflux} shows the 20-ms peak energy and photon flux 
in the 15-150 keV band.  The first column is the GRB name.  The next column shows the 
spectral model used in deriving the flux.  The next two columns are the peak energy 
flux and photon flux in the 15-150 keV band.  The unit of the flux is 
photons cm$^{-2}$ s$^{-1}$ for the peak photon flux and 10$^{-8}$ ergs cm$^{-2}$ s$^{-1}$ 
for the peak energy flux. The next column specifies the start time 
relative to the BAT trigger time for the 20-ms peak.  The last 
column contains the comments.  

The time-averaged spectral parameters are listed in Table \ref{tbl:bat_timeave_spec}.  
The first column is the GRB name.  The next three columns are the photon index, 
the normalization at 50 keV in units of 10$^{-4}$ photons cm$^{-2}$ s$^{-1}$ keV$^{-1}$, 
and $\chi^{2}$ of the fit for a PL model.  
The next four columns are the photon index, $\eop$ in units of keV, 
the normalization at 50 keV in 
units of 10$^{-3}$ photons cm$^{-2}$ s$^{-1}$ keV$^{-1}$ and $\chi^{2}$ 
of the fit in a CPL model.  The spectral parameters in 
a CPL are only shown for the bursts which meet the criteria described in Section 3.  
The number of degrees of freedom is 57 for a PL fit and 56 for a CPL fit except for GRB 060124 and 
GRB 060218 (see the comment column of the table).  The last column contains the comments. 

The time-resolved spectral parameters are listed in Table \ref{tbl:bat_timeresolved_spec}.  
The total number of time-resolved spectra is 3323.  The first column is the GRB name.  
The next two columns specify the start and the stop time relative to the BAT trigger time 
of the interval used to extract the spectrum.  The next four columns are the photon index, 
the normalization at 50 keV in units of 10$^{-4}$ photons cm$^{-2}$ s$^{-1}$ keV$^{-1}$, 
$\chi^{2}$ of the fit for a PL model and the energy flux in the 15-150 keV band in units of 10$^{-8}$ ergs 
cm$^{-2}$ s$^{-1}$.  
The next five columns are the photon index, $\eop$ in units of keV, the normalization at 50 keV 
in units of 10$^{-3}$ photons cm$^{-2}$ s$^{-1}$ keV$^{-1}$, $\chi^{2}$ of the fit in a CPL model, 
and the energy flux in units of 10$^{-8}$ ergs 
cm$^{-2}$ s$^{-1}$.  The last column contains the comment.  
As in Table \ref{tbl:bat_timeave_spec}, we only show 
CPL parameters for the fits which meet our $\Delta \chi^{2}$ criteria.  

Table \ref{tbl:bat_t90_t50_src} shows the $T_{90}$ and $T_{50}$ durations measured 
in the 140-220 keV band in the GRB rest frame for the GRBs with known redshift.  The first 
column is the GRB name.  The $T_{90}$ and $T_{50}$ durations in the GRB rest frame are 
in the second and the third columns.  The last column contains the comments.  

Table \ref{tbl:redshift} lists redshift measurements of {\it Swift} GRBs and 
their associated references.  

\subsection{Short GRBs with Extended Emission}

\begin{figure}[p]
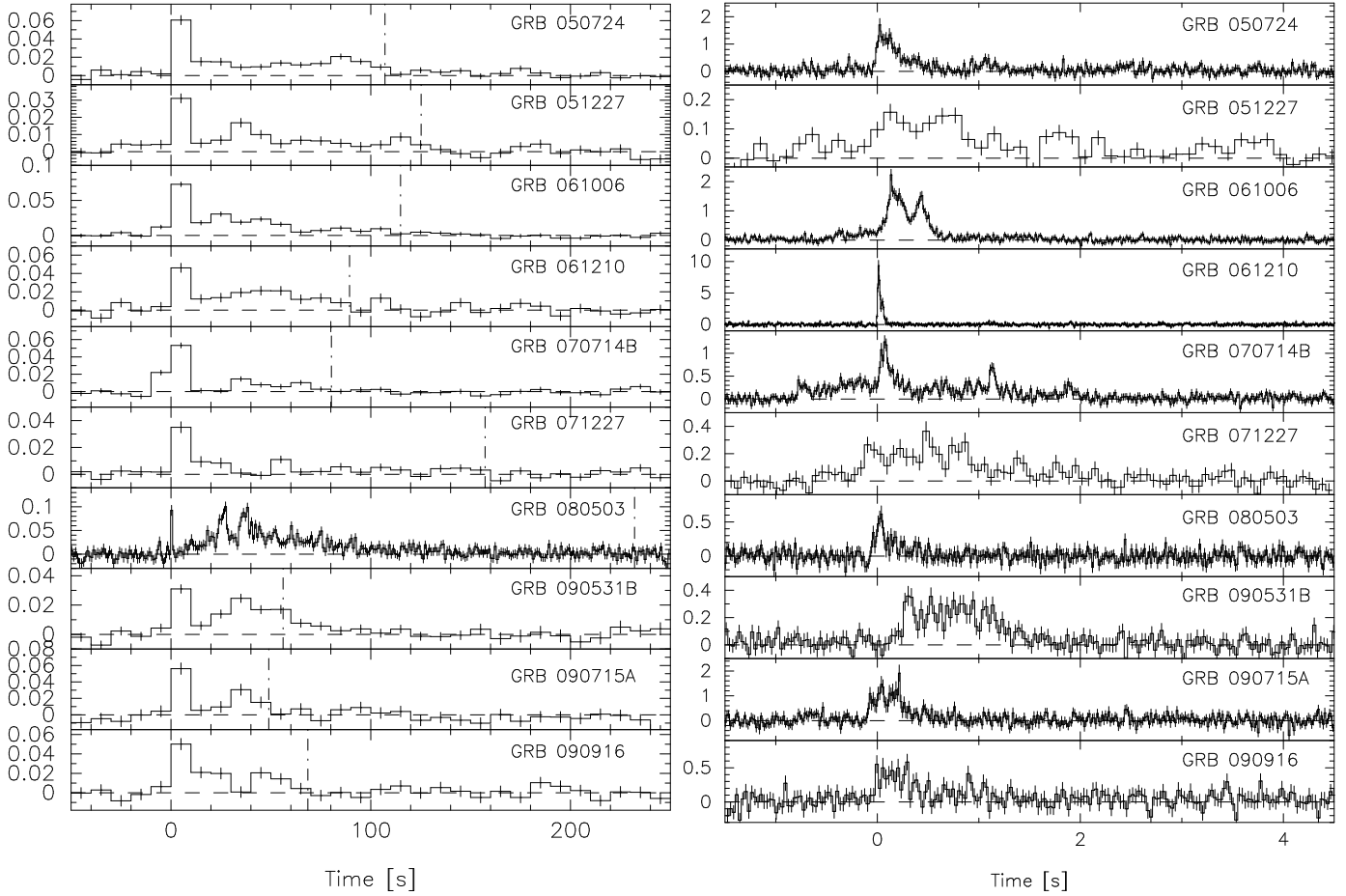

\centerline{
\includegraphics[width=9.15cm,angle=0]{fig3a.eps}
\includegraphics[width=9.0cm,angle=0]{fig3b.eps}
}
\caption{The BAT mask-weighted light curves in the 15-150 keV band 
in a coarse binning (left) and a fine binning (right) for short GRBs with 
extended emission.  The vertical dash-dotted lines in the coarse binning light 
curves show the emission end time found by {\tt battblocks}.
\label{lc_shortEE}}
\end{figure}

A distinct class of short duration GRBs (S-GRBs) has been claimed based on their prompt emission 
properties, called S-GRB with extended emission (E.E.)\citep[e.g.,][]{norris2000,barthelmy2005b}.  The 
initial short spike of a S-GRB with E.E. shows negligible spectral lag which is one of the 
strong indications that the burst is a S-GRB \citep{norris2000}.  
10 GRBs in our catalog have been classified as S-GRBs with E.E. by 
\citet{norris2010} and these are labeled as such throughout this 
paper.\footnote{GRB 060614 is classified as a L-GRB because the duration 
of the initial spike is $\sim$6 s long.  GRB 050911 is also classifed as a L-GRB because 
our standard pipeline process does not detect the significant extended emission as reported 
on \citet{norris2010}.  The $T_{90}$ of GRB 050911 based on our analysis is 16.2 s.}  
The BAT light curves of 10 S-GRBs with E.E. included in this catalog 
are shown in Figure \ref{lc_shortEE}.  
The initial short spike is usually composed of multiple pulses with a total duration of less 
than 2 seconds.  The E.E. lasts from a few tens to a few hundreds of seconds.  We distinguish 
among long GRBs (L-GRBs), S-GRBs and a S-GRBs with an E.E. throughout the paper to investigate the 
prompt emission properties of these three different classes of GRBs.  
Table \ref{statistics} summarizes the statistics of the samples in the catalog 
based on the classifications.  The fractions of L-GRBs, S-GRBs and S-GRBs with E.E. in the 
catalog are 89\%, 8\% and 2\%, respectively. 
The spectral parameters and 
the energy fluences of the initial short spike are reported in Tables \ref{tbl:shortEE_spec} and 
\ref{tbl:shortEE_fluence}.  Note that our definition 
of S-GRBs is whether $T_{90}$ is smaller than 2 s or not.  The GRBs classified as S-GRBs or 
S-GRBs with E.E. in our study are also identified in Table \ref{tbl:bat_summary}.  

\subsection{BAT GRB Position and Sky Locations}

\begin{figure}[p]
\centerline{
\includegraphics[width=12cm,angle=-90]{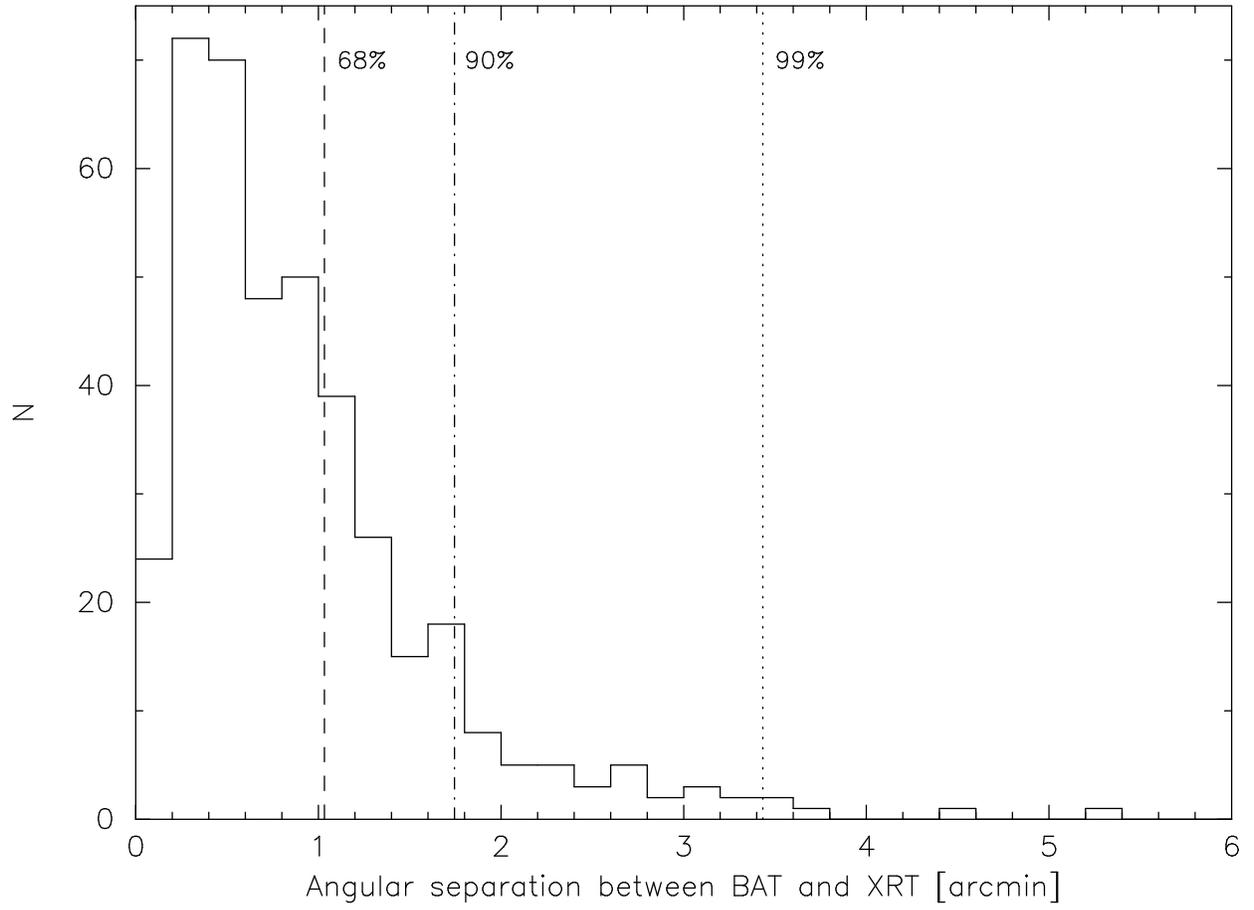}
}
\caption{The histogram of the angular difference between 
the BAT ground position and the XRT position.  
\label{fig:bat_xrt_pos_diff}}
\end{figure}

\begin{figure}[p]
\centerline{
\includegraphics[width=16cm]{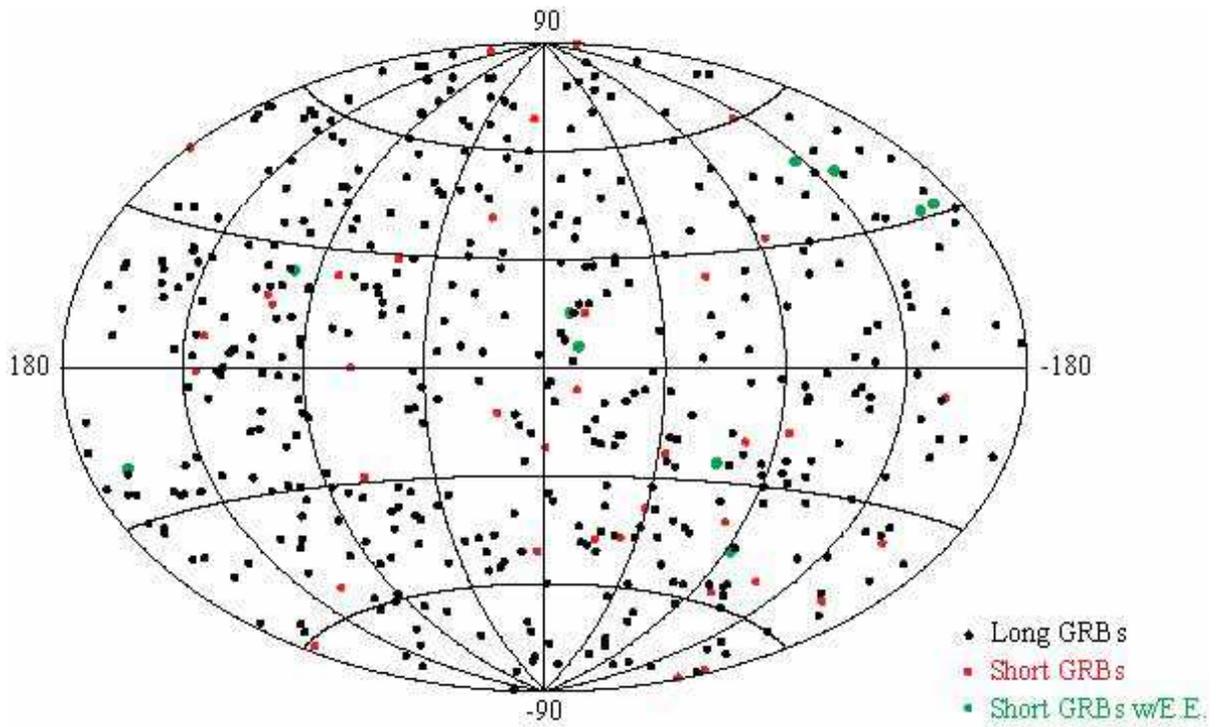}
}
\caption{Sky distribution of the 476 BAT bursts in Galactic coordinates 
with long GRBs in black, short GRBs in red and short GRBs with extended 
emissions in green. \label{bat_grb_skymap}}
\end{figure}

Figure \ref{fig:bat_xrt_pos_diff} shows the angular difference between the BAT 
ground position and the enhanced XRT position \citep{evans2009}.  The BAT ground 
position is within 1.1$^{\prime}$,  1.8$^{\prime}$ and 3.5$^{\prime}$ from the XRT 
position for 68\%, 90\% and 99\% of the bursts, respectively.  
Figure \ref{bat_grb_skymap} shows the sky map of the 476 BAT GRBs in galactic coordinates.  
L-GRBs, S-GRBs and S-GRBs with E.E. are marked in different colors.  


\subsection{Durations and Hardness}

\begin{figure}[p]
\centerline{
\includegraphics[width=12cm,angle=-90]{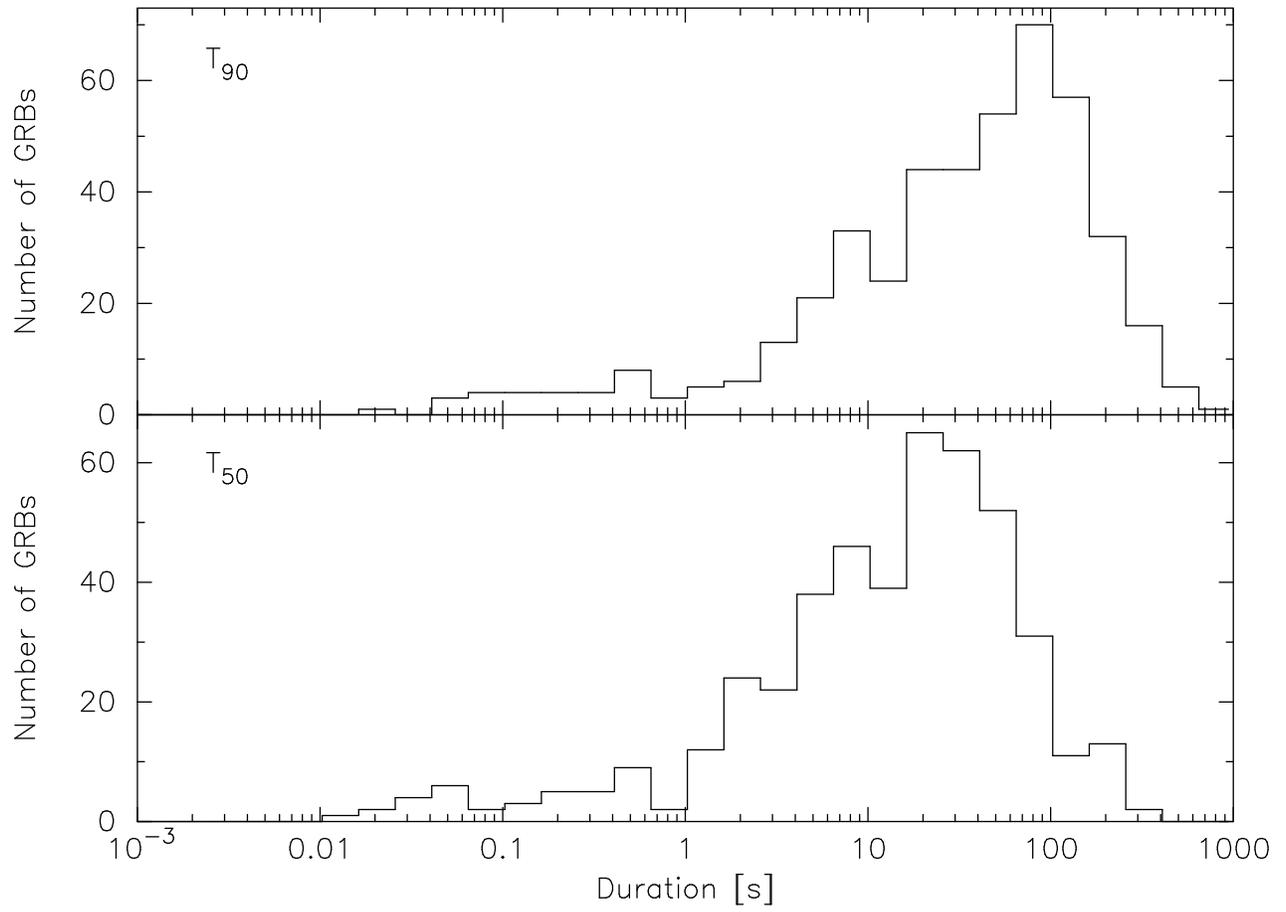}
}
\caption{$T_{90} (top)$ and $T_{50} (bottom)$ distributions from the 
BAT mask-weighted light curves in the 15-350 keV band. \label{bat_t90_t50}}
\end{figure}

\begin{figure}[p]
\centerline{
\includegraphics[width=14cm,angle=0]{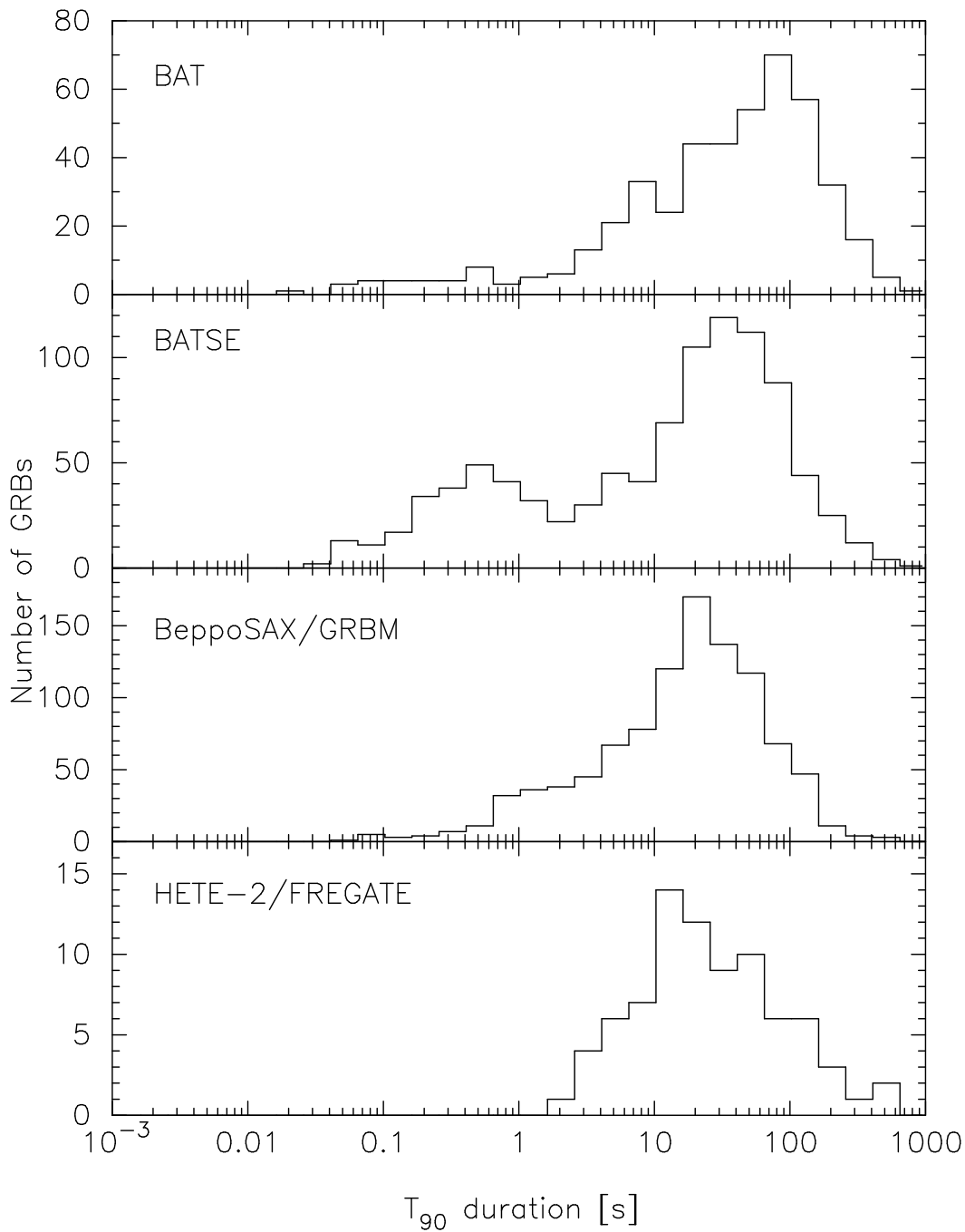}
}
\caption{From top to bottom, $T_{90}$ distribution of BAT from the mask-weighted light curves 
in the 15-350 keV band, BATSE from the light curves in the 50-350 keV band, 
{\it BeppoSAX} from the light curves of the GRBM instrument in the 40-700 keV band, 
and HETE-2 from the light curves of the FREGATE instrument in the 6-80 keV 
band. \label{bat_batse_hete_t90_t50}}
\end{figure}

The histograms of $T_{90}$ and $T_{50}$ in the 
BAT full energy band are shown in Figure \ref{bat_t90_t50}.  The average of the 
BAT $T_{90}$ and $T_{50}$ durations are 71 and 31 s, respectively.  In Figure 
\ref{bat_batse_hete_t90_t50}, we compare the $T_{90}$ distributions of BAT, 
BATSE, {\it BeppoSAX} and {\it HETE}-2.  
The BATSE $T_{90}$ are extracted from the 4B catalog \citep{4b_cat}, 
and they are measured in the 50-300 keV band.  The {\it BeppoSAX} $T_{90}$ 
are extracted from \citet{frontera2009}, and they are measured using the 
light curve of the GRBM instrument in the 40-700 keV band.  
The {\it HETE}-2 $T_{90}$ are extracted from \citet{alex_hete2},  
and they are measured using the light curve of the FREGATE instrument 
in the 6-80 keV band.  There is a clear 
shift in the peak of the L-GRB populations measured with different instruments.  
The peak of L-GRBs $T_{90}$ distribution 
from the BATSE, the {\it BeppoSAX} and the {\it HETE}-2 samples 
are around 10-30 s, whereas the BAT 
distribution is around 70 s.  It is clear from this comparison that the duration 
measurement depends upon the sensitivity of the instruments.  
Another distinct difference in the BATSE distribution compared to that of 
the BAT, the {\it BeppoSAX} and the {\it HETE}-2 distributions is a 
clear bimodality between S-GRBs and L-GRBs 
\citep[e.g.,][]{kouveliotou1993}.  The lack of S-GRBs in imaging instruments 
such as the BAT and {\it HETE}-2 is a result of the larger number of photons needed 
to ``image'' a GRB with these instruments.  
This requirement is usually difficult to achieve for S-GRBs because 
they are usually faint and their emissions are short.  
However, note that BAT has been triggering and 
localizing S-GRBs at a much higher rate than other GRB imaging instruments 
because of its large effective area and its sophisticated flight software.  
As mentioned in \citet{frontera2009}, the lack of S-GRBs in the {\it BeppoSAX} samples 
is likely to be due to the lower efficiency of the trigger system to S-GRBs.  

\begin{figure}[p]
\centerline{
\includegraphics[width=10cm,angle=-90]{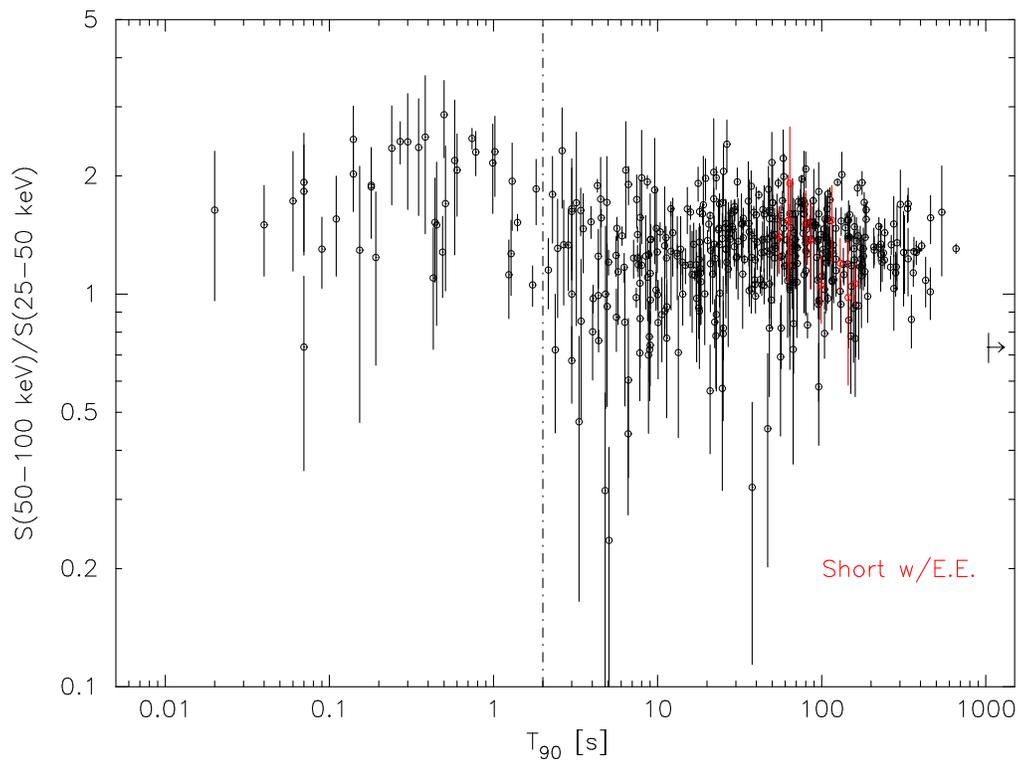}}
\vspace{1cm}
\centerline{
\includegraphics[width=10cm,angle=-90]{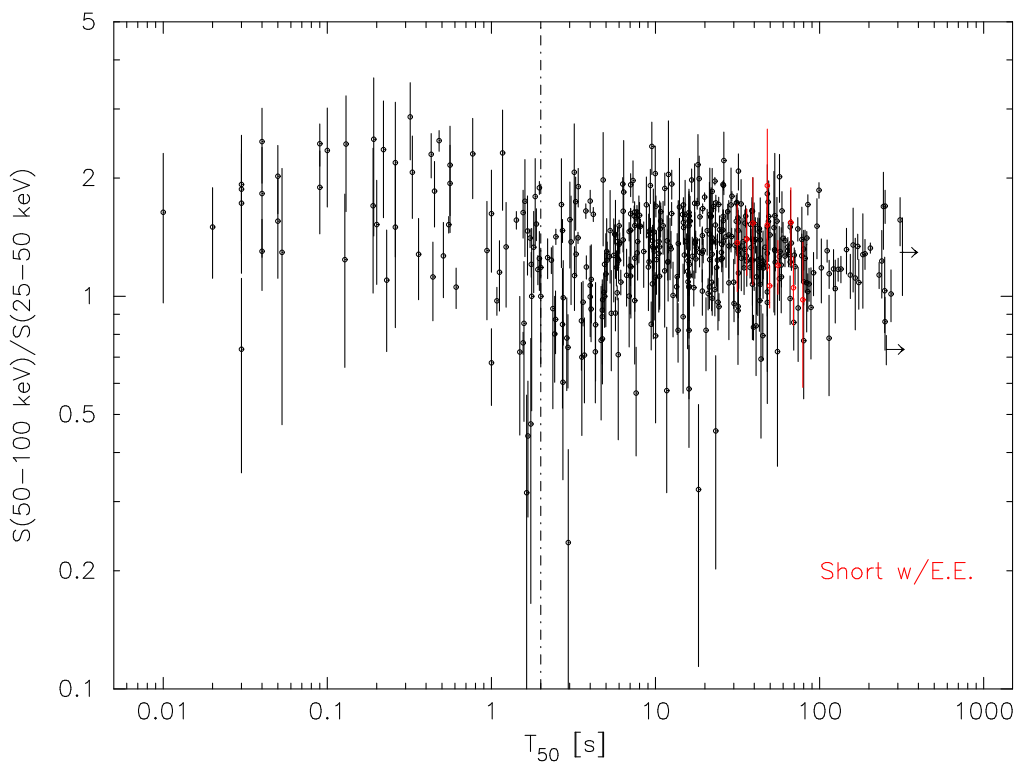}
}
\caption{The fluence ratio between the 50-100 keV and the 25-50 keV bands 
plotted with respect to $T_{90} (top)$ and $T_{50} (bottom)$. Short GRBs with an E.E. 
are shown in red.  The dashed dotted vertical line shows $T_{90}$ = 2 s.  \label{bat_hr32_dur}}
\end{figure}

Figure \ref{bat_hr32_dur} shows the fluence ratio between the 50-100 keV and the 
25-50 keV band versus the $T_{90}$ and $T_{50}$ durations of the BAT GRBs.  
It is clear from the figures 
that there is not a large number of S-GRBs that have soft spectra.  Most of the S-GRBs have 
a fluence ratio of about 2.  On the other hand, the averaged fluence ratio 
of the L-GRBs is 1.3.  
The Kolmogorov-Smironov (K-S) test probability of the fluence ratio 
between L-GRBs and S-GRBs is $8.3 \times 10^{-20}$.    
Based on this comparison, we can conclude that the S-GRBs are generally 
harder than the L-GRBs.  However, note that there is a large overlap in hardness 
between L-GRBs and S-GRBs in the BAT sample.  The S-GRBs with E.E. overlap the L-GRB samples.  

\begin{figure}[p]
\centerline{
\includegraphics[width=12cm,angle=-90]{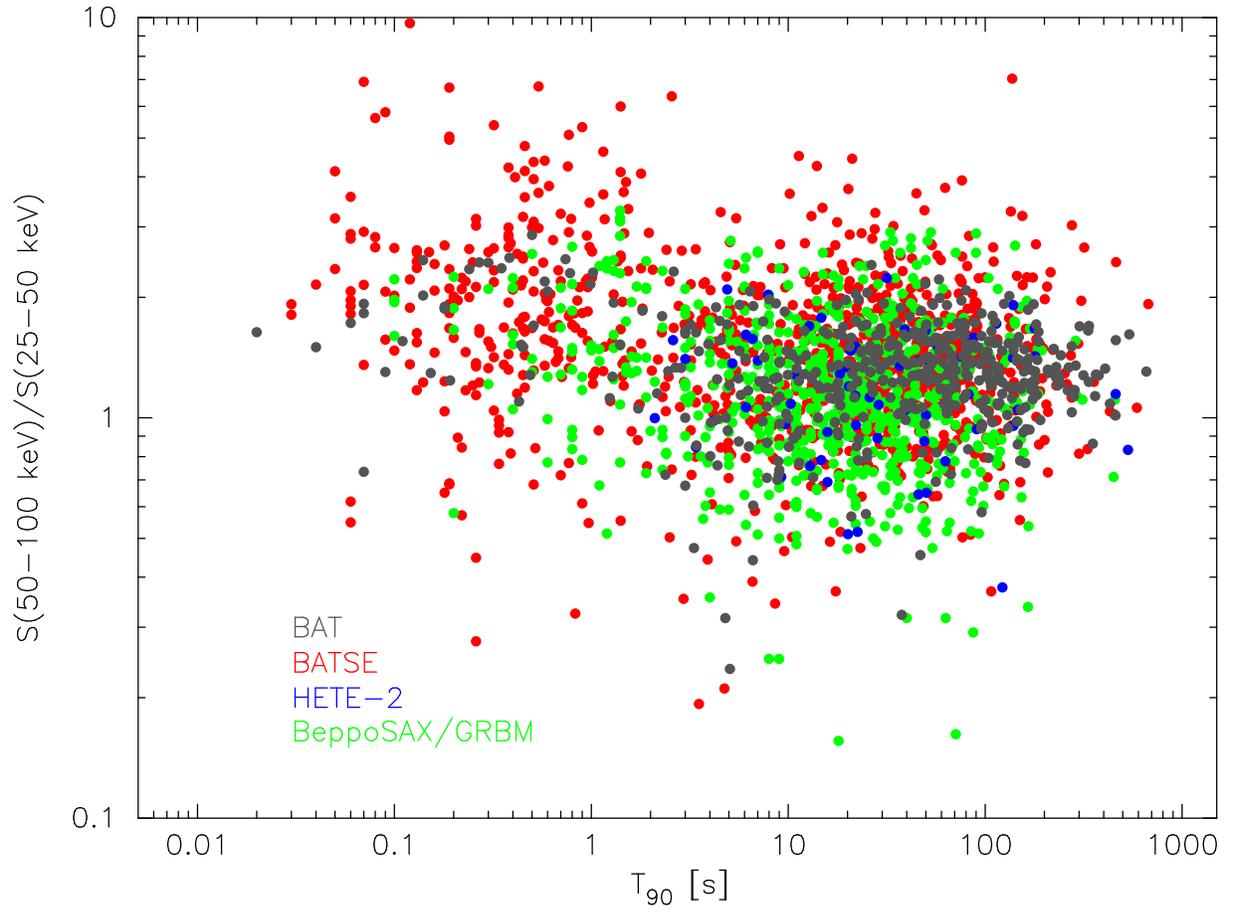}
}
\caption{The fluence ratio between the 50-100 keV and the 25-50 keV 
bands versus $T_{90}$ for BAT (dark gray), BATSE (red), {\it HETE}-2 (blue) and 
{\it BeppoSAX} (green) GRBs. 
\label{comp_hr32_dur}}
\end{figure}

The comparisons in the fluence ratio-$T_{90}$ plane for the BAT, the BATSE, the {\it BeppoSAX} 
and the {\it HETE}-2 GRBs are shown in Figure \ref{comp_hr32_dur}.  Both fluences and $T_{90}$ values 
for the BATSE sample are extracted from the 4B catalog.  For the {\it BeppoSAX} sample, 
we used the best fit simple power-law model in the catalog \citep{frontera2009} to calculate 
the fluence ratios in the 50-100 keV and the 25-50 keV band.  For the {\it HETE}-2 sample, 
we calculated the fluences in those energy bands using the spectral parameters 
reported in \citet{sakamoto2005} and \citet{alex_hete2}.  We only calculated the fluences 
for sources listed with CPL or Band parameters.\footnote{Because of this 
spectral model requirement, we are excluding a large number of X-ray flashes in the 
{\it HETE}-2 sample where a PL is the usual accepted model.}  The $T_{90}$ values of 
the {\it HETE}-2 sample are from \citet{alex_hete2}.  
As seen in Figure \ref{comp_hr32_dur}, the GRB samples of different missions are 
overlaid on each other.  

\subsection{Peak Fluxes and Fluences}

\begin{figure}
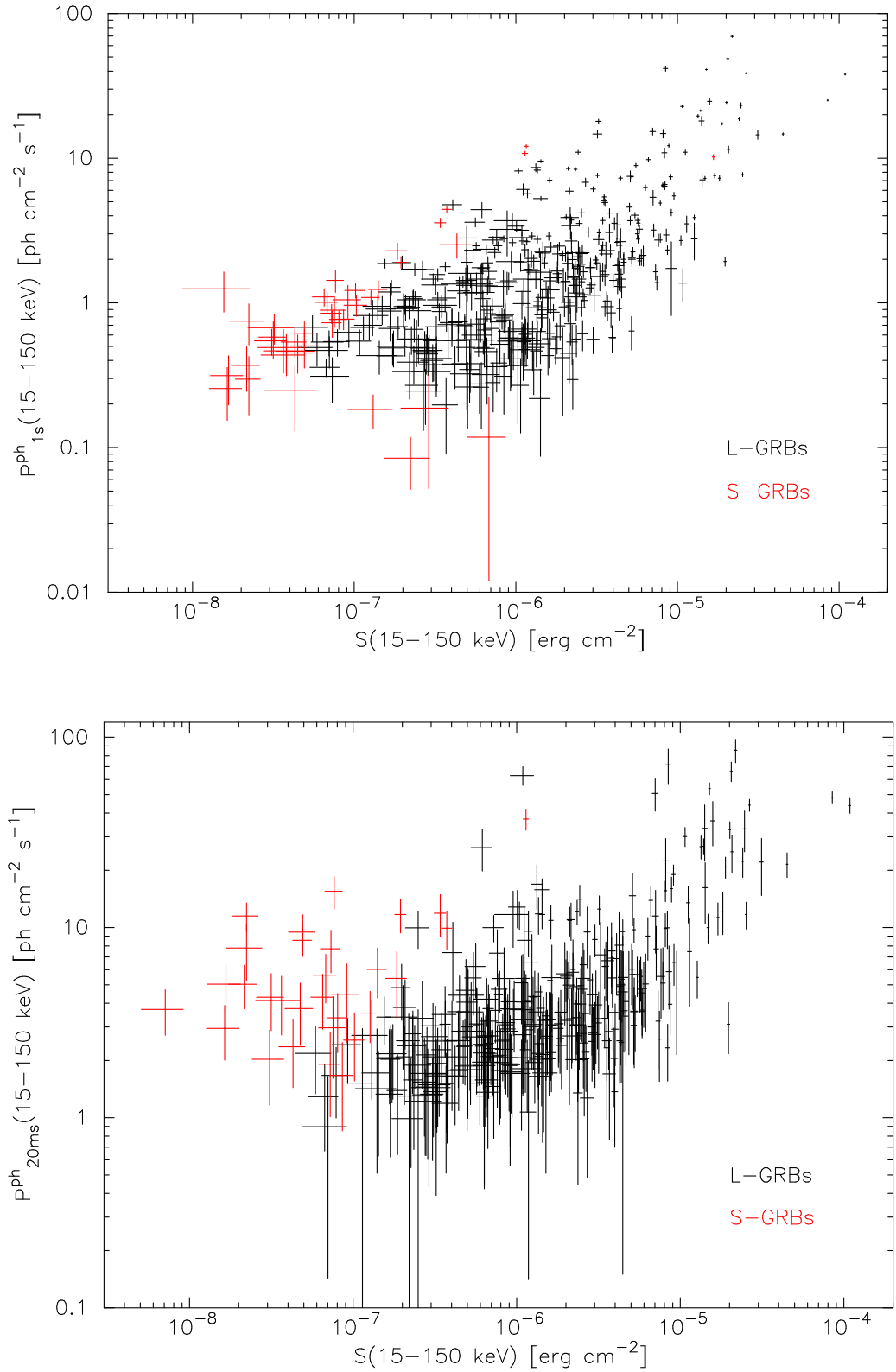

\centerline{
\includegraphics[width=10cm,angle=-90]{fig10a.eps}
}
\vspace{1cm}
\centerline{
\includegraphics[width=10cm,angle=-90]{fig10b.eps}
}
\caption{Distributions of 1 s peak photon flux (top) and 20 ms peak photon 
flux (bottom) in the 15-150 keV band plotted versus energy fluence in the 15-150 keV band.  
\label{peakflux_fluence}}
\end{figure}

Figure \ref{peakflux_fluence} shows the 1-s and the 20-ms peak photon fluxes 
versus the fluence in the 15-150 keV band.  As we showed in the BAT1 catalog, 
there is a positive correlation between peak photon flux and fluence.  
Based on the correlation between the 20 ms peak flux and the 15-150 keV fluence 
(lower panel of Figure \ref{peakflux_fluence}), it is now clear that most of the 
BAT S-GRBs populate a low fluence but high peak flux region.  
For S-GRBs, the 1-s peak flux is systematically lower than the 20-ms peak flux 
because the 1-s time window is usually predominantly larger than the actual S-GRB 
duration used for calculating the flux.  That the S-GRB population has low fluence and high 
peak flux in the BAT sample could be due to the selection effect of the 
imaging requirement in the trigger algorithm (e.g. more detected photons are needed 
to image the source).  

\begin{figure}
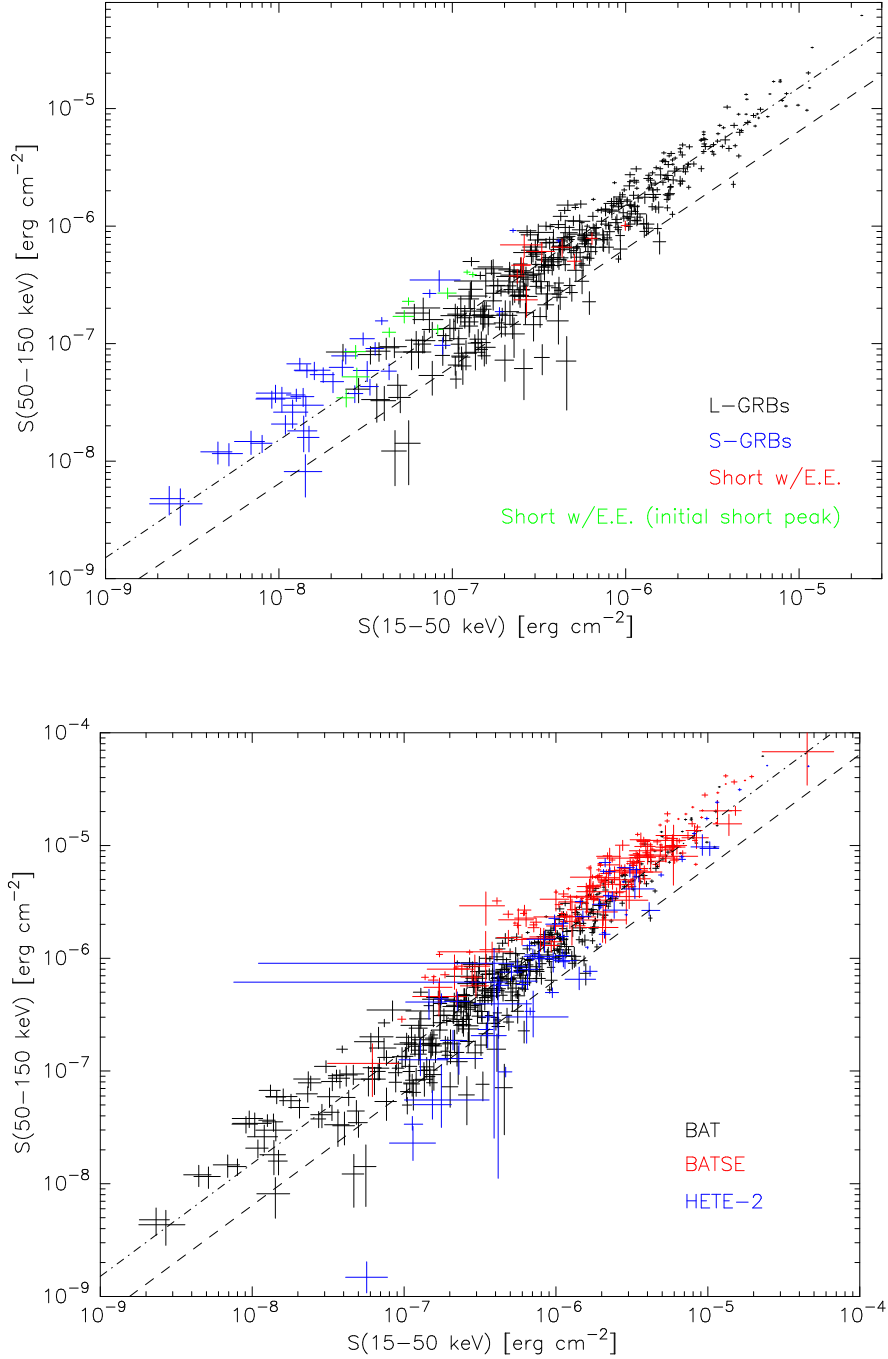

\centerline{
\includegraphics[width=8.5cm,angle=-90]{fig11a.eps}
}
\vspace{1cm}
\centerline{
\includegraphics[width=8.5cm,angle=-90]{fig11b.eps}
}
\caption{$Top$: Distribution of the energy fluence in the 50-150 keV band vs. that in 
the 15-50 keV band for the BAT GRBs.  L-GRBs are in black, S-GRBs are in blue, S-GRBs with 
E.E. are in red, and the initial short spikes of S-GRBs with E.E. are 
in green.  The dashed-dotted line is the distribution expected for the case of the Band function with 
$\alpha = -1$, $\beta=-2.5$, and $\eop = 100$ keV.  The dashed line is for the case of the Band 
function with $\alpha = -1$, $\beta=-2.5$, and $\eop = 30$ keV. $Bottom$: Distribution in the 
same plane as top among different missions.  The BAT sample is in black, the BATSE sample 
is in red, and the HETE-2 sample is in blue.  The definitions of the dashed-dotted and dashed lines 
are same as in the top panel. \label{flu-flu}}
\end{figure}

\begin{figure}[p]
\centerline{
\includegraphics[width=13cm,angle=-90]{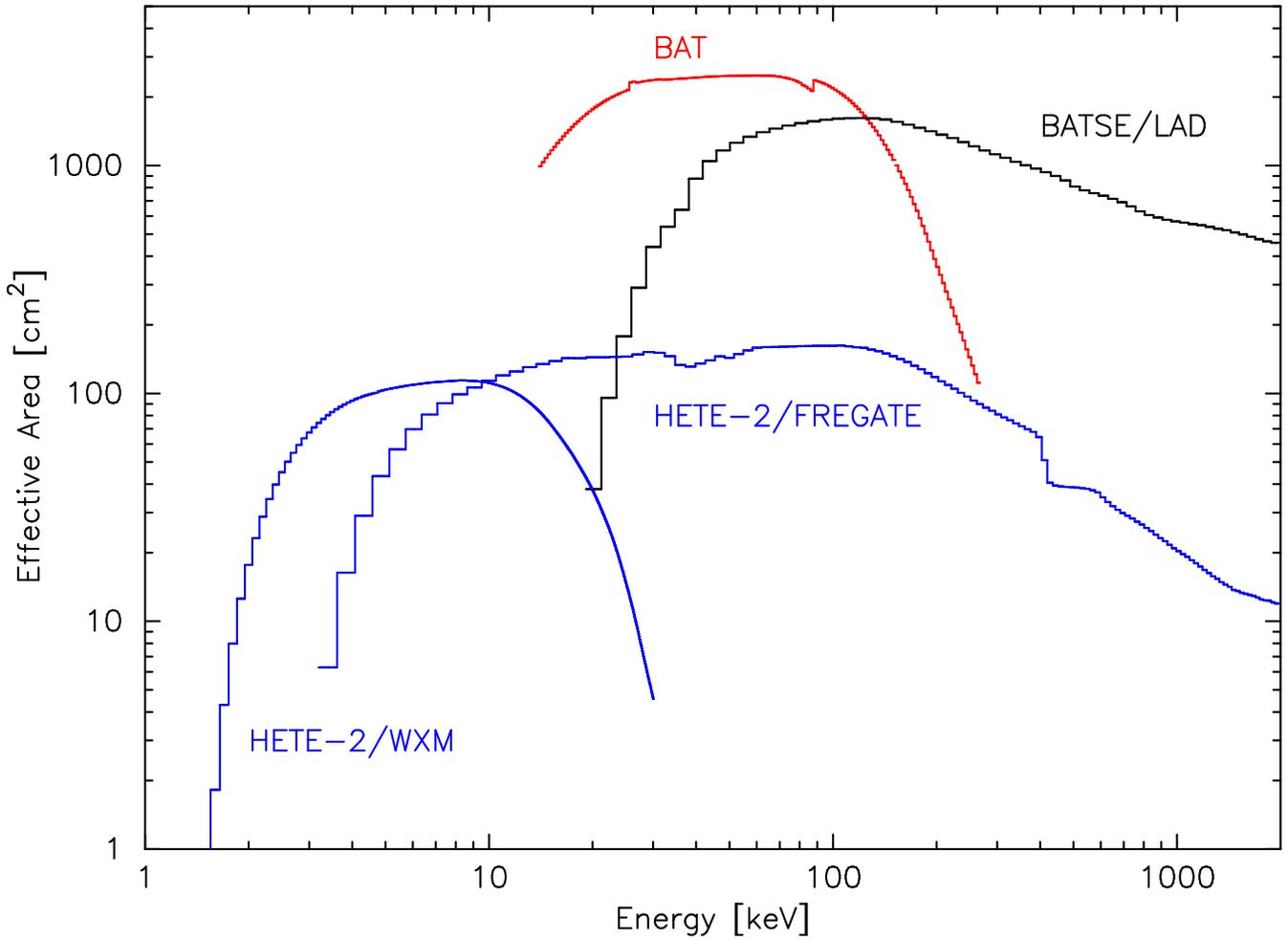}
}
\caption{Comparison of the effective area (on-axis) of BATSE/LAD (from the BATSE energy 
response file of trigger \#1123 at the CGRO/BATSE Gamma-Ray Burst Catalog: http://heasarc.gsfc.nasa.gov/W3Browse/cgro/batsegrb.html), 
Swift/BAT and HETE-2 (WXM; \citet{shirasaki2003} and FREGATE; \citet{atteia2003}).  
\label{effarea}}
\end{figure}

The fluence in the 50-150 keV band versus that in the 
15-50 keV band for the BAT GRBs is shown in the top panel of Figure \ref{flu-flu}.  
On this figure we also indicate the distribution expected for a 
Band function with a low energy photon index of $-1$ and a high energy photon 
of $-2.5$.  Comparing to these lines we see that 43\% of the BAT GRBs have $\eop$ $>$ 100 keV , 
50\% of the BAT GRBs have 30 keV $<$ $\eop$ $\leq$ 100 keV, 
and 7\% of the BAT GRBs have $\eop$ $\leq$ 30 keV.  
The S-GRBs generally have low fluences with 
hard spectra and a small overlap with the L-GRB properties.  On the other hand, 
the S-GRBs with E.E. lie in the same region as the L-GRBs.  However, the initial 
short spikes of the S-GRBs with E.E. have similar characteristics to the S-GRBs.  
The bottom of Figure \ref{flu-flu} 
compares the BAT, the BATSE, and the {\it HETE}-2 GRBs in the same fluence-fluence plane.  The 
fluences for the BATSE sample and the {\it HETE}-2 are calculated using the best fit 
spectral parameters of a CPL model and a Band function as reported in \citet{kaneko2006} 
and \citet{alex_hete2}.  The majority of the BATSE bursts have high fluences with hard spectra.  
The {\it HETE}-2 and the BAT samples have a similar characteristics in the fluences.  
The similar number of GRBs in the range $\eop$ $>$ 100 keV and 30 keV $<$ $\eop$ 
$\leq$ 100 keV is consistent with the {\it HETE}-2 GRB sample \citep{sakamoto2005}.  The 
systematically smaller number of GRBs with a soft spectrum ($\eop$ $\leq$ 30 keV) in the 
BAT is likely due to a lack of sensitivity below 15 keV for the BAT.
Differences in BAT, {\it HETE}-2 and BATSE are very likely due to the 
different response and trigger energy as shown in Figure \ref{effarea}.  

\subsection{Time-averaged Spectral Parameters}

\begin{figure}[p]
\centerline{
\includegraphics[width=14cm,angle=0]{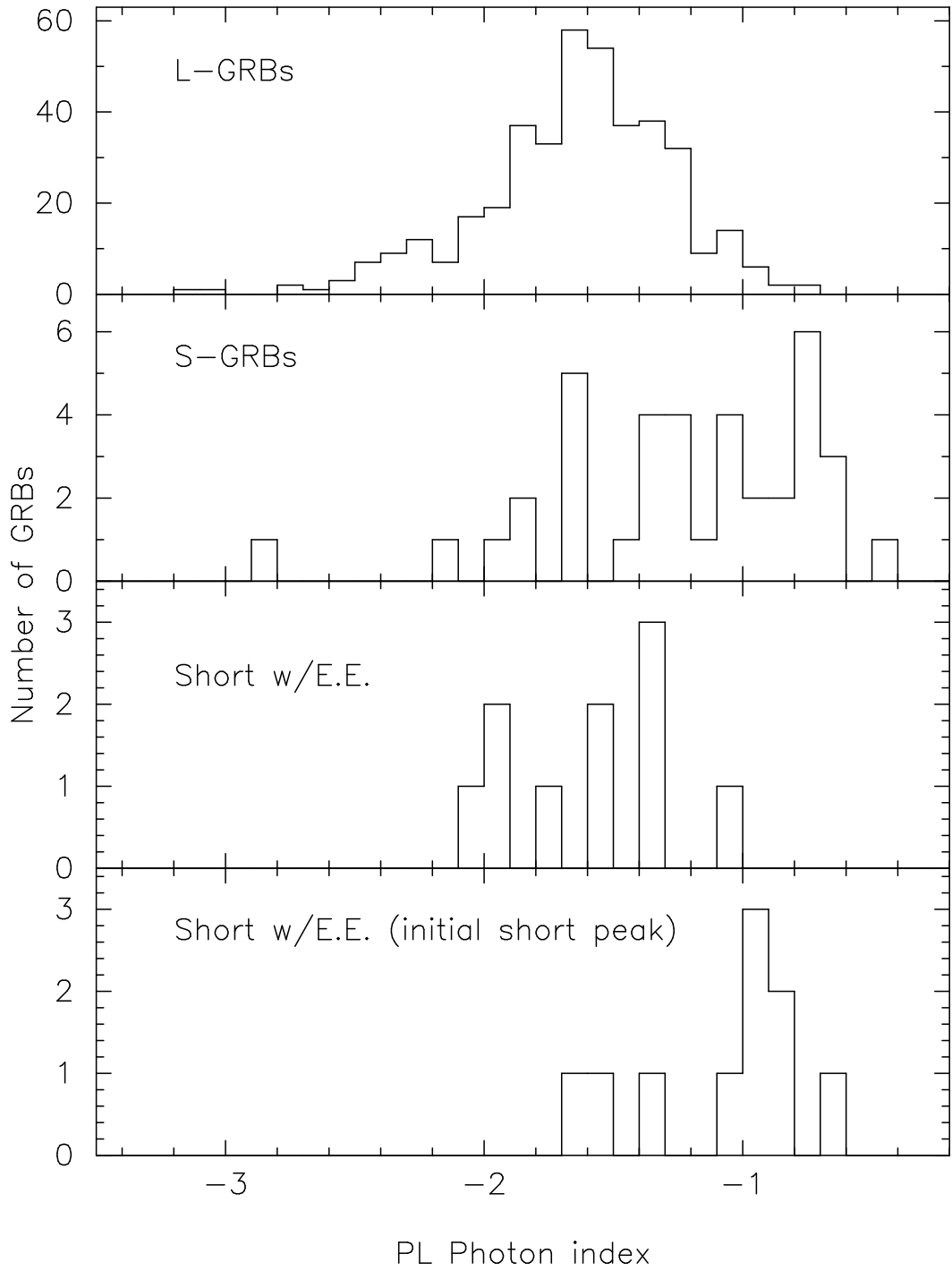}
}
\caption{From top to bottom: Histograms of the BAT time-averaged photon index for a PL fit 
for the L-GRBs, the S-GRBs, the S-GRBs with E.E., and the initial 
short spikes of the S-GRBs with E.E. \label{bat_timeave_pl_hist}}
\end{figure}

Figure \ref{bat_timeave_pl_hist} shows the histograms of the BAT time-averaged photon index 
from a PL fit for L-GRBs, S-GRBs, S-GRBs with E.E. and the initial short spikes of the S-GRBs 
with E.E.  A Gaussian fit to the histogram of 
PL photon indices of L-GRBs shows a peak at $-1.6$ with a $\sigma$ of 0.3.  This BAT photon 
index based on a PL fit is systematically steeper than the typical low-energy photon index $\alpha$ 
($\sim-1$) and also shallower than the high-energy photon index $\beta$ ($\sim-2.5$) based on a 
Band function fit 
\citep[e.g.,][]{kaneko2006}.  As demonstrated in the detailed spectral simulation study of 
\citet{ep_gamma}, the distribution of the BAT photon index in a PL fit reflects the fact 
that more than half of $\eop$ in BAT GRBs are located within the BAT energy range (15-150 keV).  
This is consistent 
with the discussion in section 4.4 that 50\% of $\eop$ in the BAT GRBs are located 30 keV 
$<$ $\eop$ $\leq$ 100 keV assuming typical spectral parameters from a fit to a Band function.  
The histogram of PL photon indices of S-GRBs has a shift toward a shallower index 
compared to that of the L-GRBs.  The time-averaged PL photon index for the S-GRBs is $-1.2$.  
We used a K-S test to check whether the L-GRBs and the S-GRBs are 
drawn from the same population.  The K-S test statistic of 10$^{-7}$ strongly indicates that 
the time-averaged PL photon index distributions of the L-GRBs and S-GRBs are different.  
We confirm in the BAT GRB sample that S-GRBs in general have a harder spectrum 
than that of L-GRBs \citep[e.g.,][]{kouveliotou1993,ghirlanda2009}.  
Although the number in the former sample is limited, the time-averaged PL photon index of S-GRBs with E.E. is 
consistent with the L-GRBs.  On the other hand, the PL photon index of the initial short spikes of the 
S-GRBs with E.E. is much more consistent with that of the S-GRBs.  

\begin{figure}[p]
\centerline{
\includegraphics[width=12cm,angle=-90]{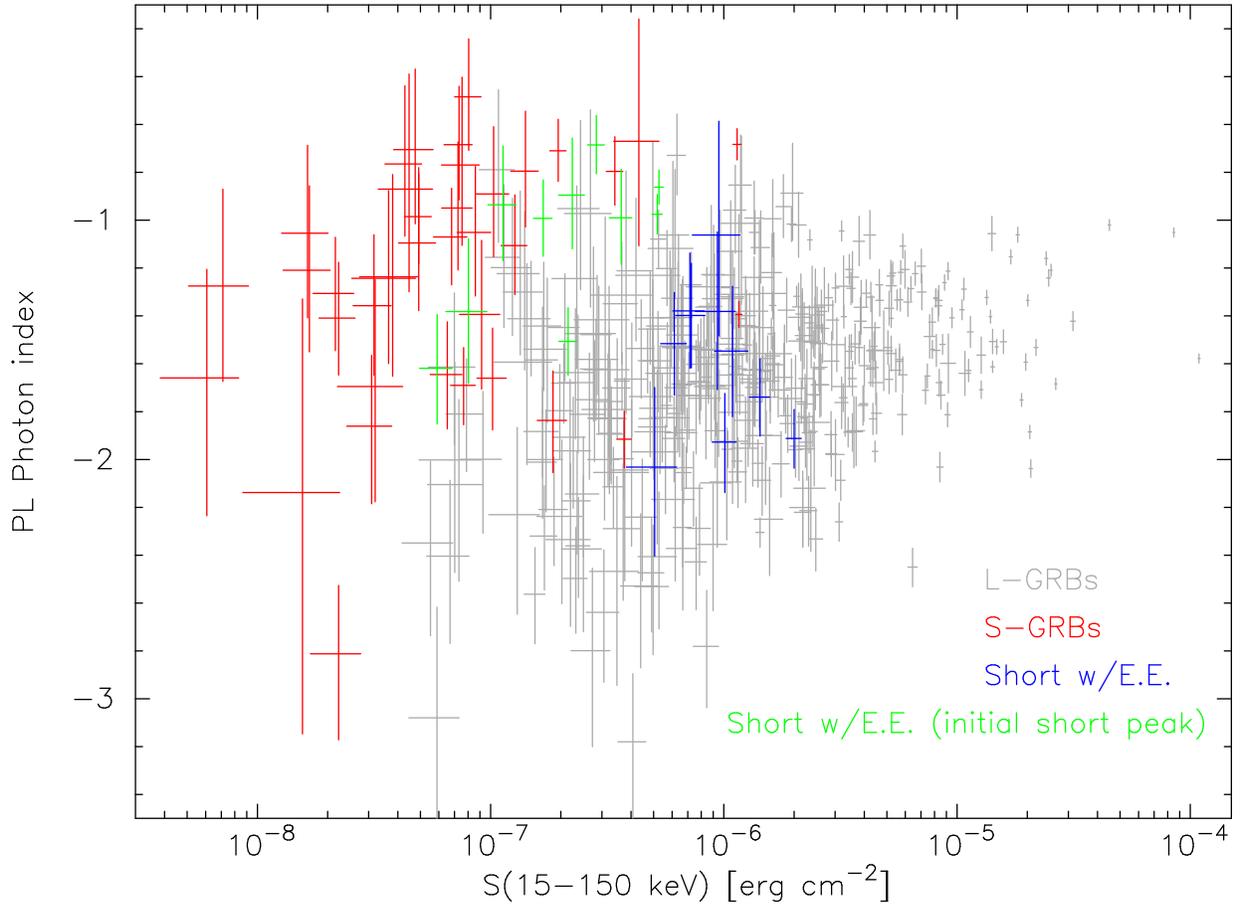}
}
\caption{Distribution of the BAT PL photon index vs. the energy fluence in 
the 15-150 keV band for the L-GRBs (light gray), the S-GRBs (red), the S-GRBs 
with E.E. (blue) and the initial short spikes of the S-GRBs with E.E.(green).  
\label{bat_timeave_pl_phindex_vs_s15_150}}
\end{figure}

Figure \ref{bat_timeave_pl_phindex_vs_s15_150} shows the BAT time-averaged 
photon index from a PL fit versus the fluence in the 15-150 keV band.  Similar to the 
trend in Figure \ref{flu-flu} and \ref{bat_timeave_pl_hist}, 
the S-GRBs are located in a lower fluence and a harder 
spectral region compared to the L-GRBs.  The time-averaged properties of the S-GRBs with E.E. 
overlap with the L-GRBs in the PL photon index - fluence plane.  The initial short spikes 
of the S-GRBs with E.E. seem consistent with the properties of S-GRBs in this plane, 
but lie in a systematically higher fluence region than the S-GRBs.  

\begin{figure}[p]
\centerline{
\includegraphics[width=12cm,angle=-90]{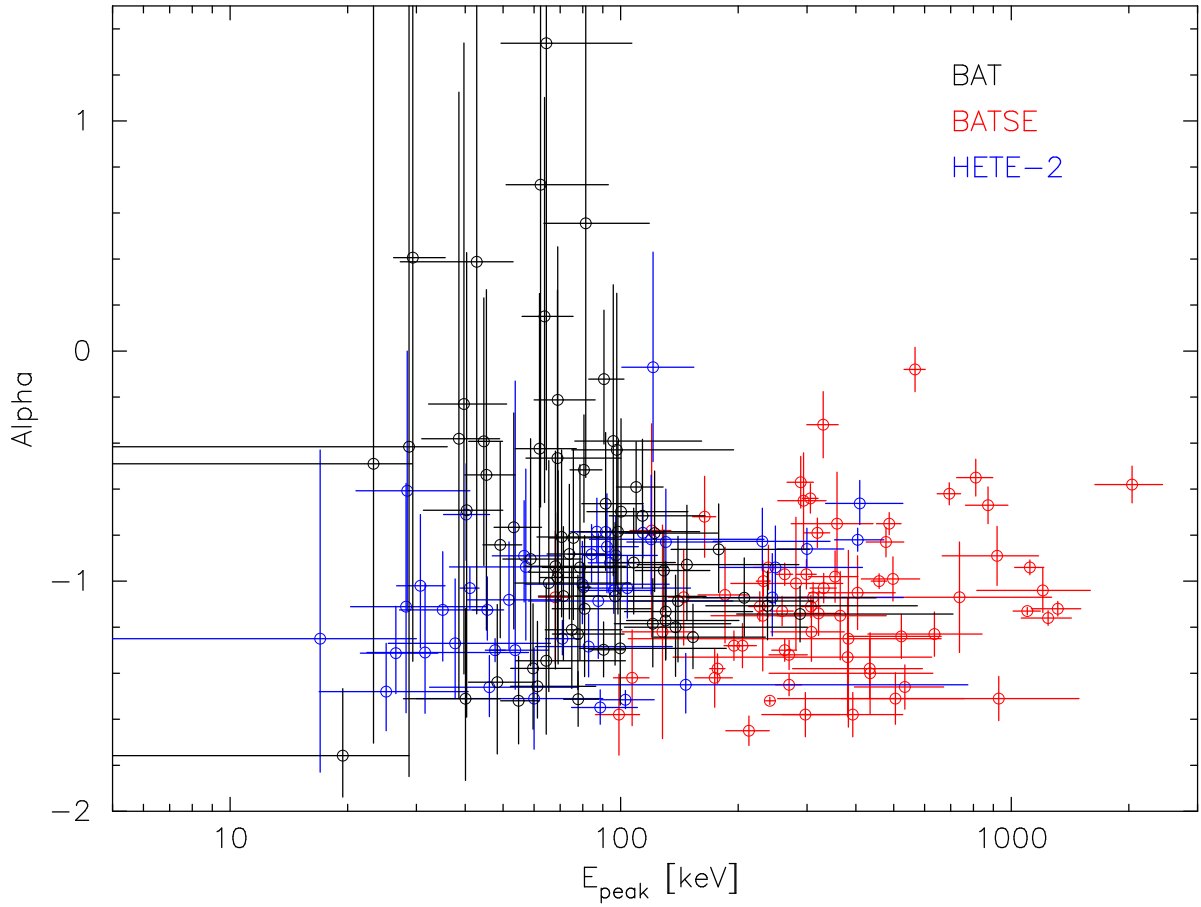}
}
\caption{Distributions of the photon index $\alpha$ and $\ep$ in 
a CPL fit for the BAT (black), the BATSE 
(red) and the HETE-2 (blue) GRBs. \label{timeave_cpl_alpha_ep}}
\end{figure}

\begin{figure}
\centerline{
\includegraphics[width=12cm,angle=-90]{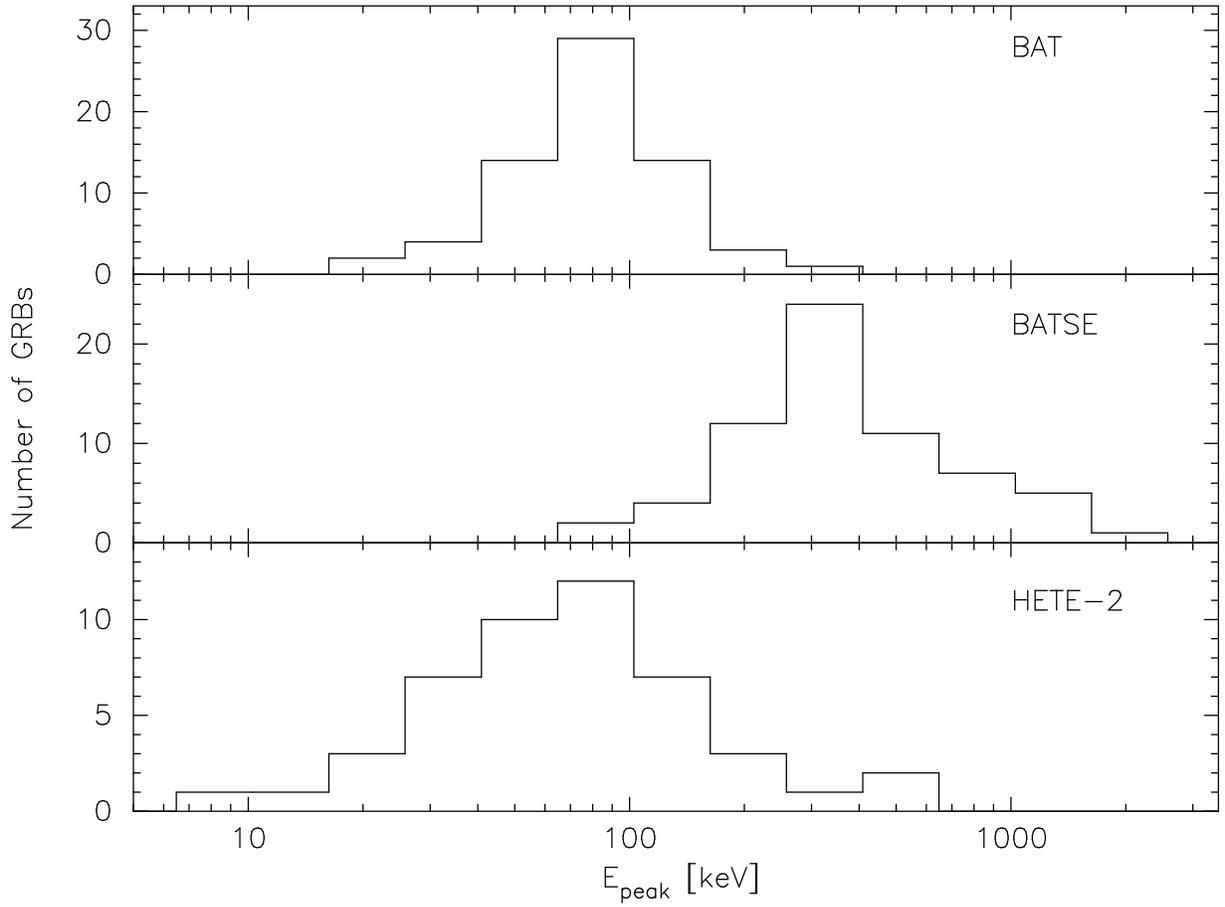}
}
\caption{Histograms of $\ep$ in a CPL fit for the BAT (top), the BATSE 
(middle) and the HETE-2 (bottom) GRBs.  \label{timeave_cpl_ep_hist}}
\end{figure}

The low-energy photon index $\alpha$ and $\eop$ from a CPL fit are shown for 
the BAT, the BATSE and the {\it HETE}-2 GRBs in Figure \ref{timeave_cpl_alpha_ep}.  77 out of 
456 time-averaged spectra (17\%) in the BAT sample show a significant improvement 
of $\chi^{2}$ in a CPL fit over a PL fit ($\Delta \chi^{2} >$ 6).  The BATSE 
and the {\it HETE}-2 spectral parameters from a CPL fit are taken from \citet{kaneko2006} 
and \citet{alex_hete2} respectively.  The low-energy photon index $\alpha$ is consistent 
among all the instruments.  However, most of the $\eop$ 
of the BATSE GRBs are larger than 100 keV, whereas the majority of $\eop$ in 
the BAT and the {\it HETE}-2 GRBs 
are less than 100 keV.  This is more clearly represented in the histograms of $\eop$ shown in 
Figure \ref{timeave_cpl_ep_hist}.  The Gaussian fits to the log-normal $\eop$ histograms of the 
BAT, the BATSE and the {\it HETE}-2 samples 
reveal the peaks to be at 79 
keV, 320 keV, and 65 keV with $\sigma$ in $\log(\eop)$ of 0.18, 0.22, and 0.31 respectively.  
The K-S test probabilities in the log-normal $\eop$ distributions 
for the BAT and the BATSE bursts, and the BAT and the {\it HETE}-2 
bursts are $9 \times 10^{-21}$ and 0.15 respectively.  
As clearly seen in these histograms, although the peaks and the widths of the $\eop$ distributions 
differ among the instruments, they are overlapping.  Moreover, there is no sign of a bimodal 
$\eop$ distribution in the measurements from a single instrument.  
Therefore, we may conclude that ``true'' $\eop$ has a single broad log-normal 
distribution.  The difference of the $\eop$ distributions among the GRB instruments 
is very likely due to an instrumental selection effect.  

\subsection{Time-resolved Spectral Parameters}

\begin{figure}
\centerline{
\includegraphics[width=12cm,angle=-90]{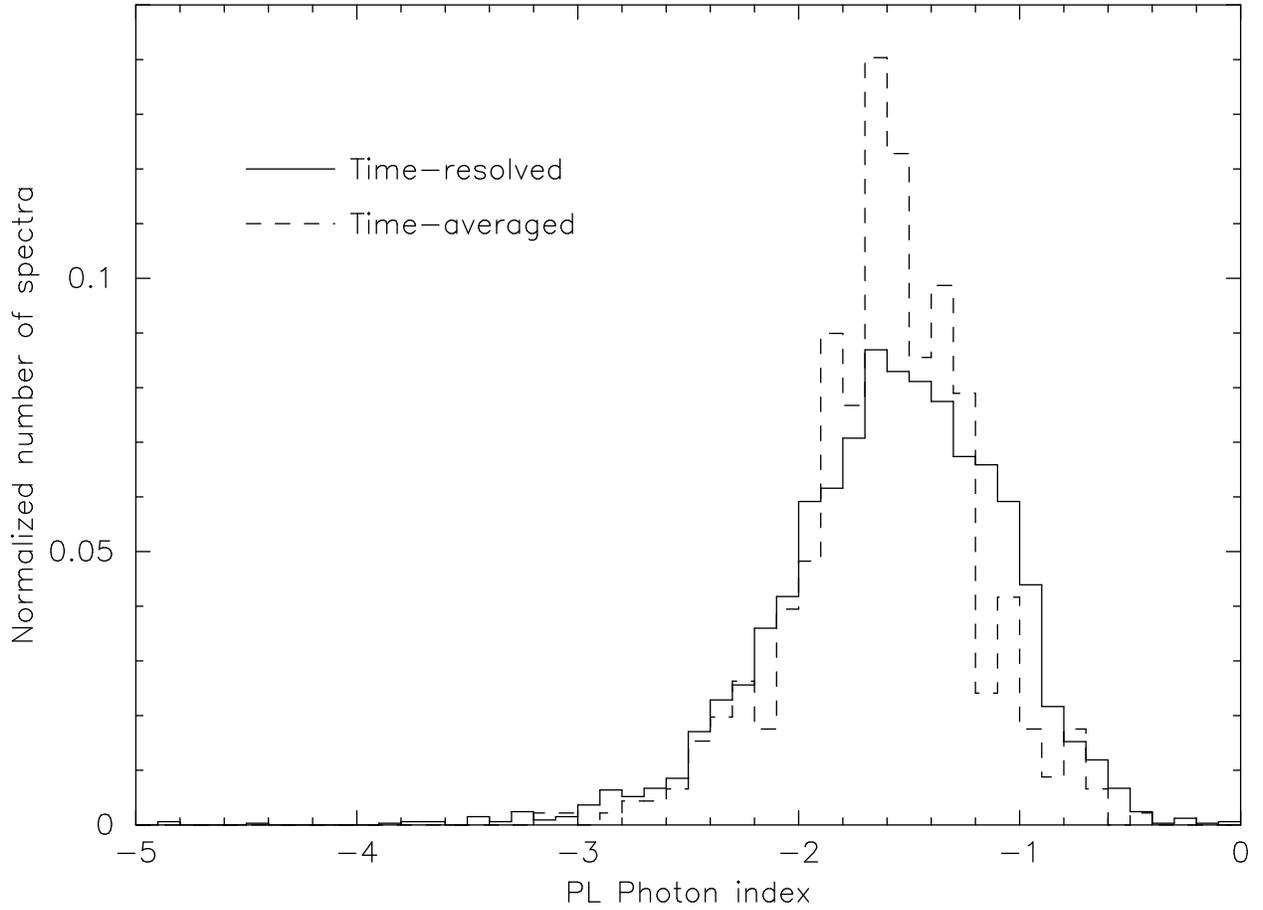}
}
\caption{Histograms of the BAT photon index in a PL fit for 
the time-resolved (solid) and the time-averaged (dash) spectra.  The 
histograms are normalized by the total number of spectra in each sample. \label{time_resolved_pl_phindex_hist}}
\end{figure}

We have chosen 3284 out of 3323 time-resolved spectra, for further study, based on the goodness of fit 
($\chi^{2}_{\nu} < 2$) and also the constraints on the spectral parameters.  
Figure \ref{time_resolved_pl_phindex_hist} compares the distributions of the photon 
index in a PL fit between the time-resolved and the time-averaged spectral samples.  
Fitting a Gaussian to each of the histograms gives, for the time-resolved sample a peak 
at $-1.53$ with $\sigma$ of 0.47, and for the time-averaged sample a peak at $-1.57$ with $\sigma$ of 0.32.  
The K-S test probability for the time-resolved 
and the time-averaged PL photon index distribution is $10^{-4}$.  Therefore, 
there is a marginal difference in the photon index of a PL fit between the time-resolved and 
the time-averaged spectra, especially in the widths of their distributions.  

Figure \ref{time_resolved_cpl_ep_hist} shows the differences in $\eop$ from the CPL fit 
between the time-resolved and the time-averaged spectra.  472 out of 3284 time-resolved 
spectra show 
a significant improvement of $\chi^{2}$ in a CPL fit over a PL fit ($\Delta \chi^{2} > 6$).  
The Gaussian fits to these log-normal $\eop$ histograms show a peak at 68 keV 
with $\log(\eop)$ of 0.23 for the time-resolved sample and a peak at 77 keV with $\log(\eop)$ of 0.19 
for the time-averaged spectra.  We found a K-S probability 
of 0.04 for the comparison in $\eop$ distributions between the time-resolved and the 
time-averaged spectra.  Therefore, there is no significant difference in $\eop$ based 
on a CPL fit between the BAT time-resolved and the time-averaged spectra.  

\begin{figure}
\centerline{
\includegraphics[width=12cm,angle=-90]{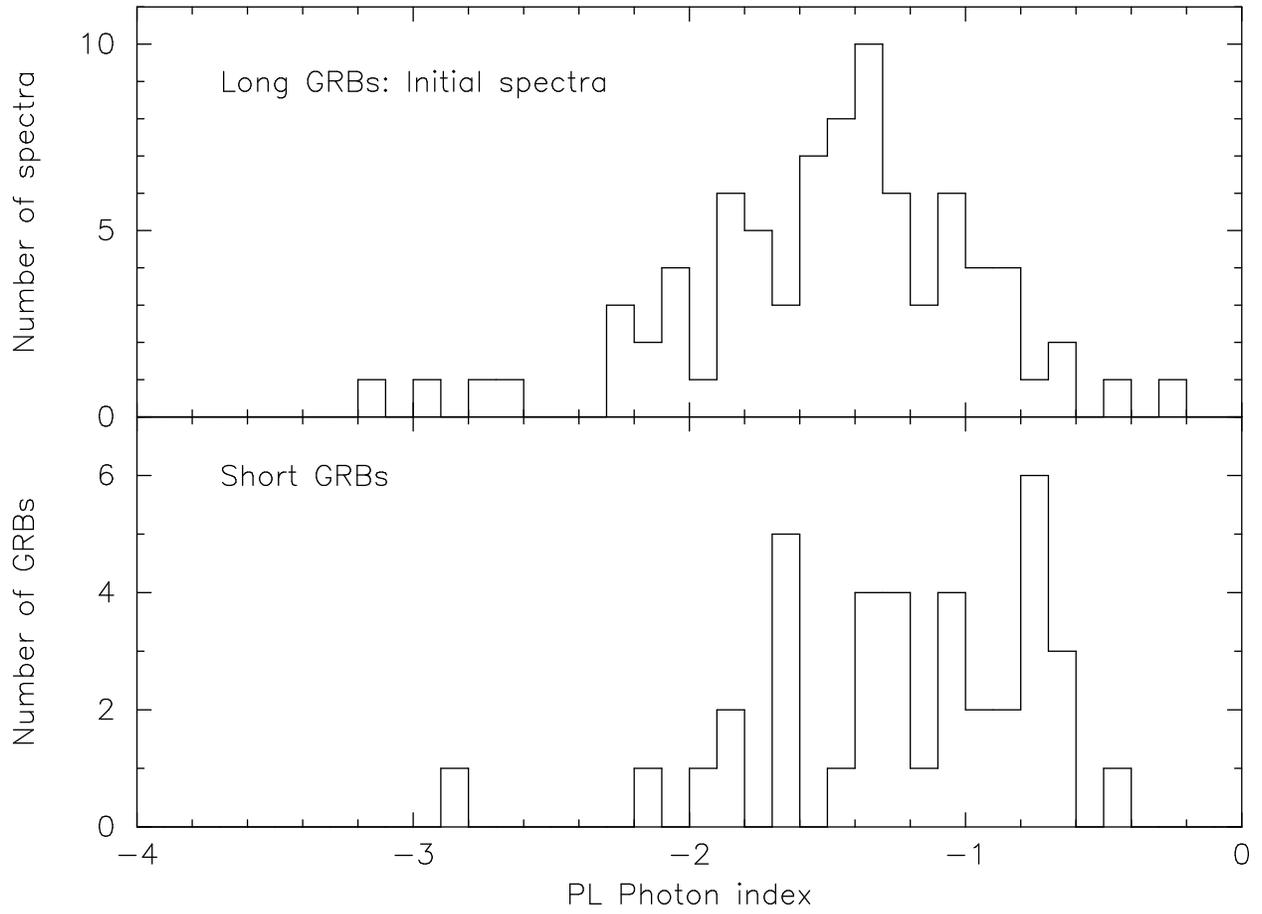}
}
\caption{Histograms of the BAT photon index in a PL fit for 
initial spectra of long GRBs (top) and for short GRBs (bottom).   
\label{time_resolved_initial}}
\end{figure}

\citet{ghirlanda2009} suggested a possible resemblance between the low-energy photon index 
$\alpha$ of the S-GRBs and the initial part of the spectrum of the L-GRBs in the BATSE GRBs.  
Figure \ref{bat_timeave_pl_hist} shows a significant difference 
in the time-averaged photon index for a PL fit in the BAT data between the L-GRBs and the S-GRBs.  
Therefore, it is worth investigating this difference 
using our time-resolved spectral results.  Figure \ref{time_resolved_initial} 
shows the histograms of the BAT photon index in a PL fit for the initial 
spectra of the L-GRBs and the time-averaged photon index of the S-GRBs.  
The initial spectrum is chosen to be the first spectrum of the time-resolved spectra for 
each burst.  
The K-S test probability 
comparing the initial PL photon indices of the L-GRBs and the time-averaged photon indices 
of the S-GRBs is 0.02.  
Therefore, the K-S test shows no clear indication that the PL photon 
indices in the initial intervals of the L-GRBs and the time-averaged PL photon indices of 
the S-GRBs are drawn from the same parent population in the BAT data.  

\begin{figure}
\centerline{
\includegraphics[width=12cm,angle=-90]{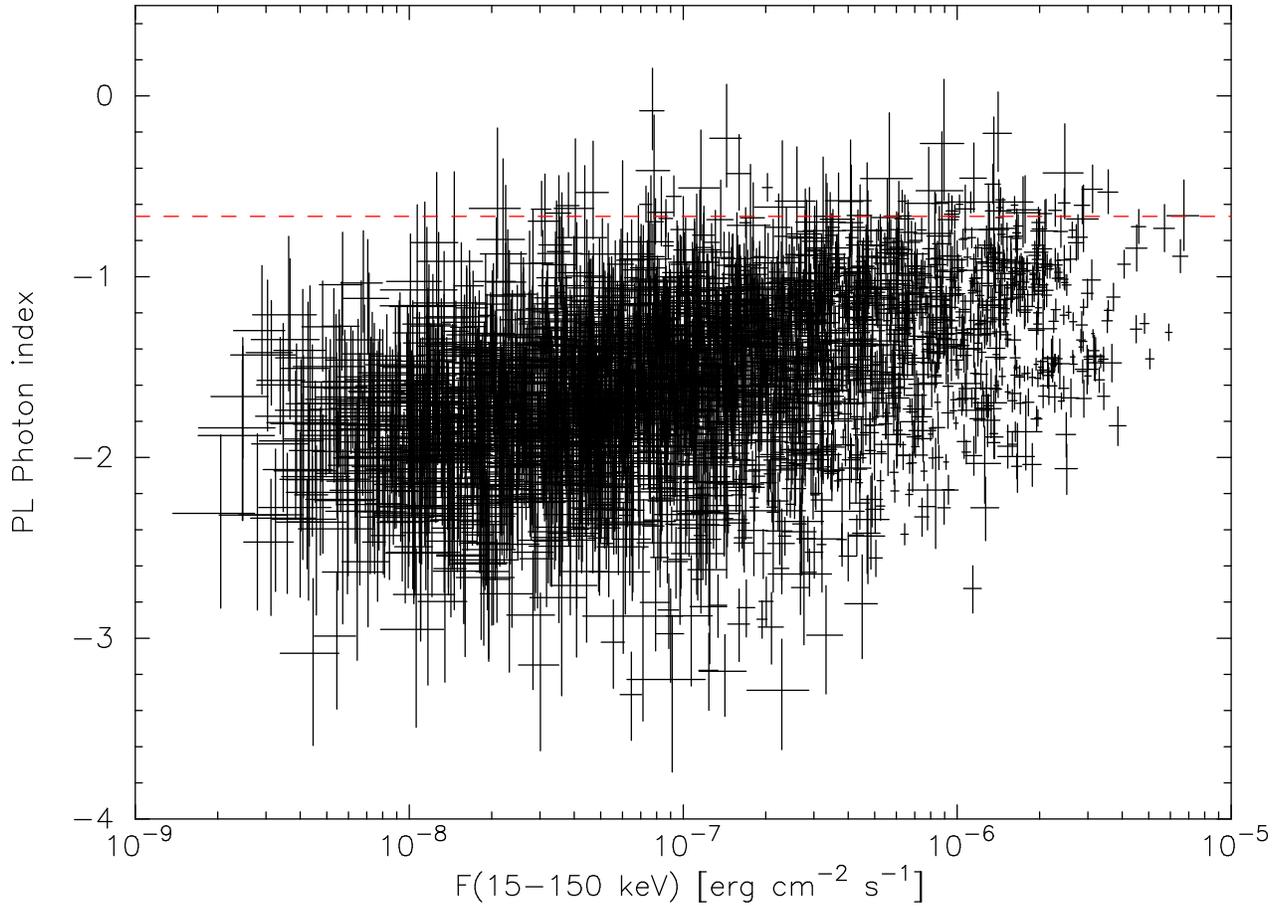}
}
\caption{Distribution of the BAT photon index in a PL fit vs. energy flux 
in the 15-150 keV band.  The red dashed line shows the photon index of -2/3. 
The total number of time-resolved spectra in this plot is 2968.
\label{time_resolved_pl_phindex_vs_flux}}
\end{figure}

\begin{figure}
\centerline{
\includegraphics[width=12cm,angle=-90]{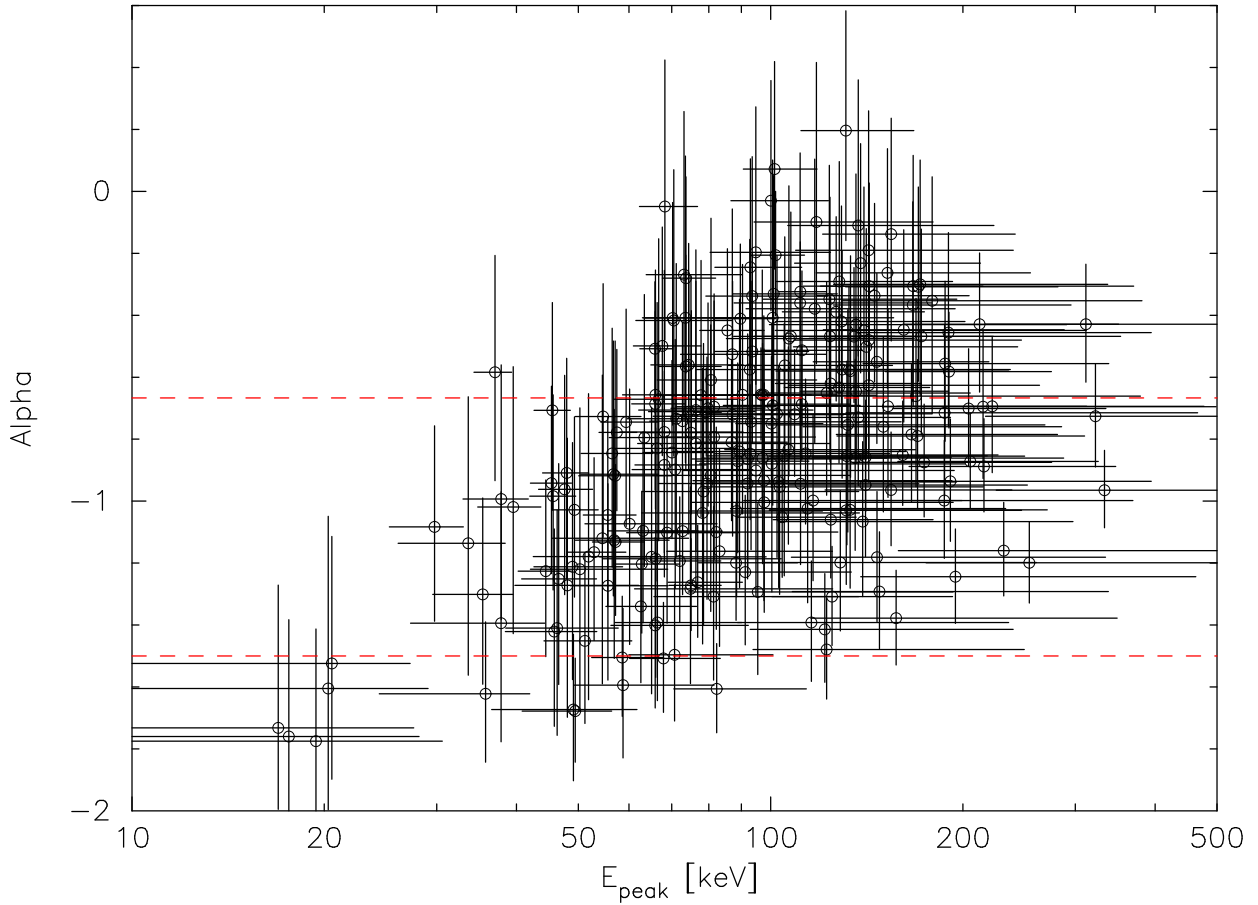}
}
\caption{Distribution of the BAT photon index $\alpha$ versus $\ep$ in a CPL 
fit.  The red dashed lines show the photon index region from $-3/2$ to $-2/3$.
The total number of time-resolved spectra in this plot is 234.  \label{time_resolved_cpl_phindex_vs_ep}}
\end{figure}

\begin{figure}[p]
\centerline{
\includegraphics[width=6cm,angle=0]{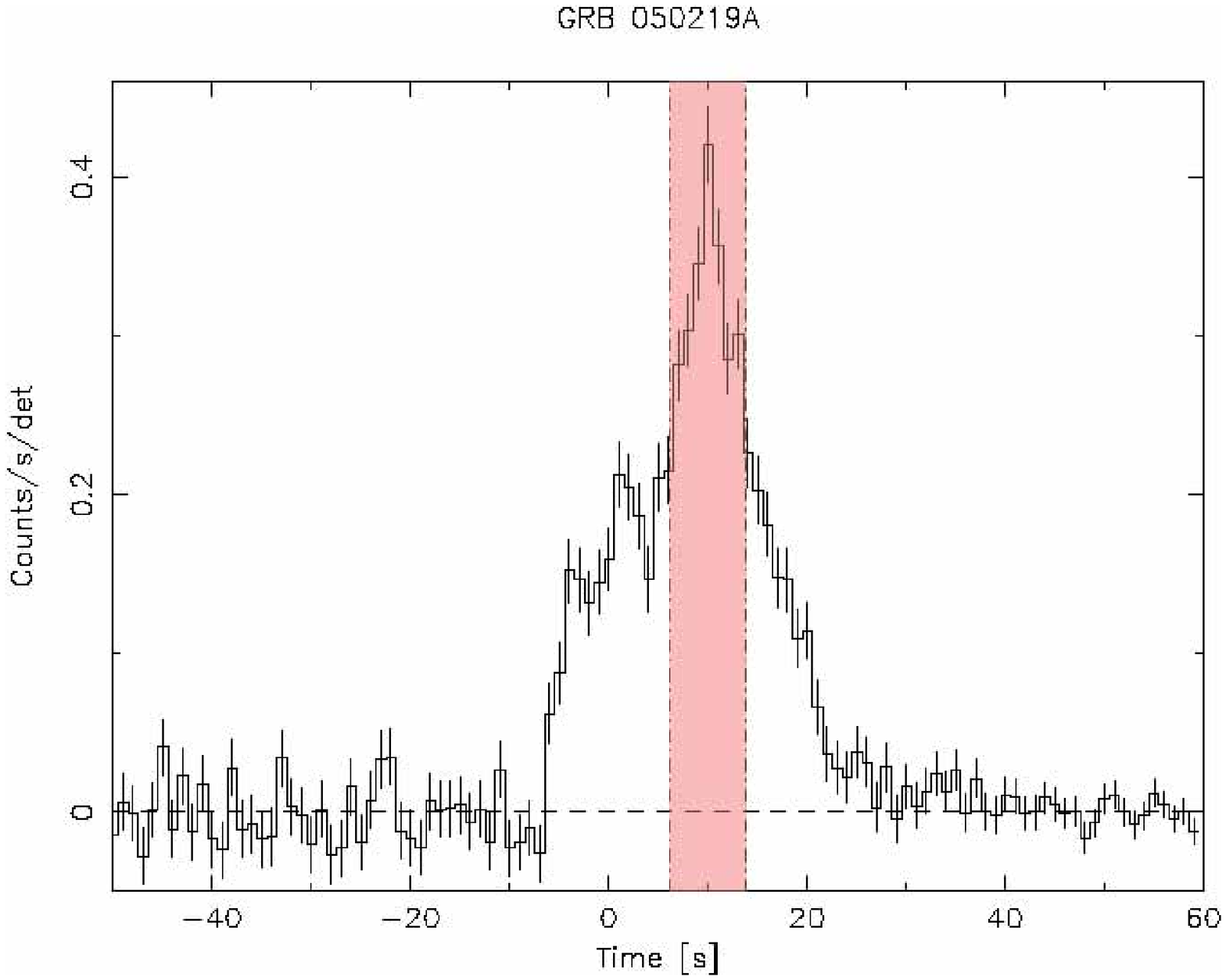}
\hspace{0.5cm}
\includegraphics[width=6cm,angle=0]{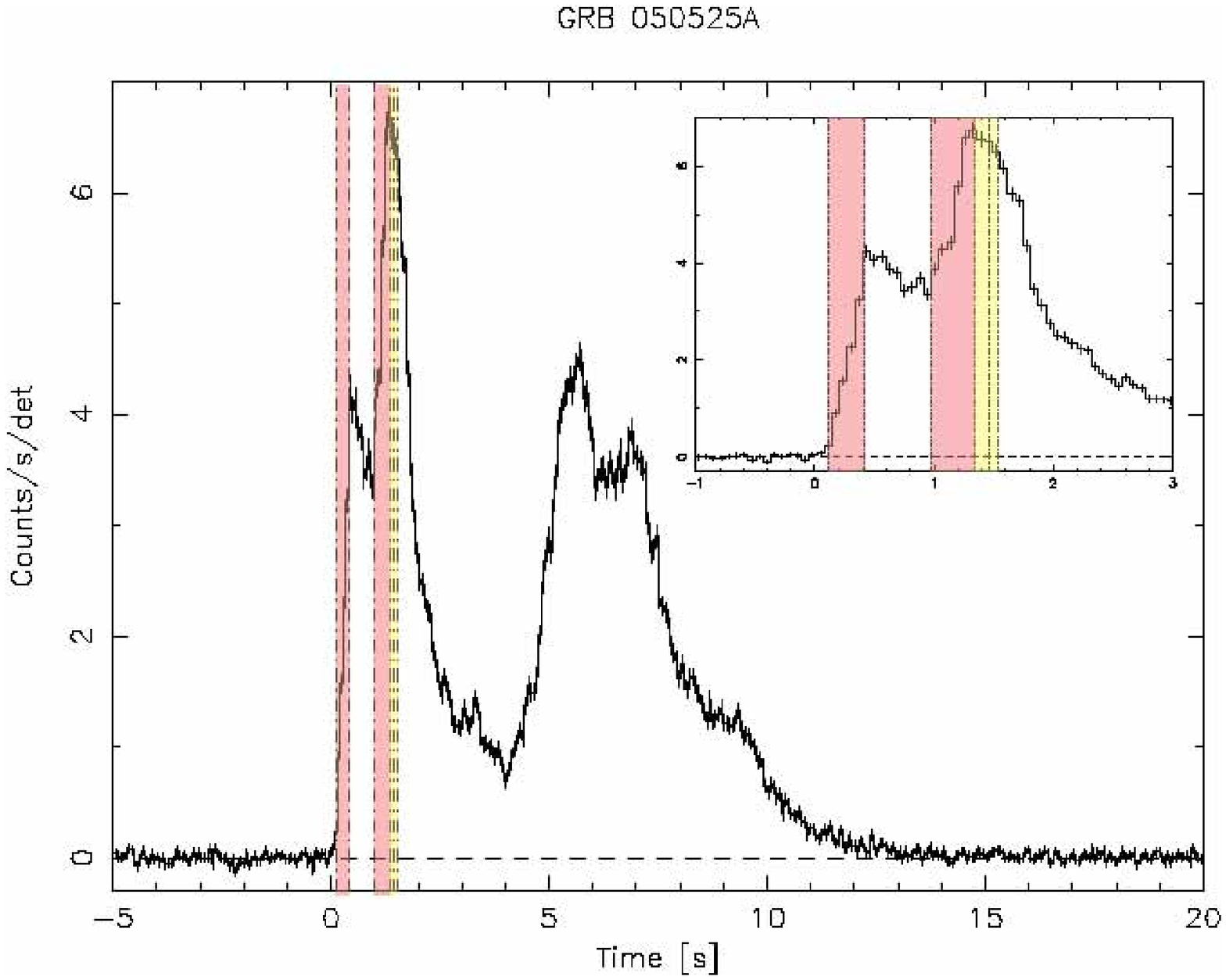}
}
\vspace{0.5cm}
\centerline{
\includegraphics[width=6cm,angle=0]{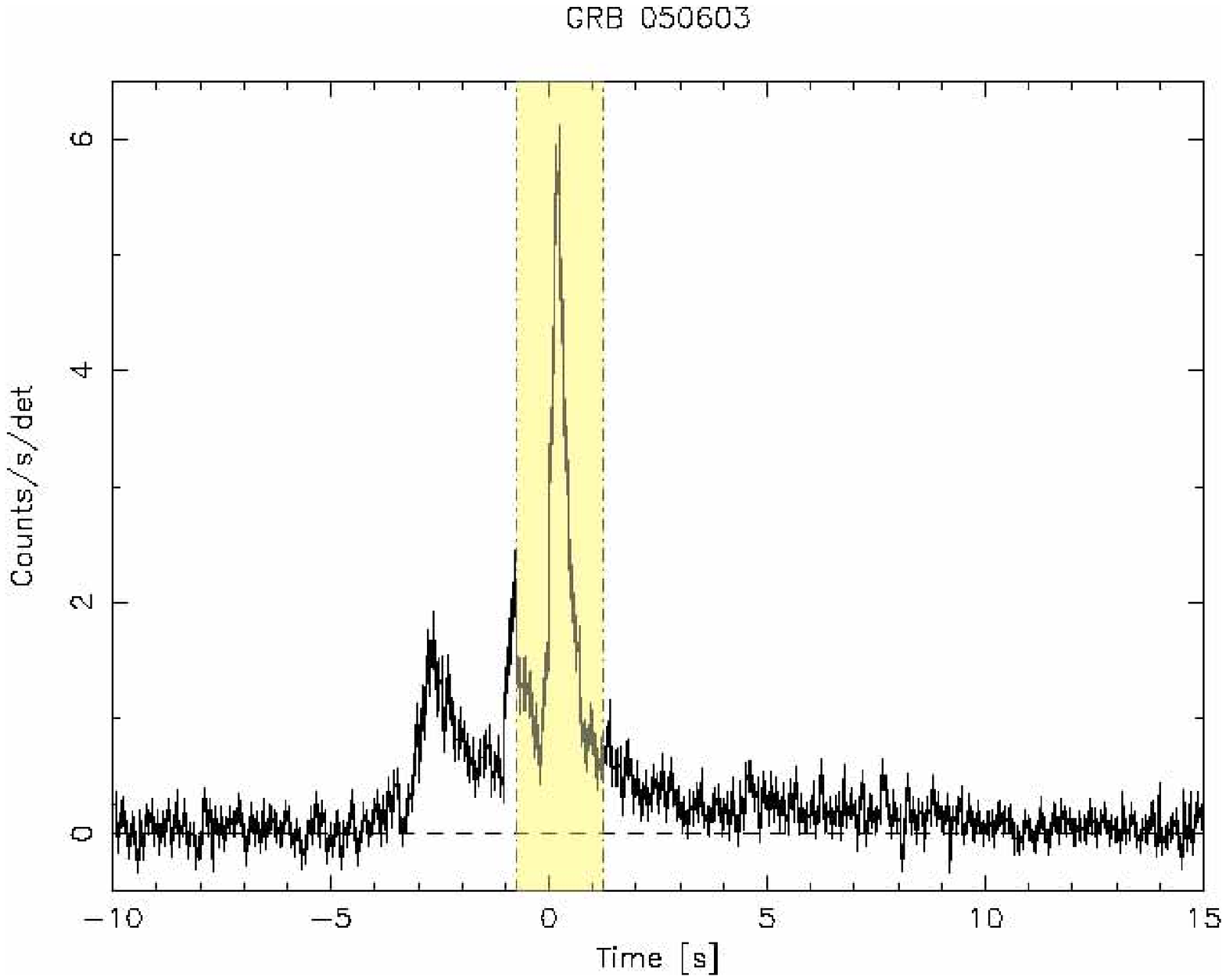}
\hspace{0.5cm}
\includegraphics[width=6cm,angle=0]{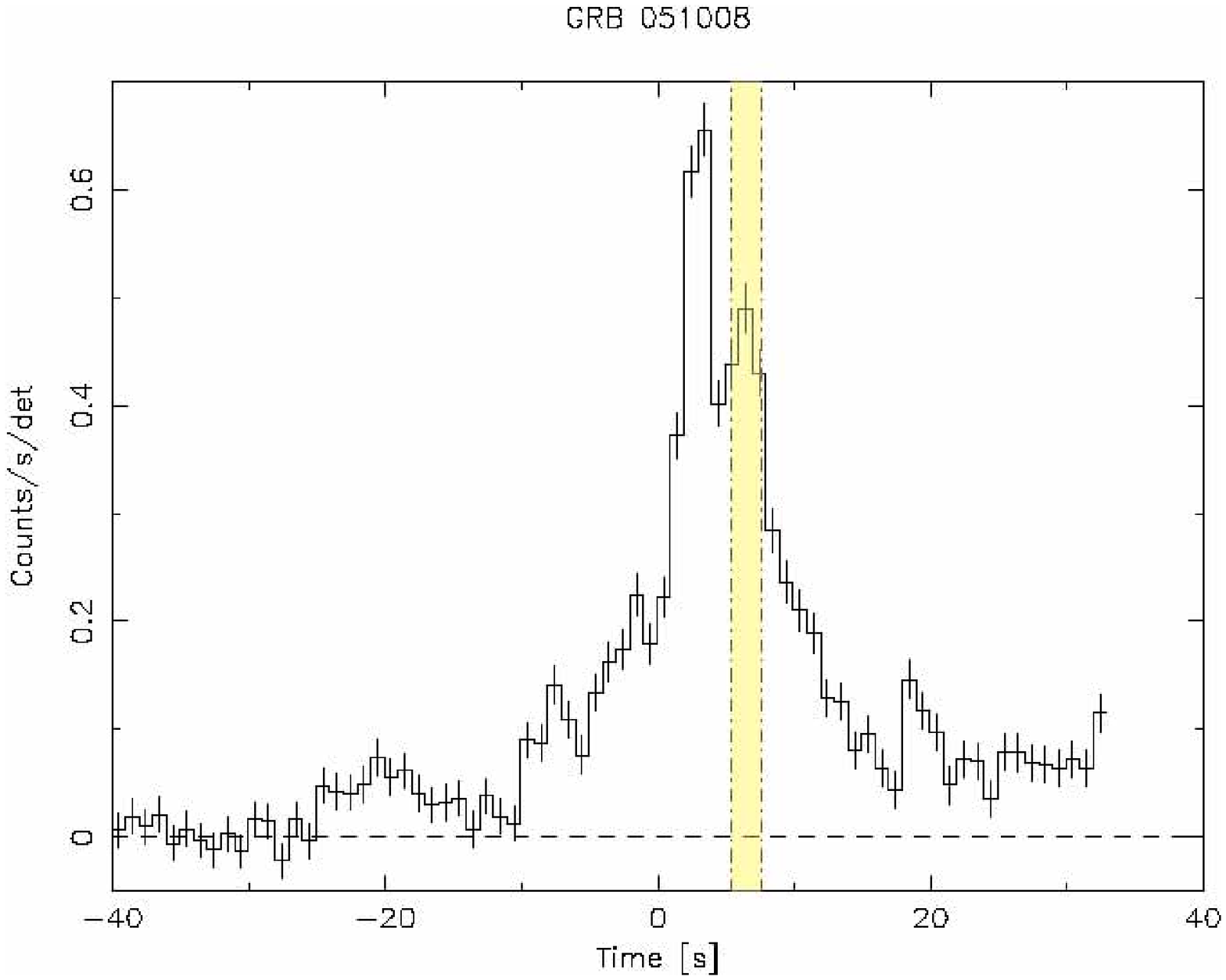}
}
\vspace{0.5cm}
\centerline{
\includegraphics[width=6cm,angle=0]{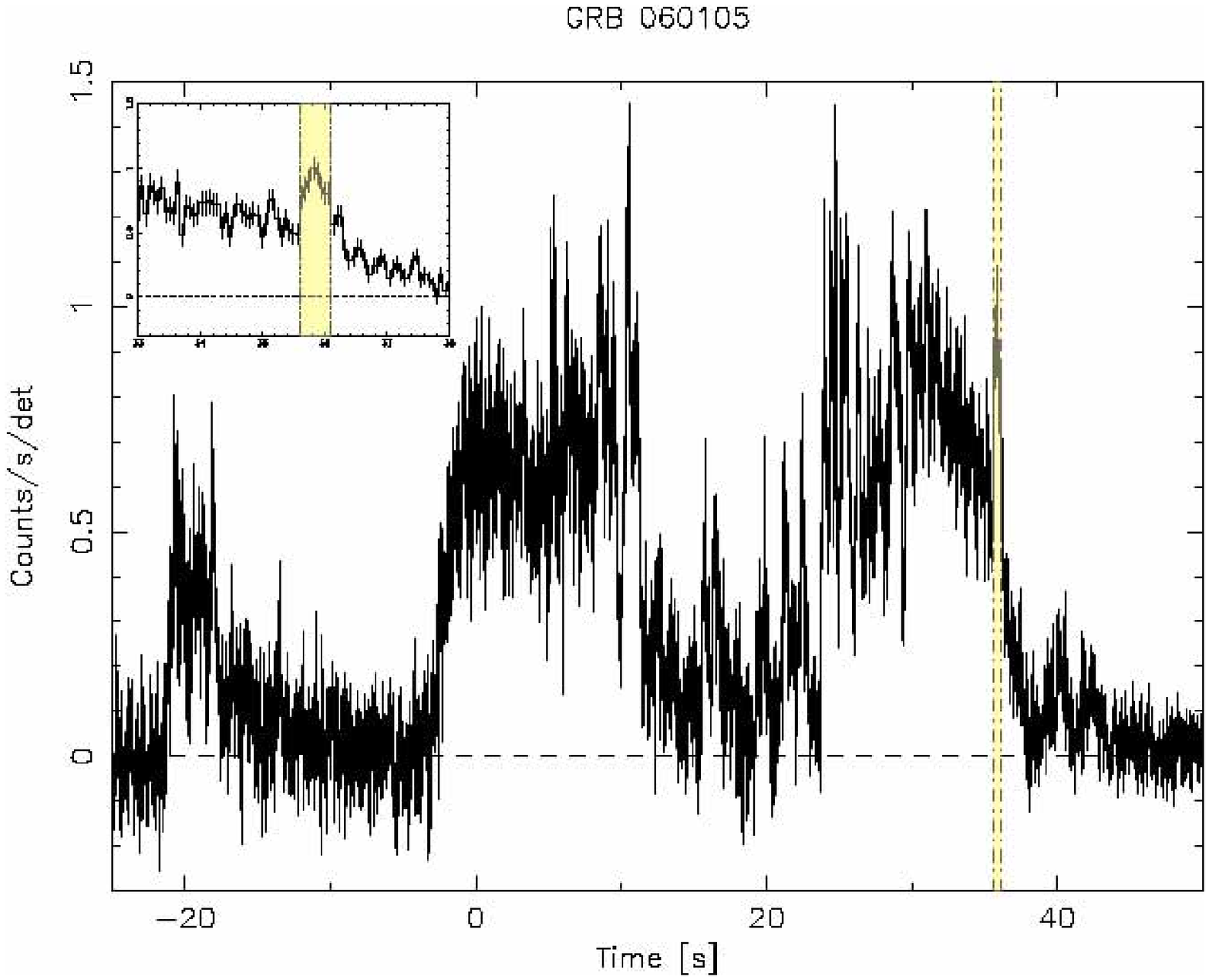}
\hspace{0.5cm}
\includegraphics[width=6cm,angle=0]{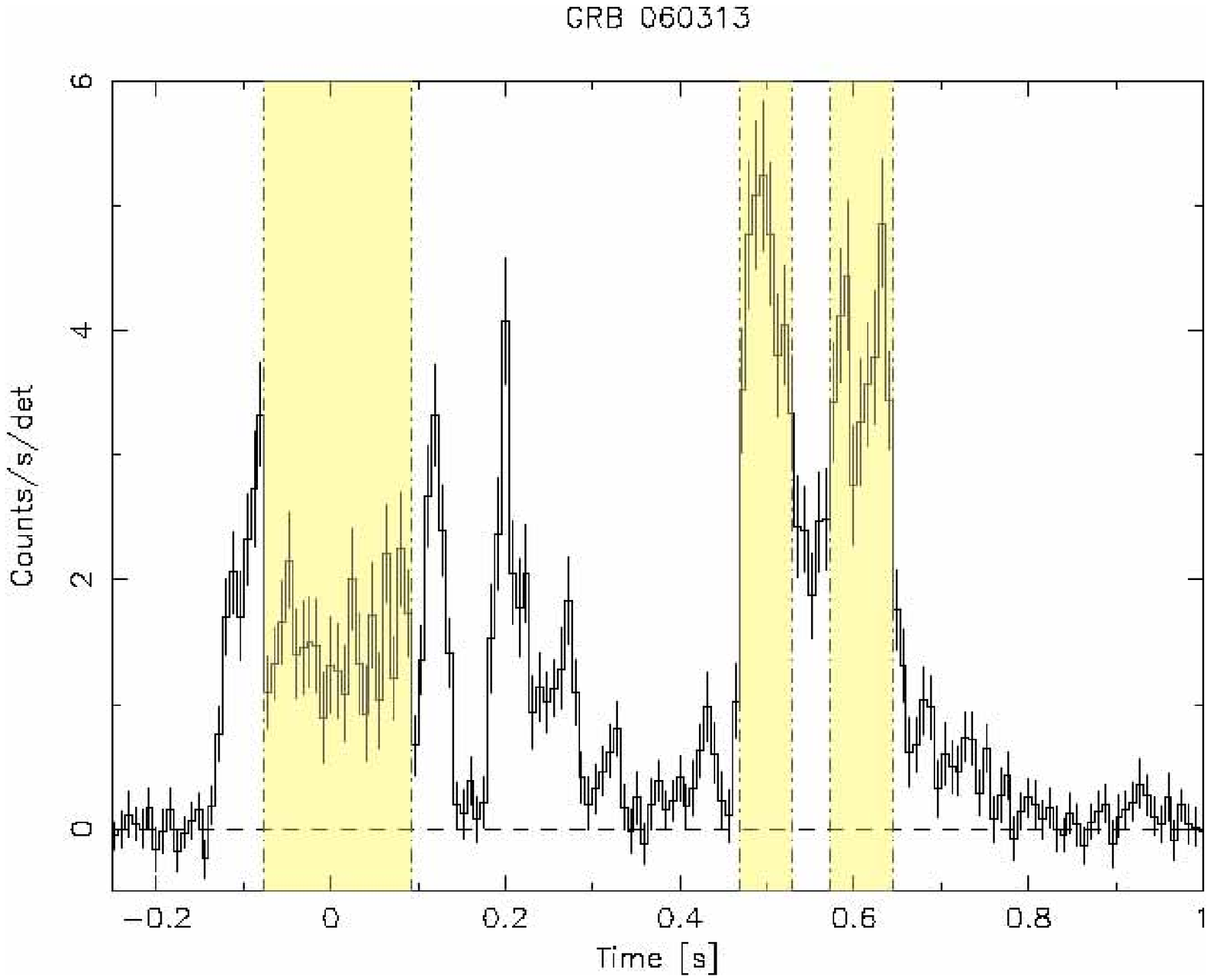}
}
\vspace{0.5cm}
\centerline{
\includegraphics[width=6cm,angle=0]{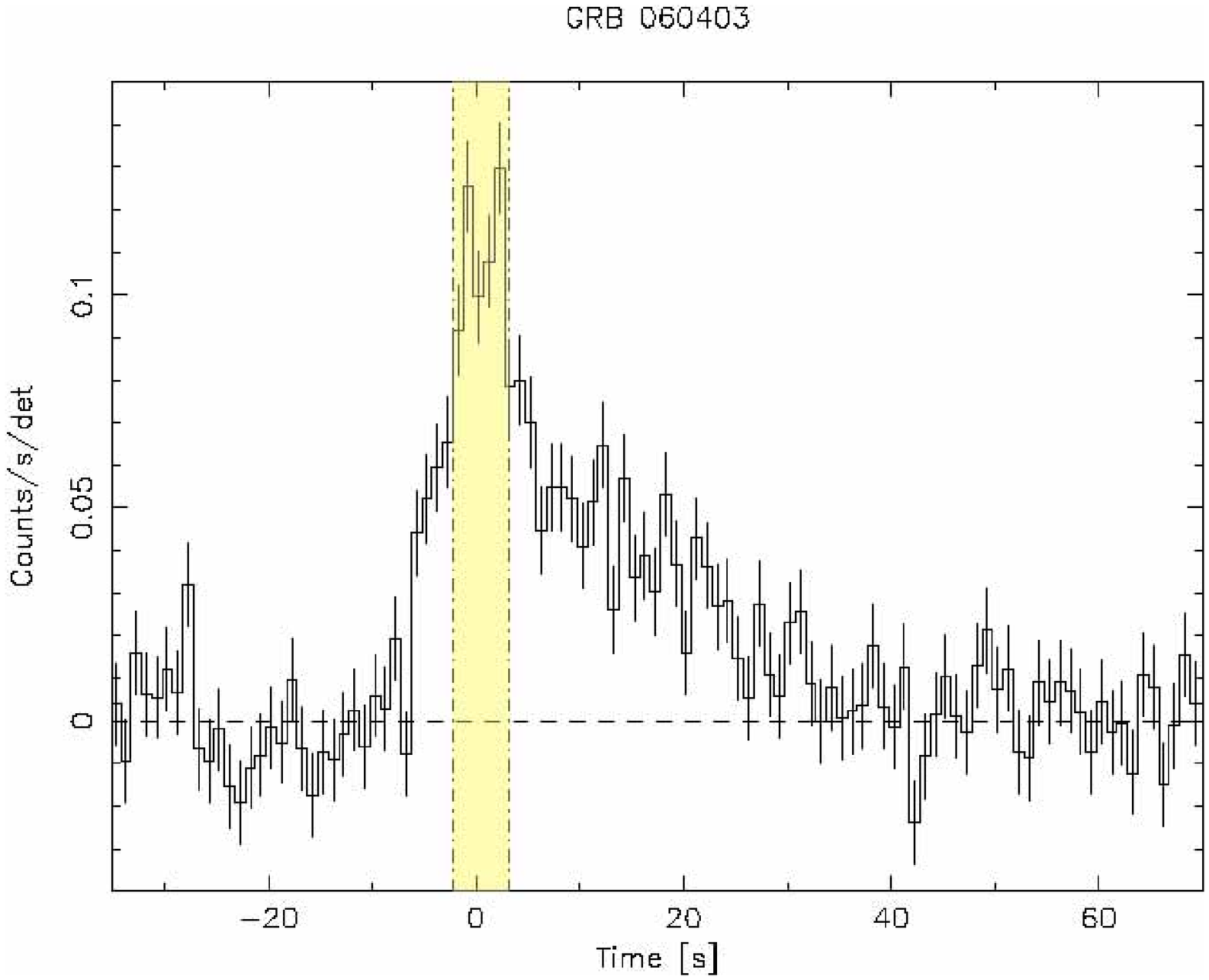}
\hspace{0.5cm}
\includegraphics[width=6cm,angle=0]{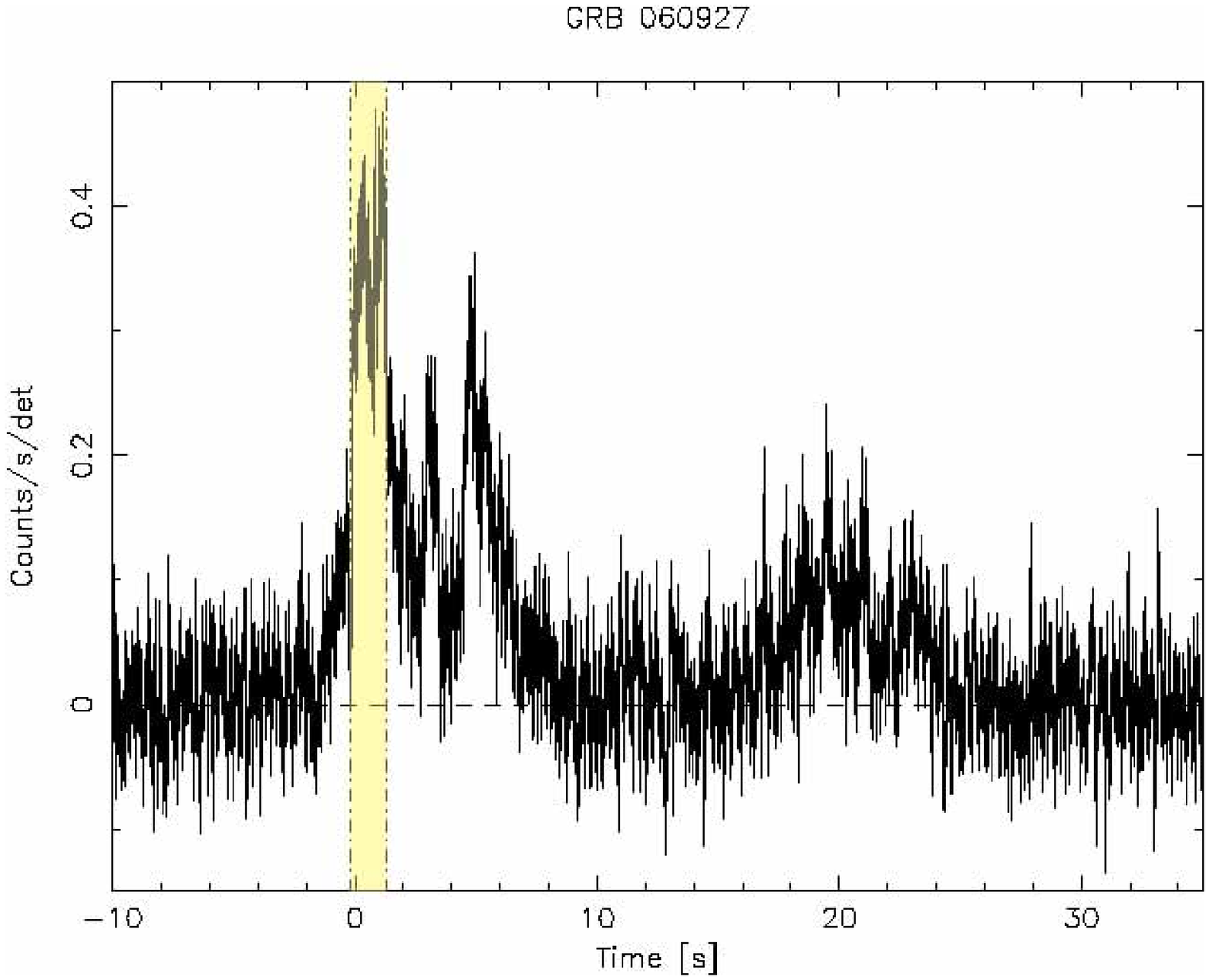}
}
\caption{The BAT light curves with the intervals for which the spectra exceed 
the limits of the SSM as shaded.  The red shading indicates intervals 
exceeding the limit by $>3.2$ $\sigma$.  The yellow shading indicates intervals 
exceeding the limit by $>1.6$ $\sigma$.  \label{lod_lc1}}
\end{figure}

\begin{figure}[p]
\centerline{
\includegraphics[width=6cm,angle=0]{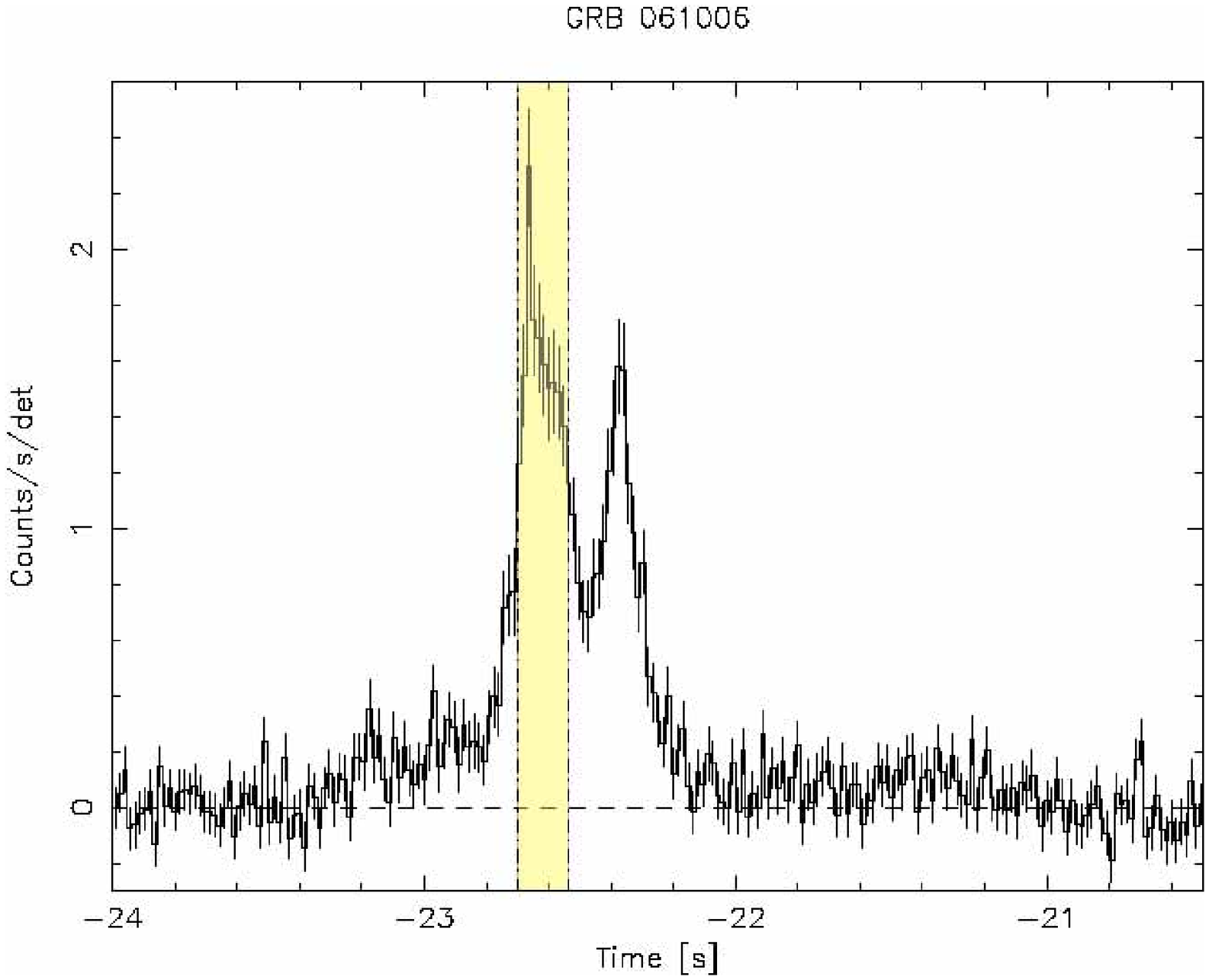}
\hspace{0.5cm}
\includegraphics[width=6cm,angle=0]{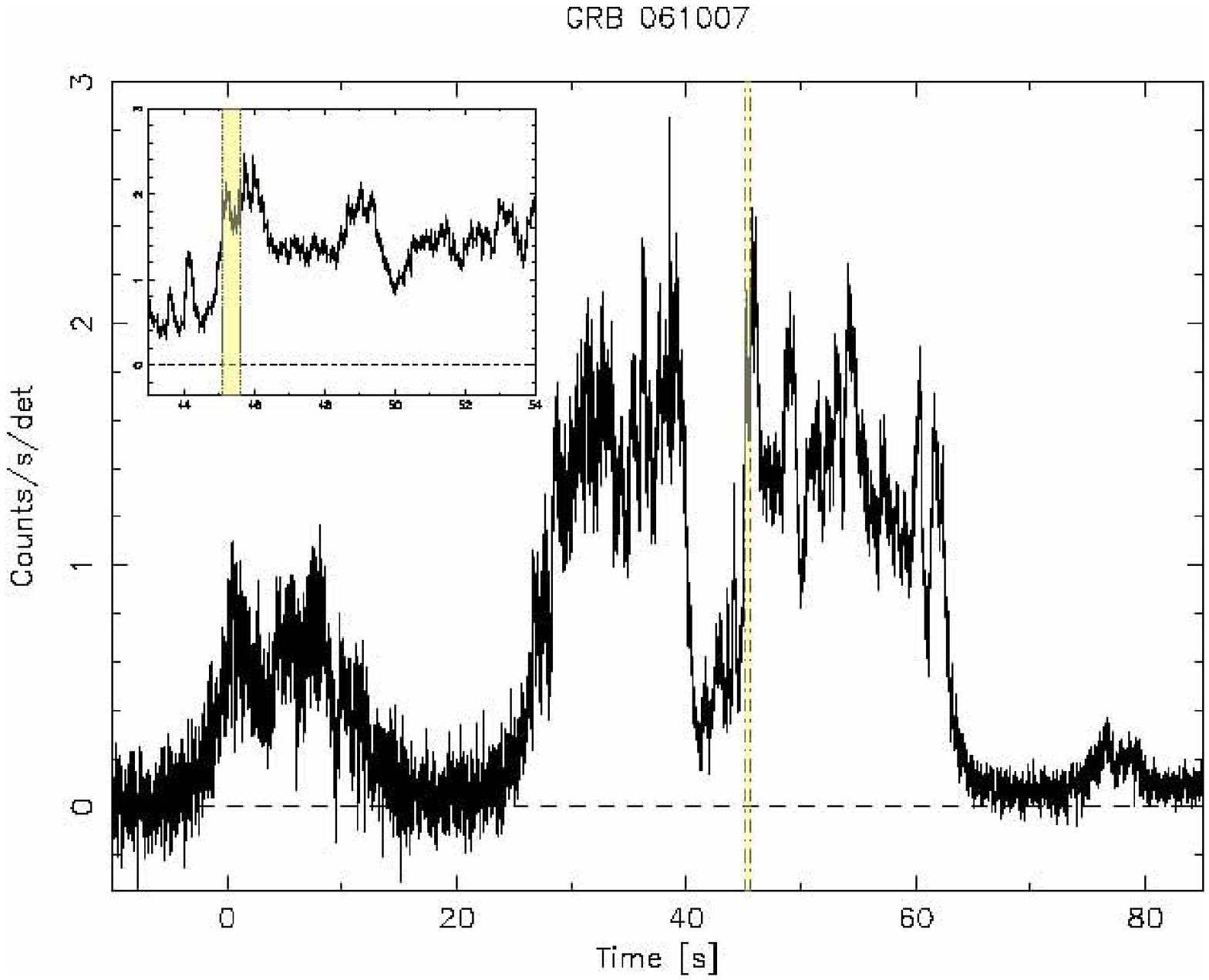}
}
\vspace{0.5cm}
\centerline{
\includegraphics[width=6cm,angle=0]{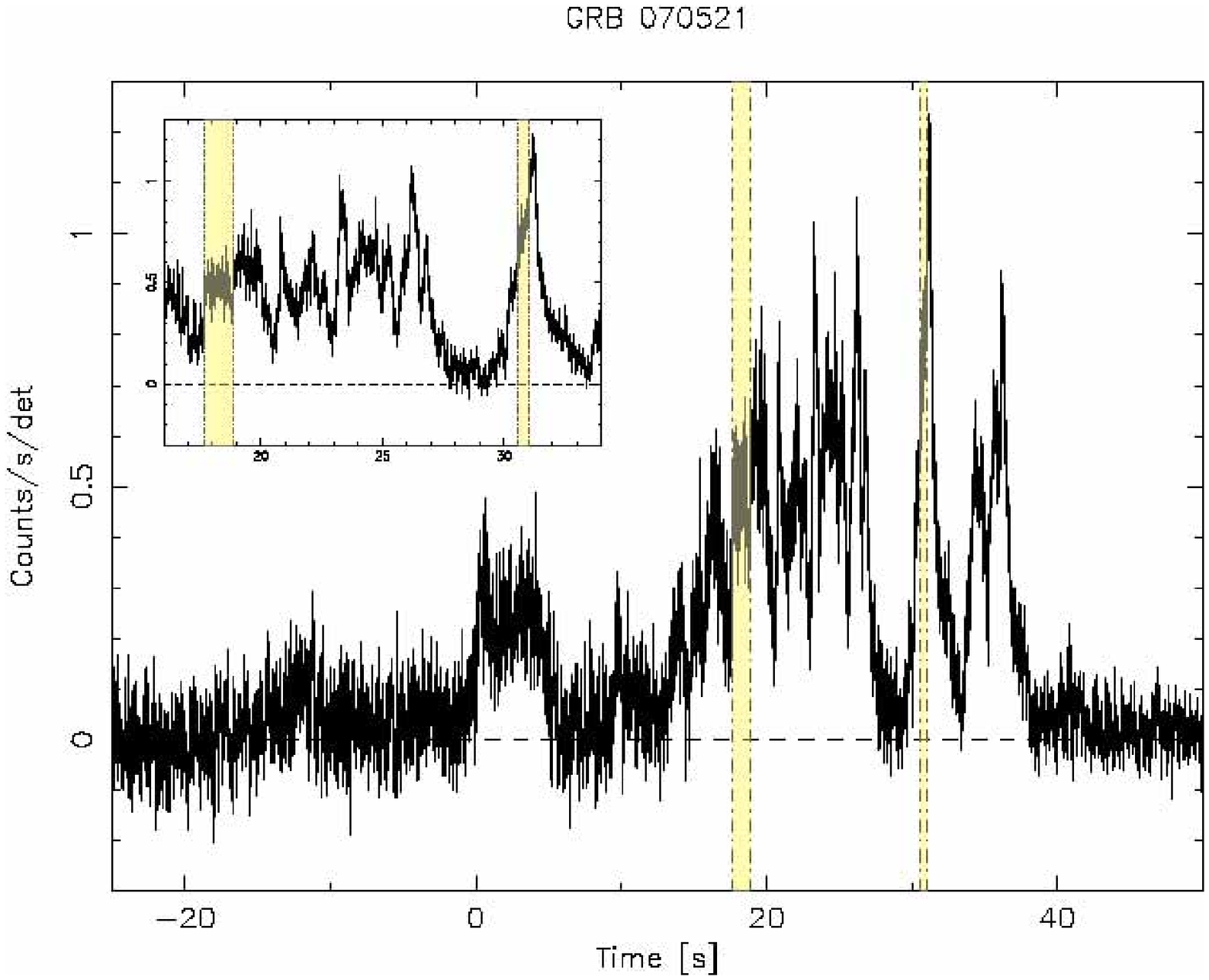}
\hspace{0.5cm}
\includegraphics[width=6cm,angle=0]{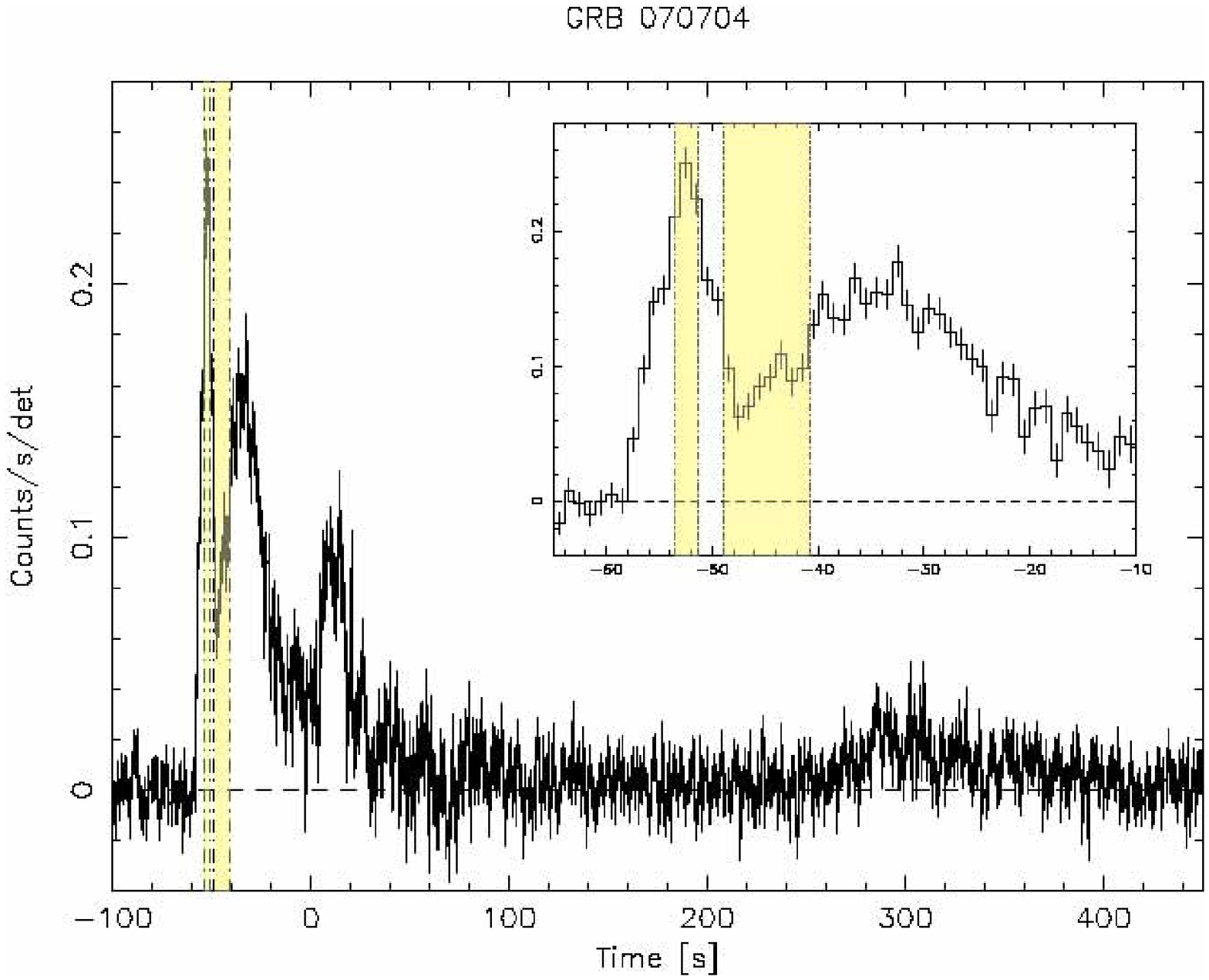}
}
\vspace{0.5cm}
\centerline{
\includegraphics[width=6cm,angle=0]{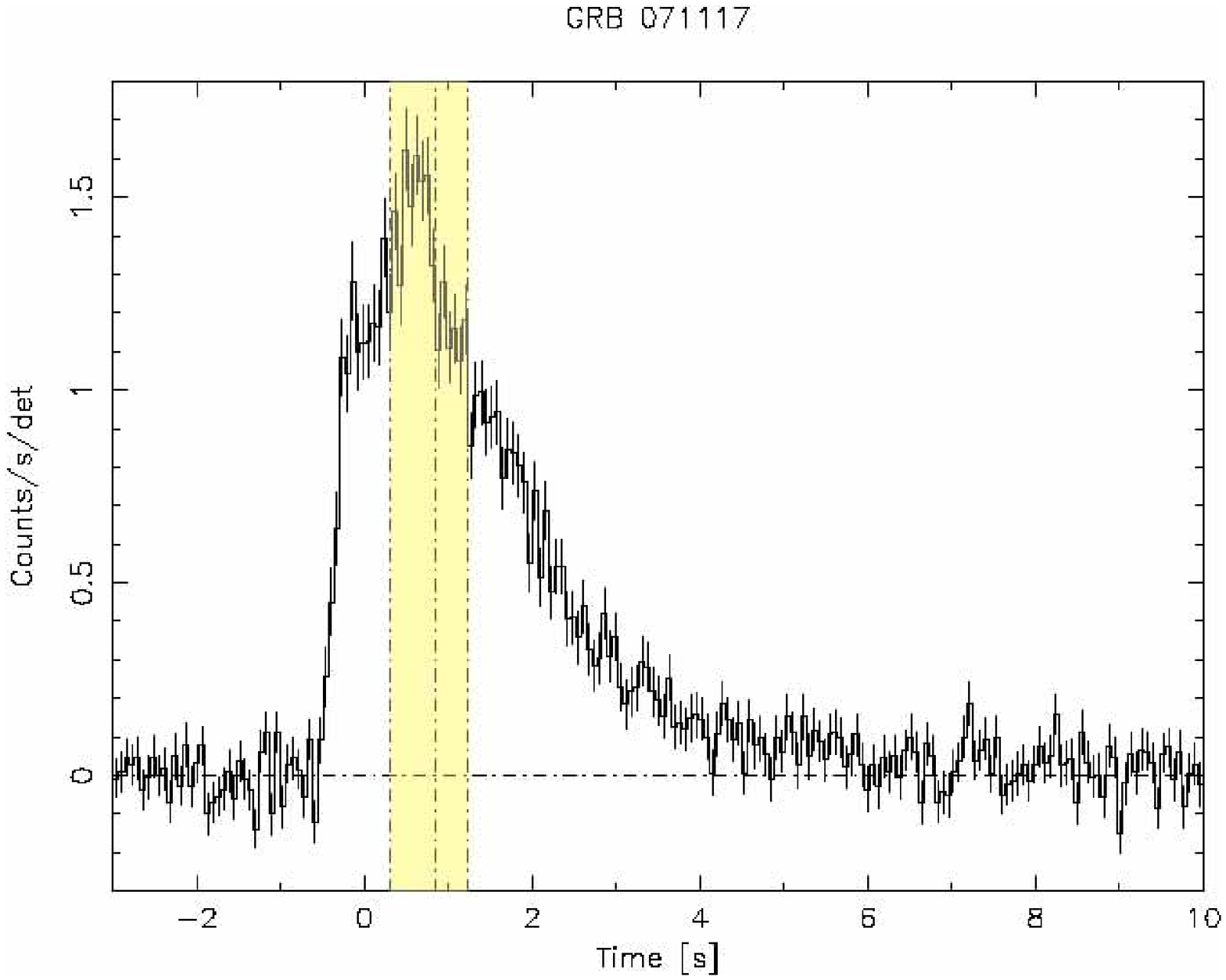}
\hspace{0.5cm}
\includegraphics[width=6cm,angle=0]{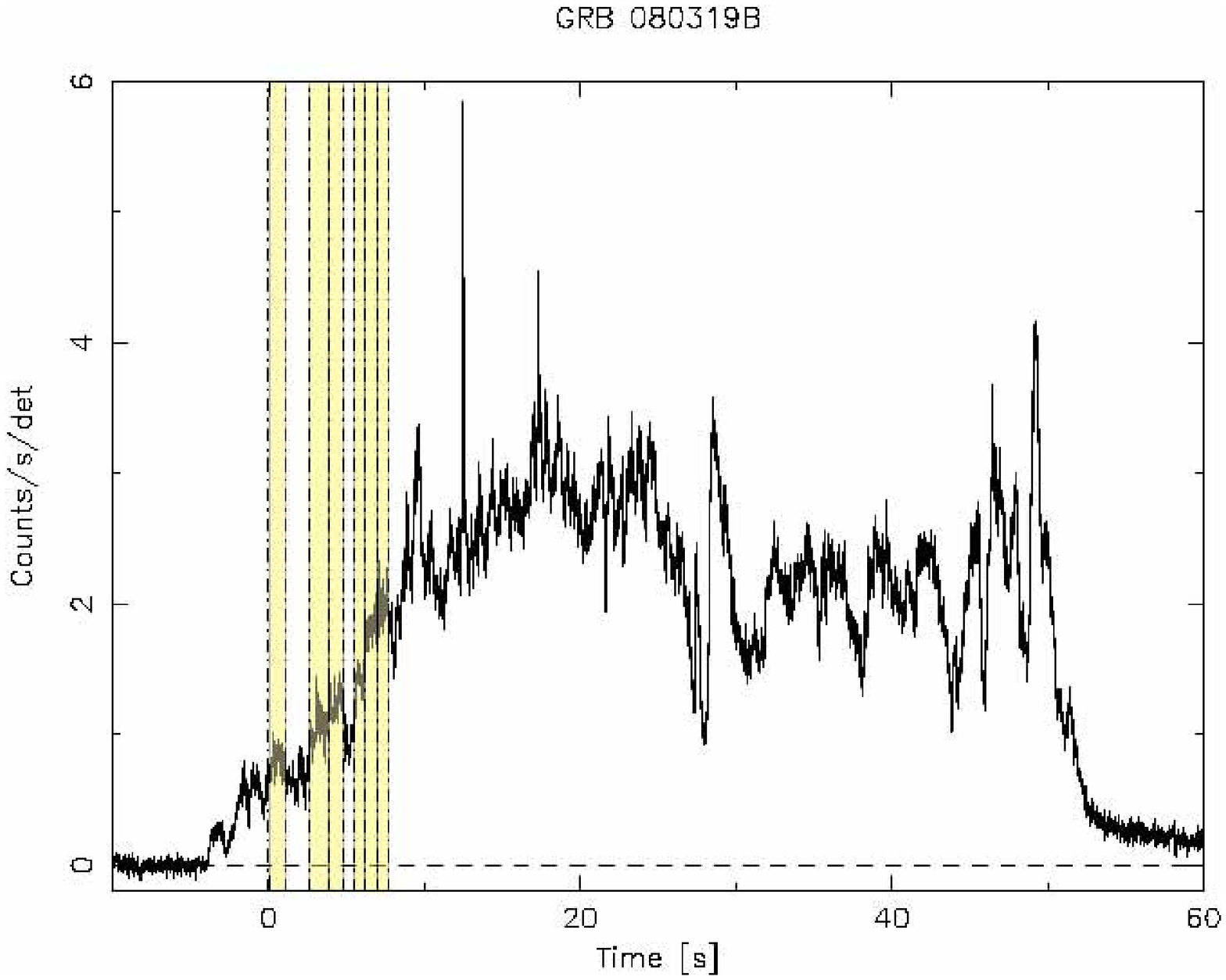}
}
\centerline{
\includegraphics[width=6cm,angle=0]{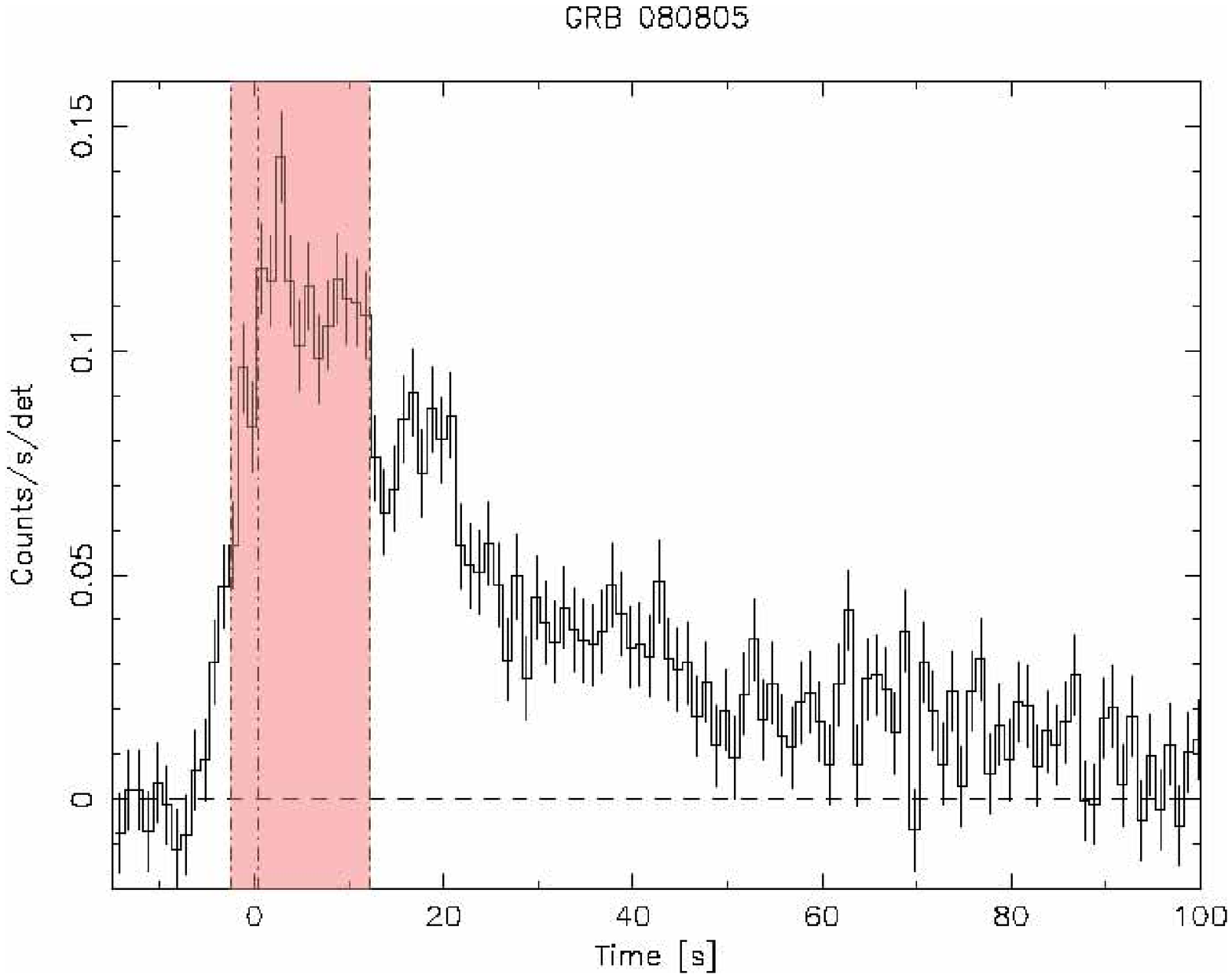}
\hspace{0.5cm}
\includegraphics[width=6cm,angle=0]{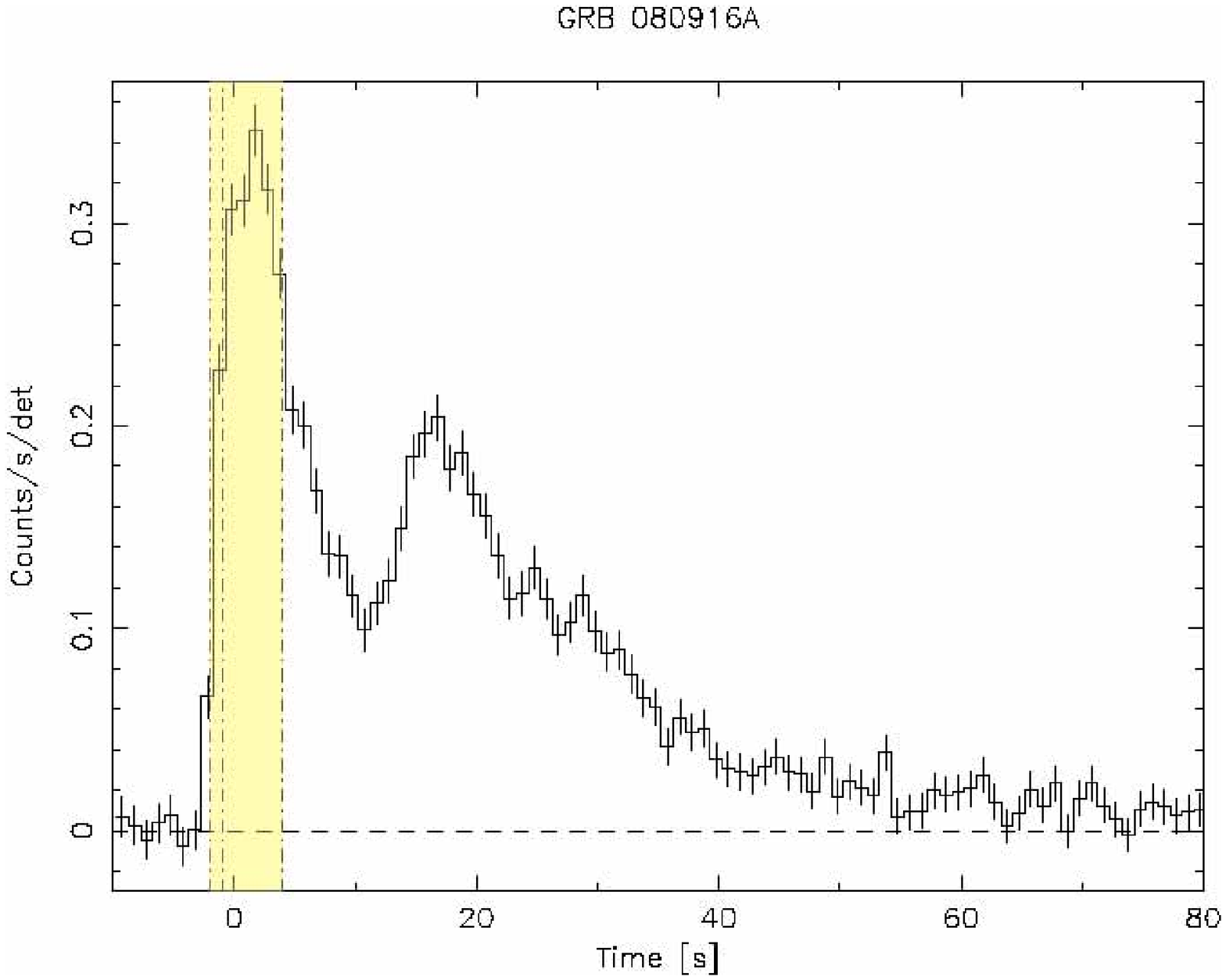}
}
\caption{Continue\label{lod_lc2}}
\end{figure}

\begin{figure}[p]
\centerline{
\includegraphics[width=6cm,angle=0]{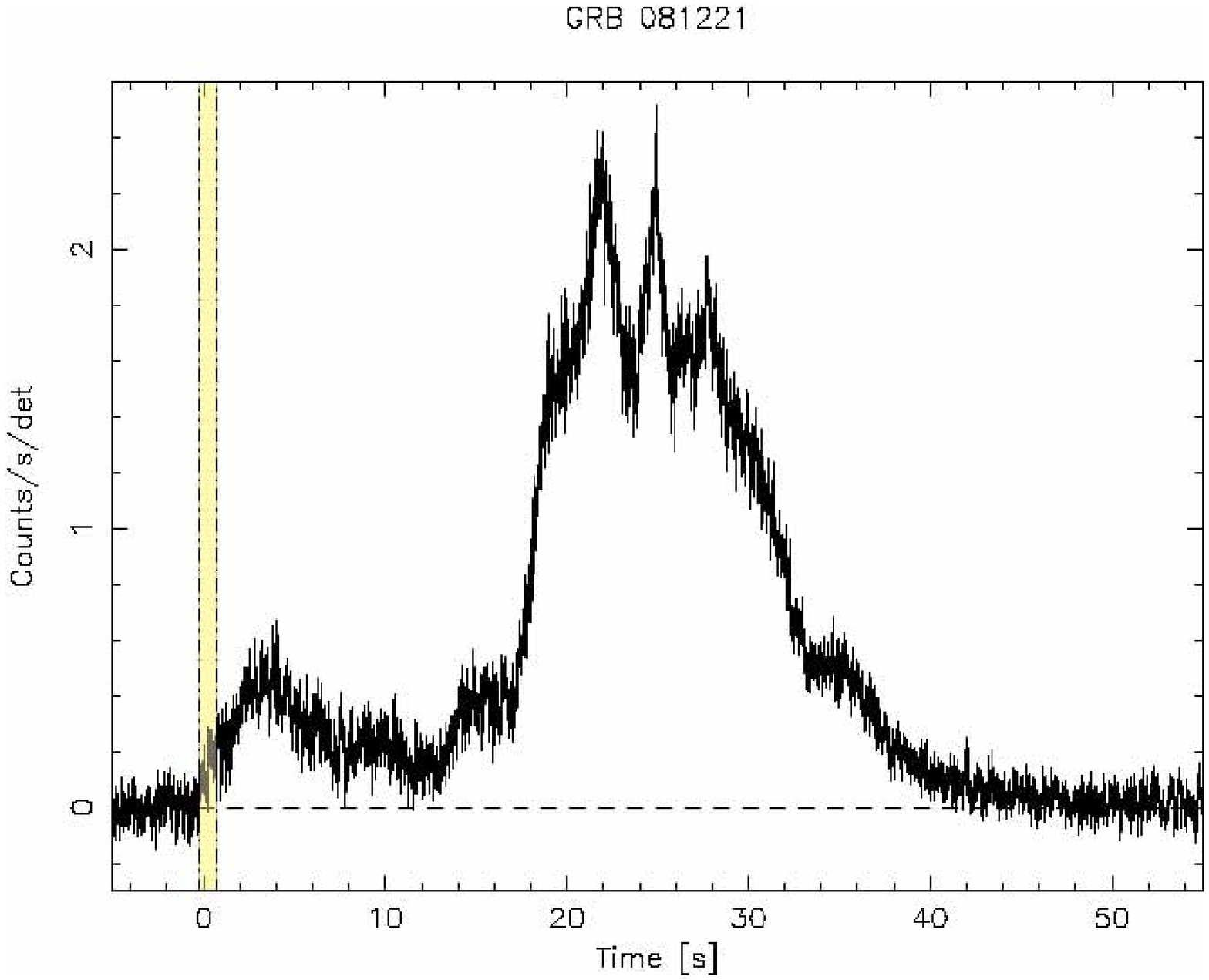}
\hspace{0.5cm}
\includegraphics[width=6cm,angle=0]{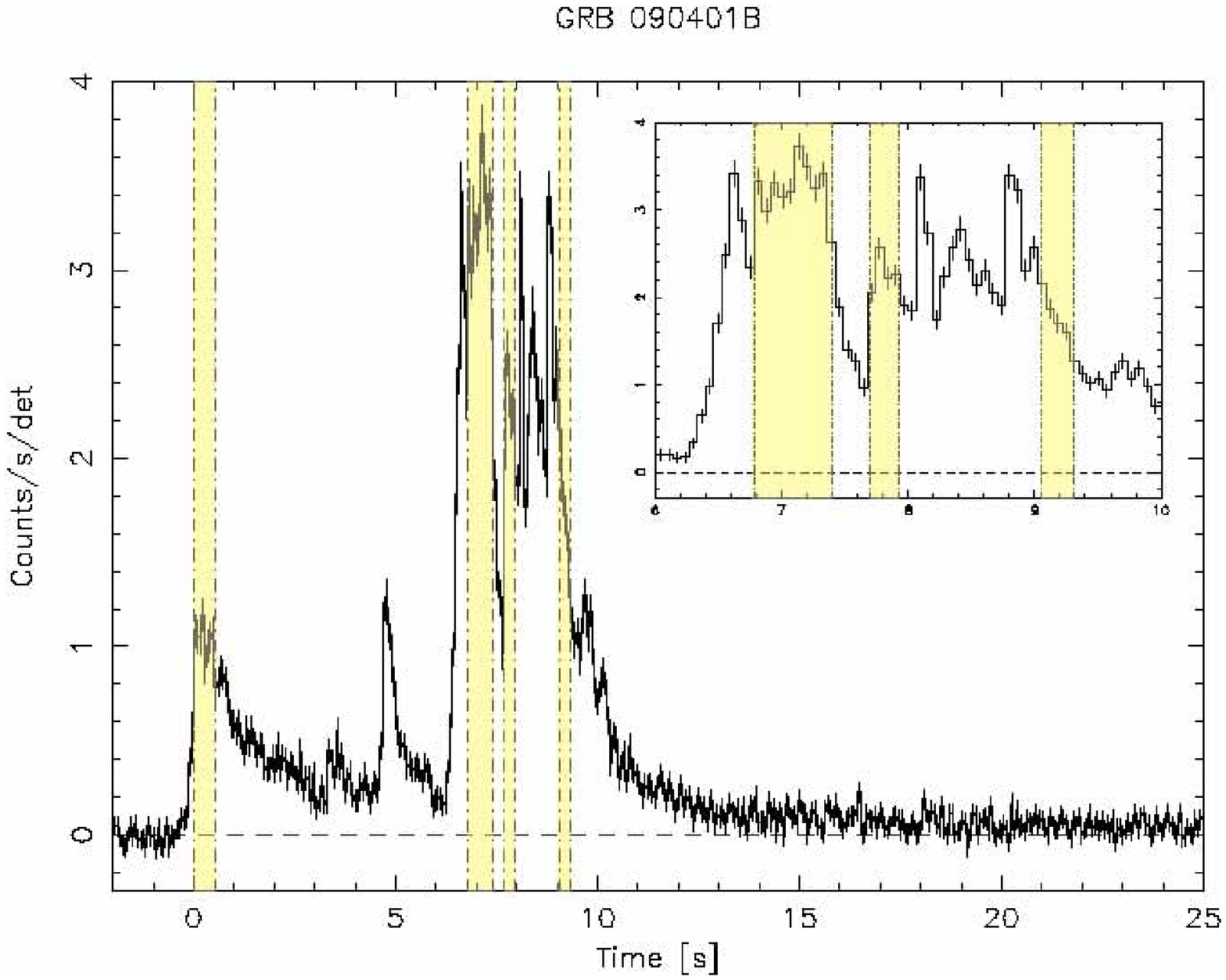}
}
\vspace{0.5cm}
\centerline{
\includegraphics[width=6cm,angle=0]{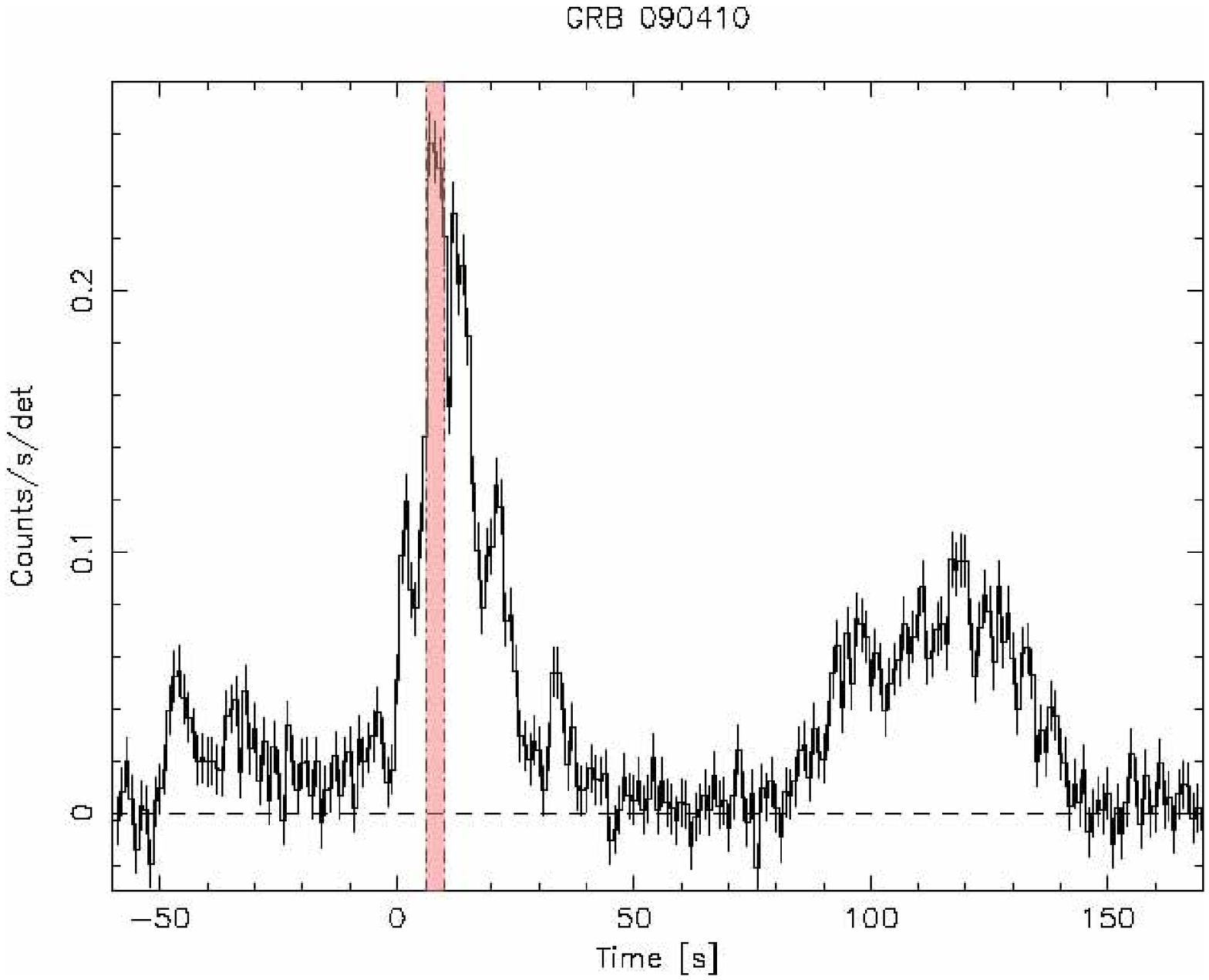}
\hspace{0.5cm}
\includegraphics[width=6cm,angle=0]{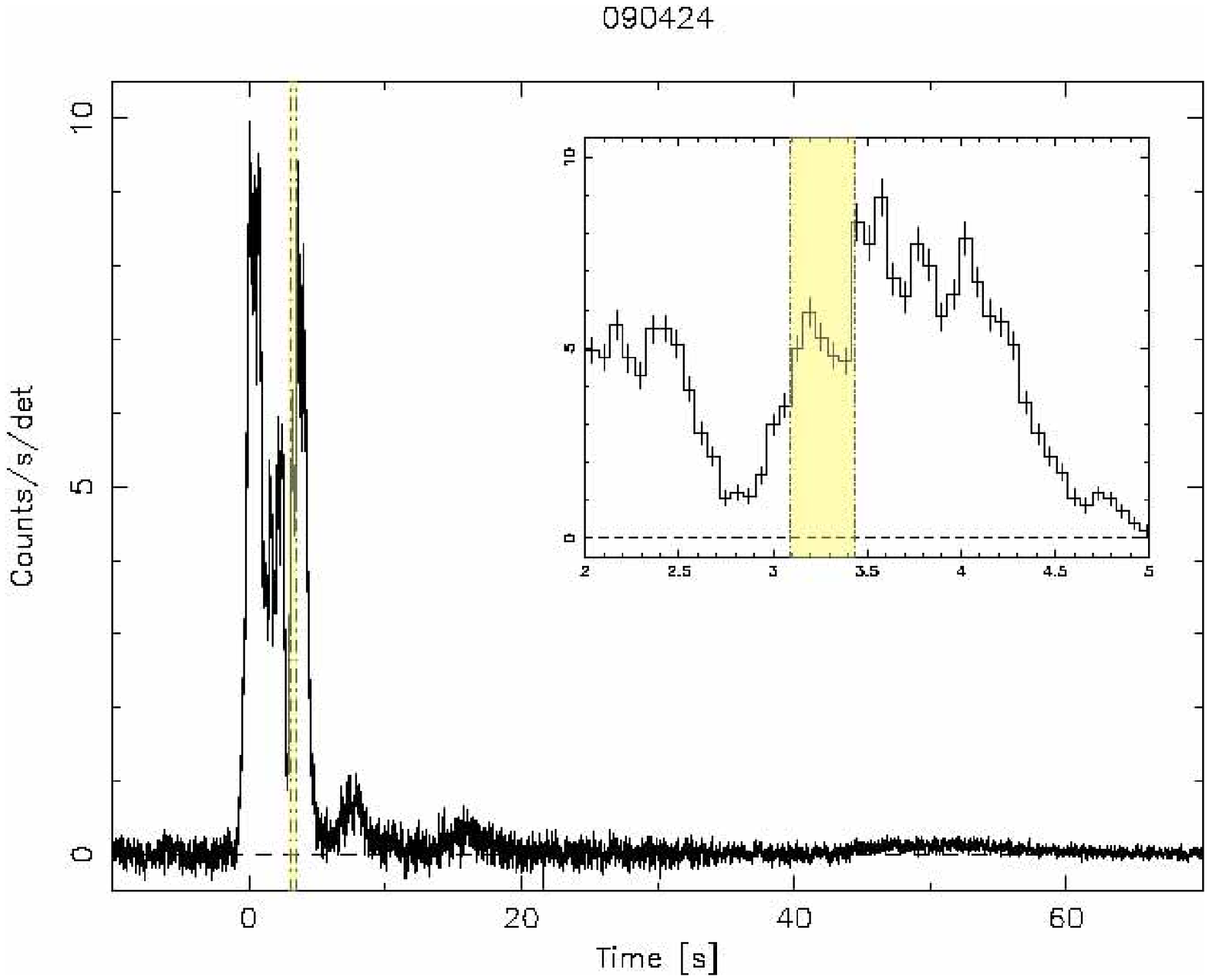}
}
\vspace{0.5cm}
\centerline{
\includegraphics[width=6cm,angle=0]{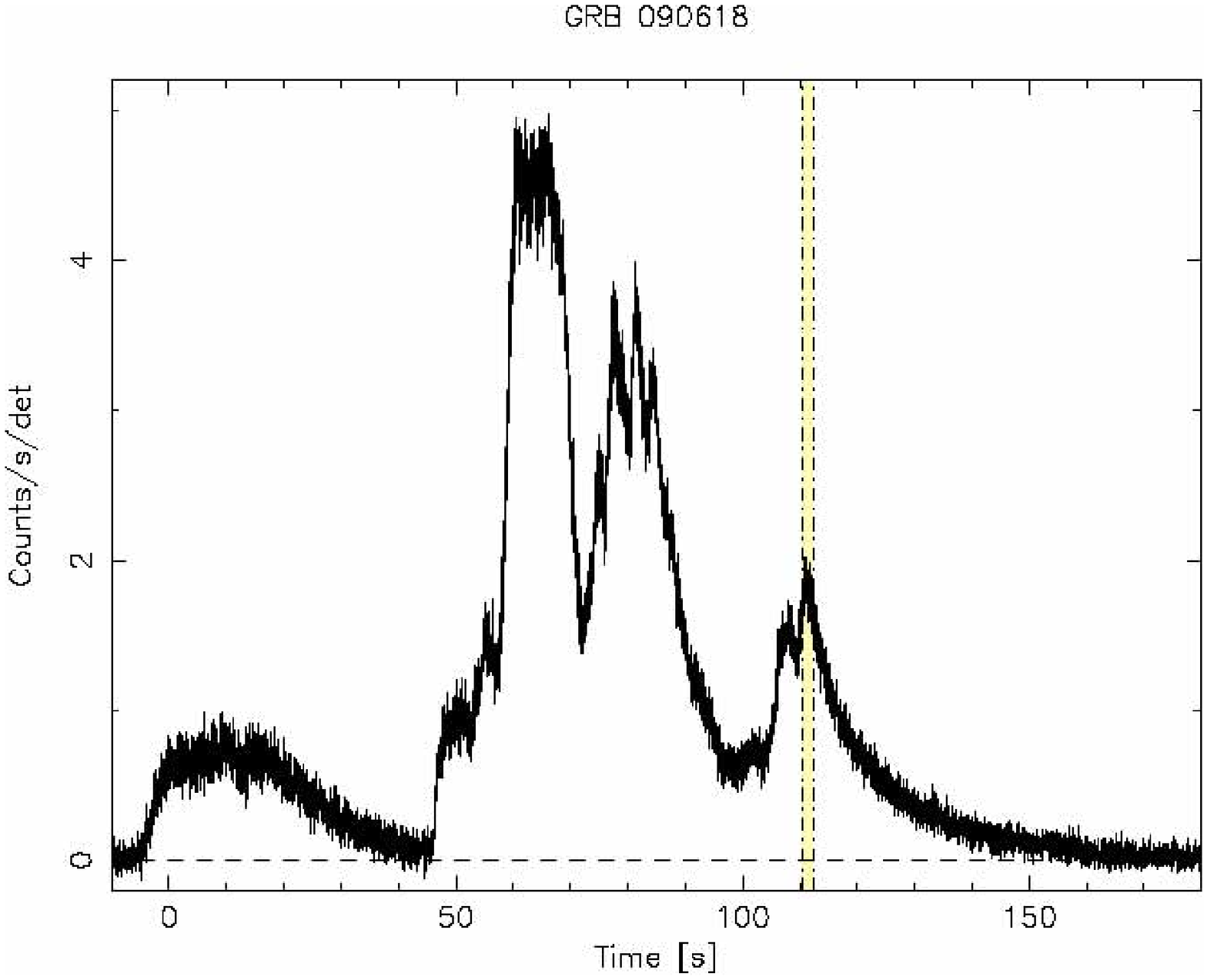}
\hspace{0.5cm}
\includegraphics[width=6cm,angle=0]{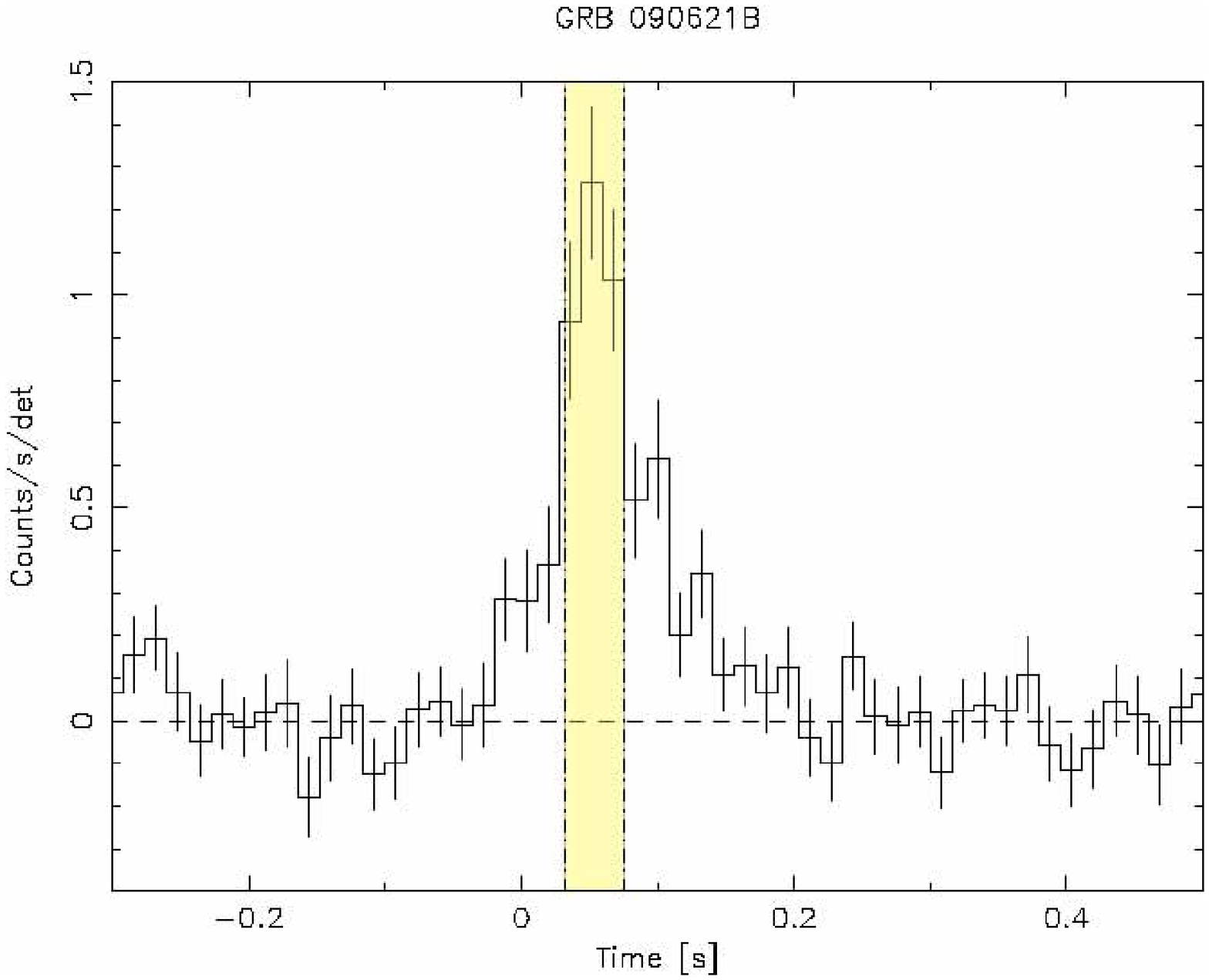}
}
\vspace{0.5cm}
\centerline{
\includegraphics[width=6cm,angle=0]{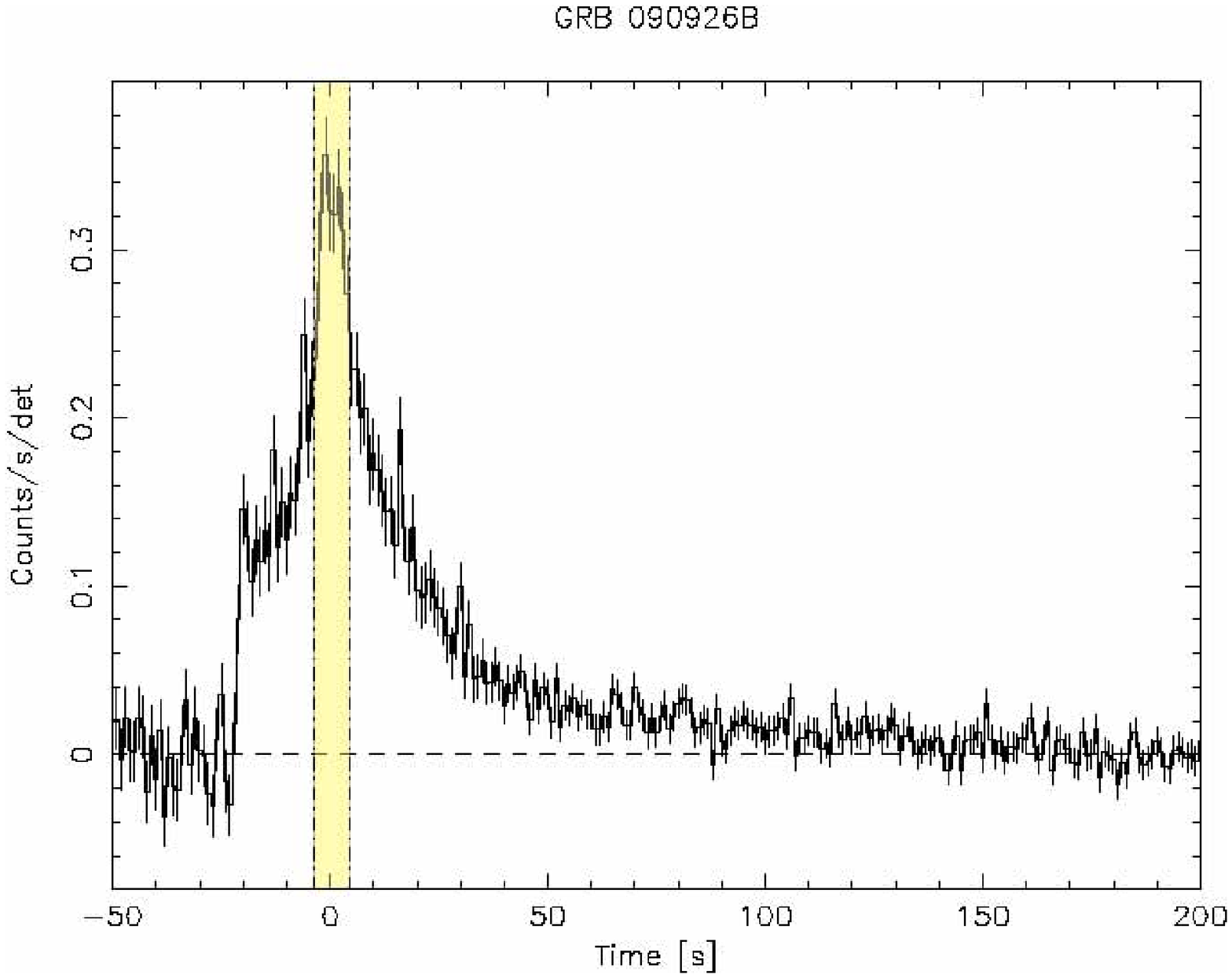}
\hspace{0.5cm}
\includegraphics[width=6cm,angle=0]{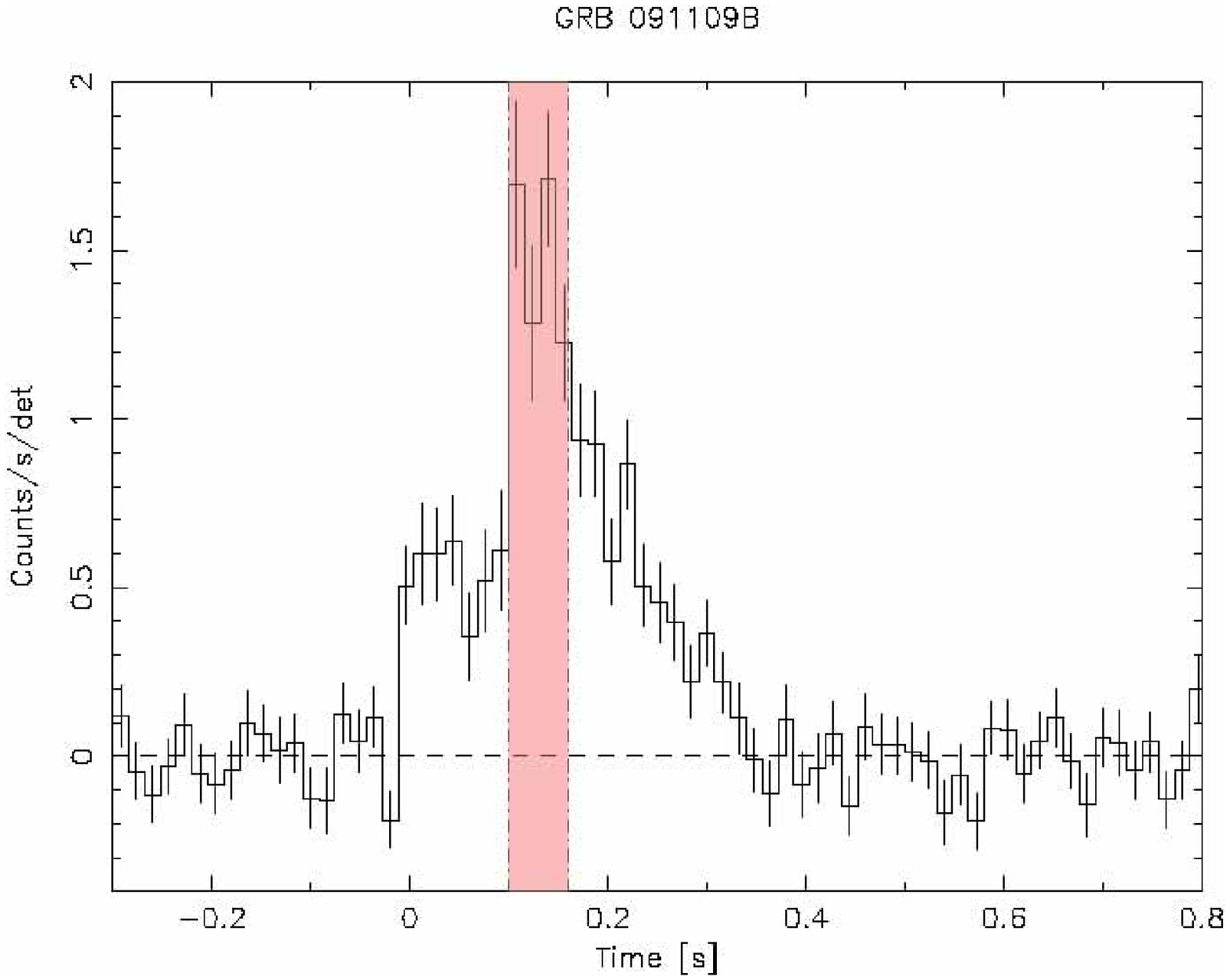}
}
\caption{Continue\label{lod_lc3}}
\end{figure}

One of the important questions about prompt emission from GRBs is whether or not 
the observed spectrum is correctly represented by the synchrotron shock model (SSM) 
\citep{reesmeszaros1992,sari1996}.  
According to \citet{preece1998}, 23\% of the BATSE time-resolved 
spectra have low-energy photon indices which violate the SSM limit 
from $-3/2$ to $-2/3$ (the so called ``line of death'' problem).  
Although the BAT photon index based on 
a PL fit does not represent the low-energy photon index $\alpha$ of the Band function 
if $\eop$ is located inside the BAT energy range (see section 4.5), 
the BAT photon index should be the actual low energy photon index $\alpha$ if $\eop$ is 
located above the BAT energy range.  
Therefore, 
the photon index derived from a PL fit in the BAT time-resolved spectra is the interesting 
dataset to investigate the line of death problem.  
Figure \ref{time_resolved_pl_phindex_vs_flux} shows the 
BAT photon index versus the energy flux in the 15-150 keV band using time-resolved spectra 
with a PL fit.  We only used the results for which the 90\% error of the photon index has been 
constrained within $\pm$0.5.  
Out of 2968 points in the figure, 18 spectra (0.6\%) exceed the 
hard side of the line of death ($>$$-2/3$) by $>$1.6 $\sigma$ level.  
Figure \ref{time_resolved_cpl_phindex_vs_ep} shows the distribution 
of the low-energy photon index $\alpha$ versus $\eop$ of the BAT time-resolved 
spectra for which there is a significant improvement of $\chi^{2}$ in a CPL fit over a PL fit.  
Just as in the PL fit samples, we only used spectra where the low-energy photon index 
$\alpha$ is constrained within $\pm$0.5 at the 90\% confidence level.  The low-energy 
photon index of 23 out of 234 spectra are harder than $-$2/3 by $>$1.6 $\sigma$ level.  
There is only one time-resolved spectrum (GRB 090618 from 110.5 s 
to 112.272 s from the trigger time) which violates the softer side 
of the line of death ($<$$-$3/2) at the significance level of $>$1.6 $\sigma$.  
Therefore, the total fraction of the BAT spectra 
in the CPL samples which violate either line of death is $\sim$10\%.  
Although there are the spectra which violate the line of death in the BAT data, 
it is likely that the number of the BAT spectra which violate the limit is a factor of two 
smaller than the BATSE result.  
Figures \ref{lod_lc1}, \ref{lod_lc2} and \ref{lod_lc3} show the BAT light curves 
with shading of the time intervals exceeding the line of death.  
We notice that the intervals exceeding the line of death are mostly in bright spikes 
in the light curves.  Furthermore, in some bursts, 
the rising part of the light curve (e.g. GRB 080319B) exceeds the limit.  
Although it is hard to conclude which part of the light curve 
violates the line of death 
because of the 
small number of samples in our study, we do see a general trend that 1) a bright peak 
and 2) a rising part of a peak in the light curve tend to exceed the line 
of death limit. This is consistent with the BATSE results that the initial part of 
the FRED pulses tend to violate the line of death \citep{lu2010}.  
However, we do want to stress that majority of the photon indices derived from the BAT 
spectra are consistent with the expectation of the SSM.  

\begin{figure}
\centerline{
\includegraphics[width=12cm,angle=-90]{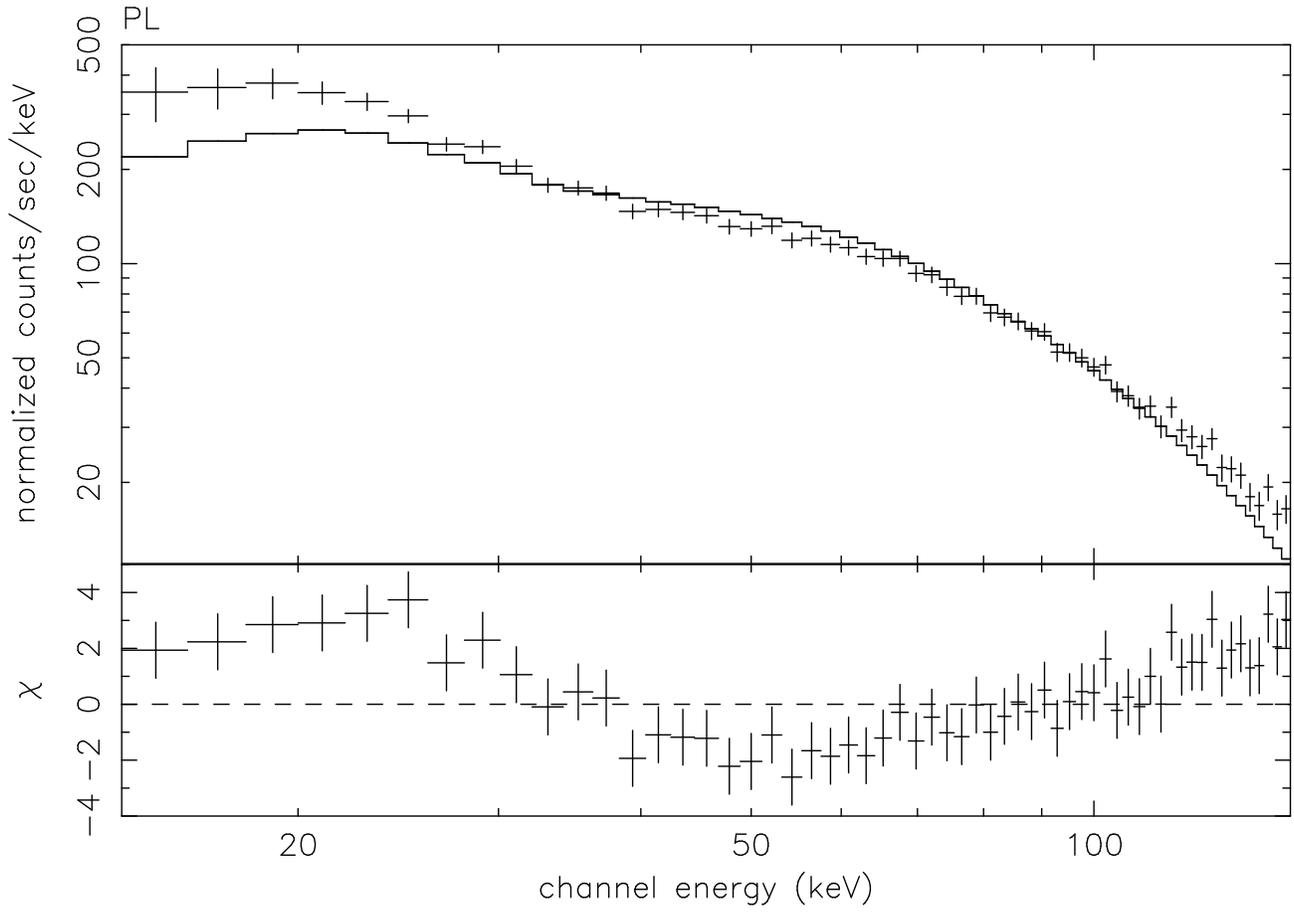}}
\caption{A simulated BAT spectrum with a Band function plus an additional 
power-law component (see text).  The solid line shows a fit to a PL model. 
\label{bat_sim_spec}}
\end{figure}

\begin{figure}
\centerline{
\includegraphics[width=12cm,angle=-90]{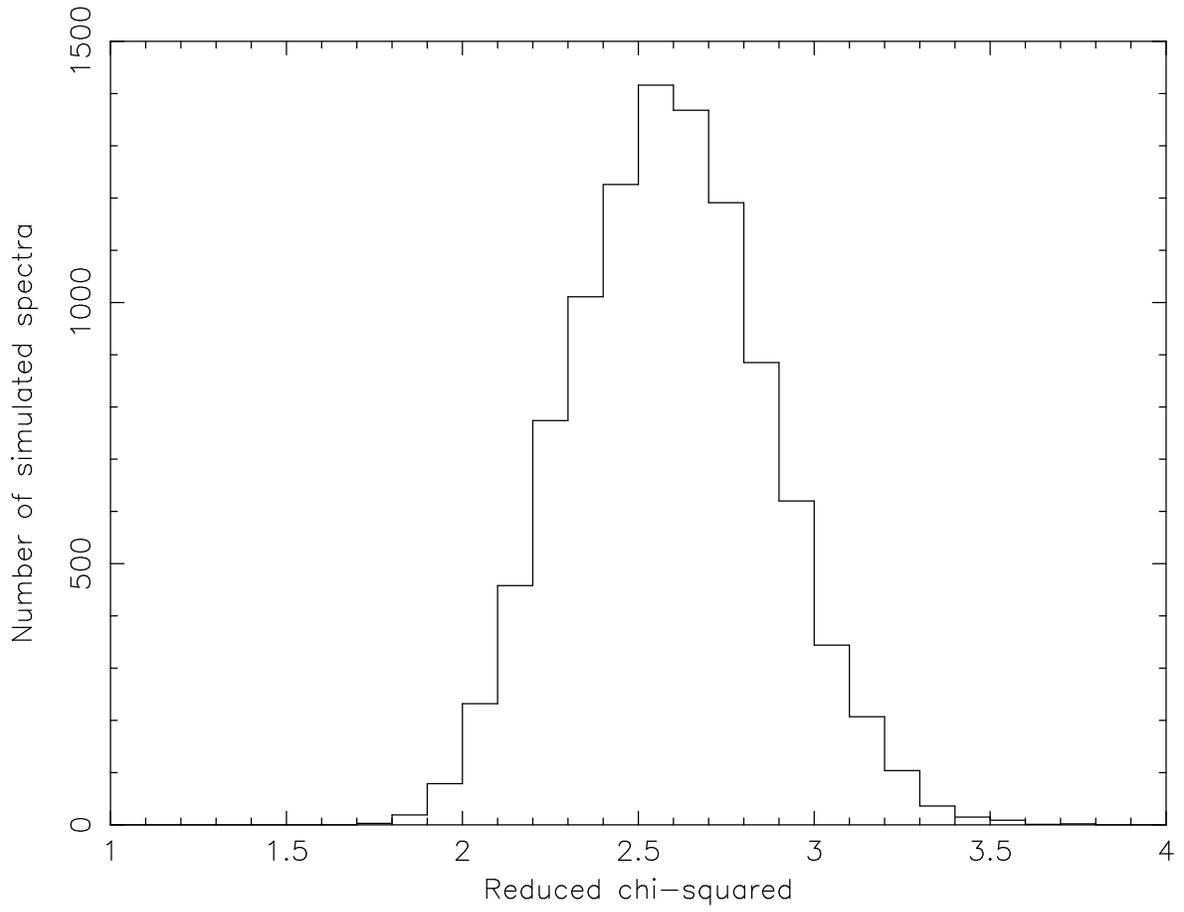}
}
\caption{Histogram of reduced $\chi^{2}$ in a PL fit of 10,000 BAT simulated spectra 
of the {\it Fermi} GRB 090902B interval b spectral parameters as the input spectrum.  
\label{bat_sim_spec_chi2_hist}}
\end{figure}

\begin{figure}[p]
\centerline{
\includegraphics[width=12cm,angle=-90]{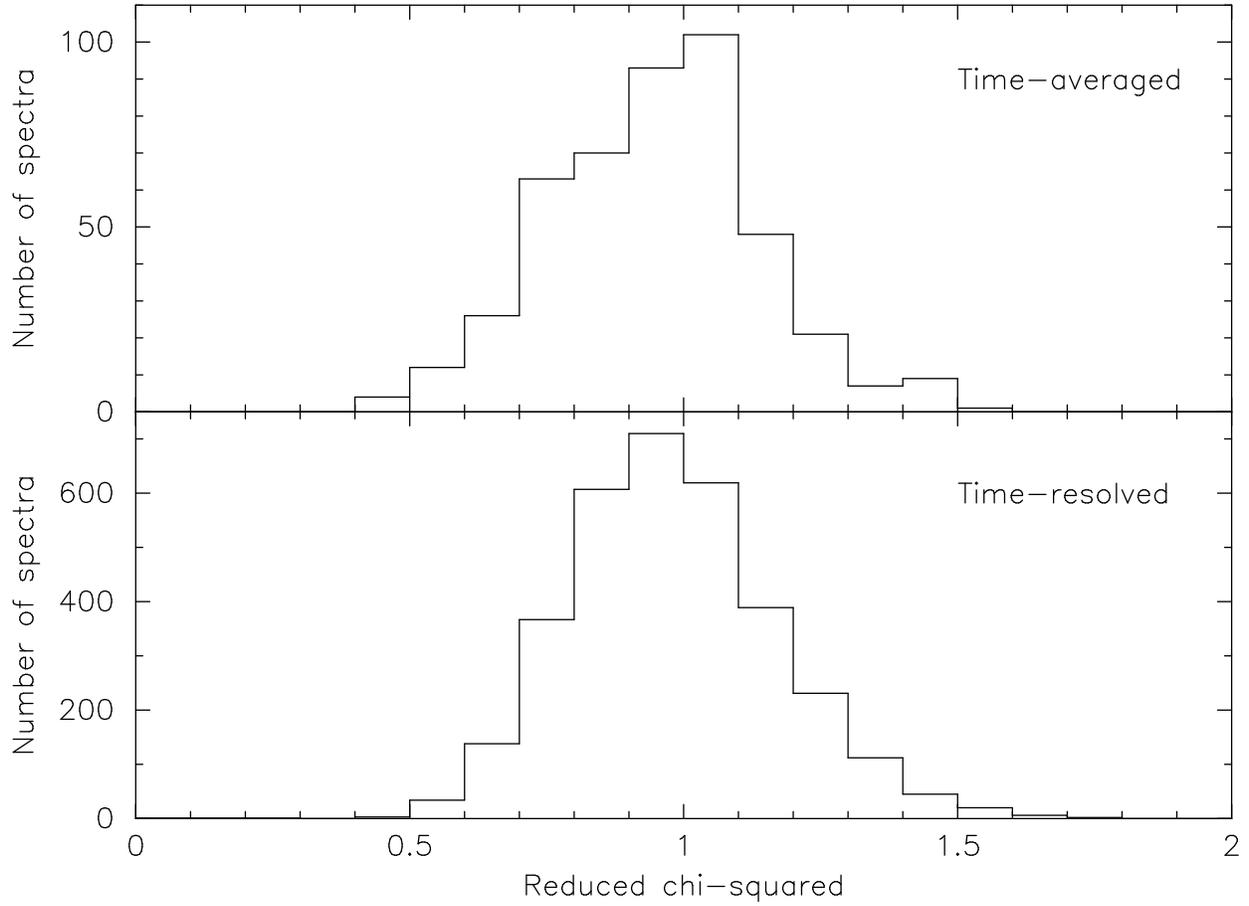}
}
\caption{Histogram of reduced $\chi^{2}$ in the best fit model (either 
a PL or a CPL model) for the real time-averaged spectra (top) and the 
time-resolved spectra (bottom).  The Gaussian fits to these histograms shows,  
respectively, peaks of 0.95 with a $\sigma$ of 0.19 and 0.96 with a $\sigma$ 
of 0.18. \label{chi2_hist}}
\end{figure}

An interesting result from the {\it Fermi} mission is that in at least one burst, 
GRB 090902B, \citep{grb090902b} which has a strong GeV detection by the Large Area Telescope 
shows the presence of an underlying power-law component in addition to a Band function 
component in the time-resolved spectra.  The extra power-law 
component emerges not only in the GeV energy range but also in the $<$50 keV energy range.   
We decided to investigate if such a feature is present as well in the BAT time-resolved spectra.  
To do this, we perform BAT spectral simulations using the spectral parameters 
of the interval b of GRB 090902B (from 4.6 s to 9.6 s) as reported in \citet{grb090902b}: 
the Band function parameters of $\alpha = 0.07$, 
$\beta = -3.9$ and $\eop = 908$ keV, and the extra power-law photon index of $-1.94$.  
We simulate 10,000 spectra using the BAT energy response of 30$^{\circ}$ off-axis with the 
{\tt xspec fakeit} command.   The background from the real data is included in the simulation.  
When then checked to see how well the two models, PL and CPL, fit the simulated spectra.  
Figure \ref{bat_sim_spec} shows one 
example of a simulated spectrum fitted by a PL model.  Since the input spectrum has 
a steep power-law (photon index of $-1.94$) and then breaks to a very flat power-law 
(photon index of 0.07) around 50 keV, the BAT simulated spectrum shows a significant 
deviation from a PL fit.  In this example, the reduced $\chi^{2}$ in a PL fit is 2.93.  
Figure \ref{bat_sim_spec_chi2_hist} shows the reduced $\chi^{2}$ distribution of all the 
BAT simulated spectra when fitted by a PL model. 
A Gaussian fit to this histogram yields a peak of 2.58 with $\sigma$ of 0.28.  
99.97\% of the simulated spectra show a reduced $\chi^{2}$ $\geq$ 1.7.  None of the CPL fits 
to the simulated spectra shows a significant improvement over a PL fit.  
By comparison when we look at the real BAT spectra, we find that 
the reduced $\chi^{2}$ distributions of the time-averaged and 
the time-resolved spectra fitted either by a PL or a CPL model are well 
centered around 1 (Figure \ref{chi2_hist}).  There are only two time-resolved spectra 
(out of 3284) which show a reduced $\chi^{2}$ $>$ 1.7.  
For these two spectra, 
the residuals from the best fit models are not similar to the 
residuals seen in Figure \ref{bat_sim_spec_chi2_hist}.  
For the time-averaged spectra, the poorest 
reduced $\chi^{2}$ is 1.54.  We can therefore conclude that we do not confirm the 
existence of an extra power-law component which extends below 50 keV in the BAT GRB spectra.  
Although GRB 090902B and other {\it Fermi} GRBs for which detections of the extra 
power-law component are claimed could be a special type of GRB which has never been observed by 
the BAT, we believe that confirmations from other GRB instruments are required to 
validate the existence of the extra power-law component, and especially its presence in 
the hard X-ray range of the spectrum. 

\subsection{Observed Properties vs. Redshifts}
\begin{figure}
\centerline{
\includegraphics[width=12cm,angle=-90]{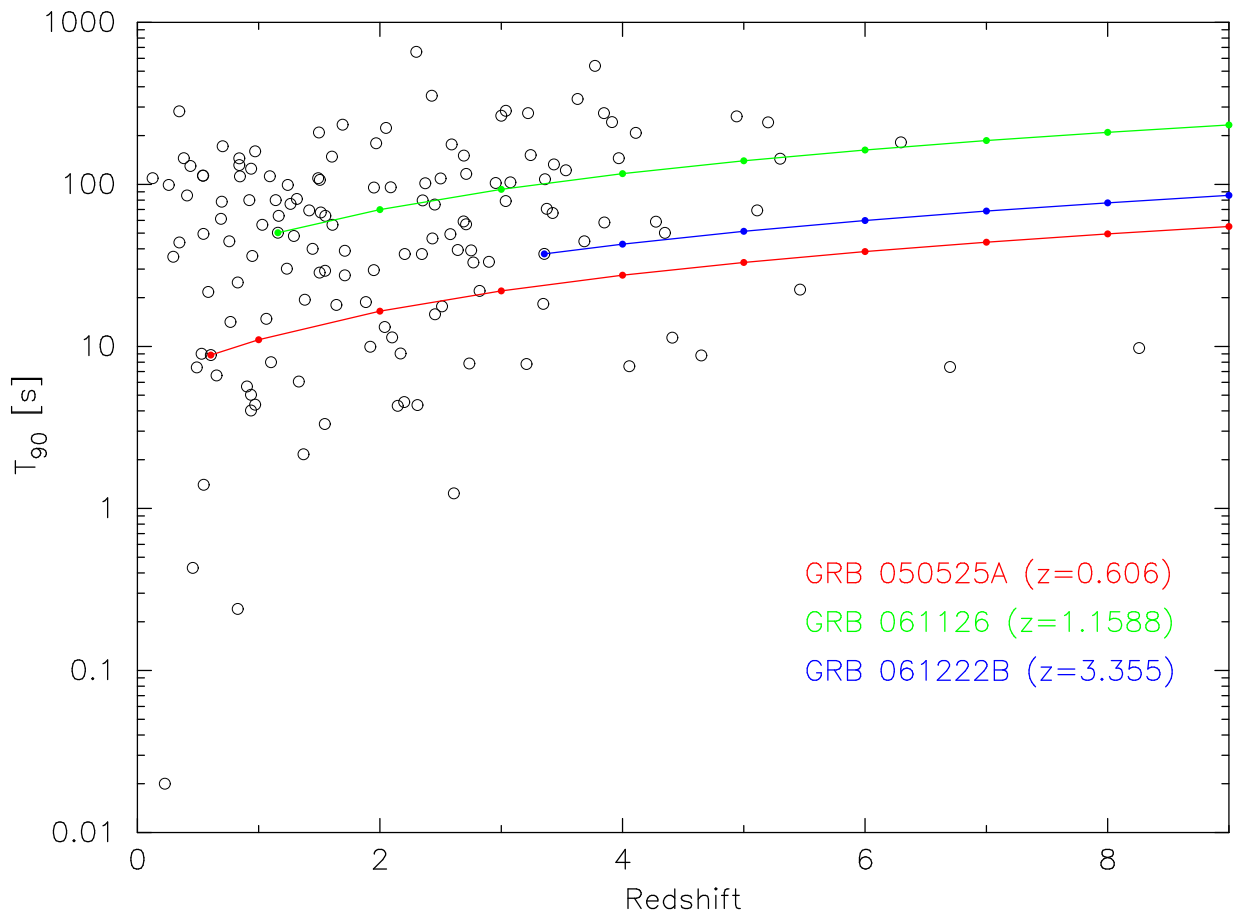}
}
\caption{Distribution of the BAT observed $T_{90}$ vs. redshift.  For three known redshift 
GRBs, GRB 050525A (z=0.606), GRB 061126 (z=1.1588) and GRB 061222B (z=3.355), 
we calculated the trajectories of estimated observed $T_{90}$ as a function of redshift 
assuming that the duration changes only by the time-dilation effect.  \label{t90_obs_z}}
\end{figure}

Figure \ref{t90_obs_z} shows the observed BAT $T_{90}$ duration 
versus redshift.  We also plot the calculated $T_{90}$ that we would 
observe if three particular GRBs, GRB 050525A (z=0.606), GRB 061126 (z=1.1588) and GRB 
061222B (z=3.355) had occurred at different redshifts.  If high-z bursts have 
the same duration distribution in the rest frame as low-z bursts, then 
we would expect the duration of a high-redshift GRB tends to be longer as 
shown in the trajectories.  
Due to the intrinsic scatter in duration and also the relatively small 
number of high-redshift GRBs, it is difficult to conclude whether we see a clear 
indication of the time-dilation effect in the data.  

\begin{figure}
\centerline{
\includegraphics[width=12cm,angle=-90]{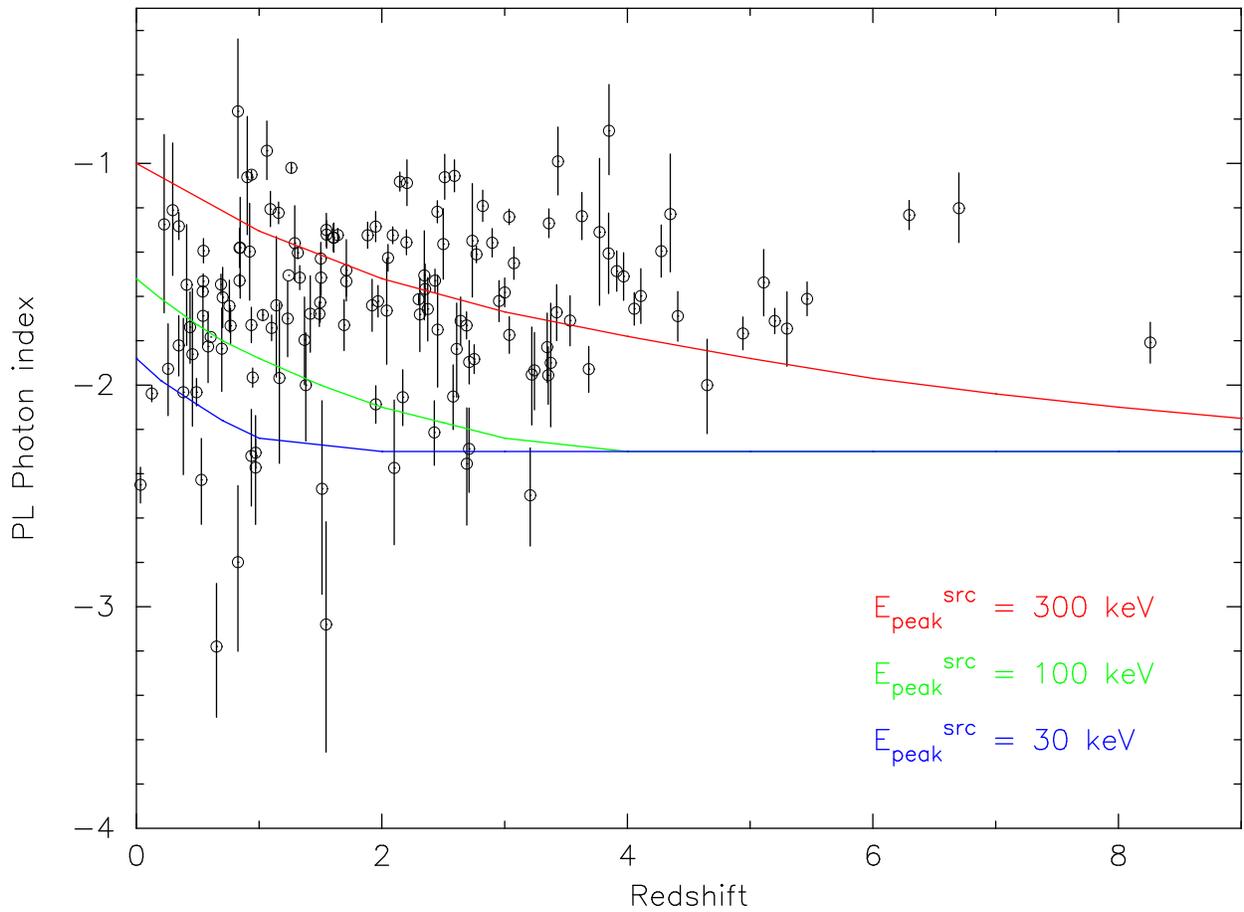}
}
\caption{Distribution of the BAT observed PL photon index vs. redshift.  The overlaid 
curves are the estimates of the BAT observed PL photon index as a function 
of redshift assuming the intrinsic spectrum is the redshifted spectrum of the 
typical Band function shape ($\alpha=-0.87 \pm 0.33 (1\sigma)$ and 
$\beta=-2.36 \pm 0.31 (1\sigma)$; see \citet{ep_gamma})
with the rest-frame $\ep$ of 300 keV (red), 100 keV (green) 
and 30 keV (blue).  
\label{pl_phindex_vs_z}}
\end{figure}

Figure \ref{pl_phindex_vs_z} shows the observed photon 
index of the time-averaged spectra in a PL fit versus redshift.  If there is an 
intrinsic spectral shape in the GRB rest frame, we would expect the observed spectra 
to be softer when the redshift is higher.  The overlaid curves in the figure 
show the expected BAT observed PL photon index as a function of redshift for a model burst with 
$\esp$ of 300 keV, 100 keV and 30 keV.  We used the $\ep$-$\Gamma$ relation \citep{ep_gamma} 
to convert $\eop$ into the BAT PL photon index for these curves.  
There is no obvious observed trend because of the intrinsic scatter in the 
data and relatively small sample of high-redshift GRBs.  

\begin{figure}
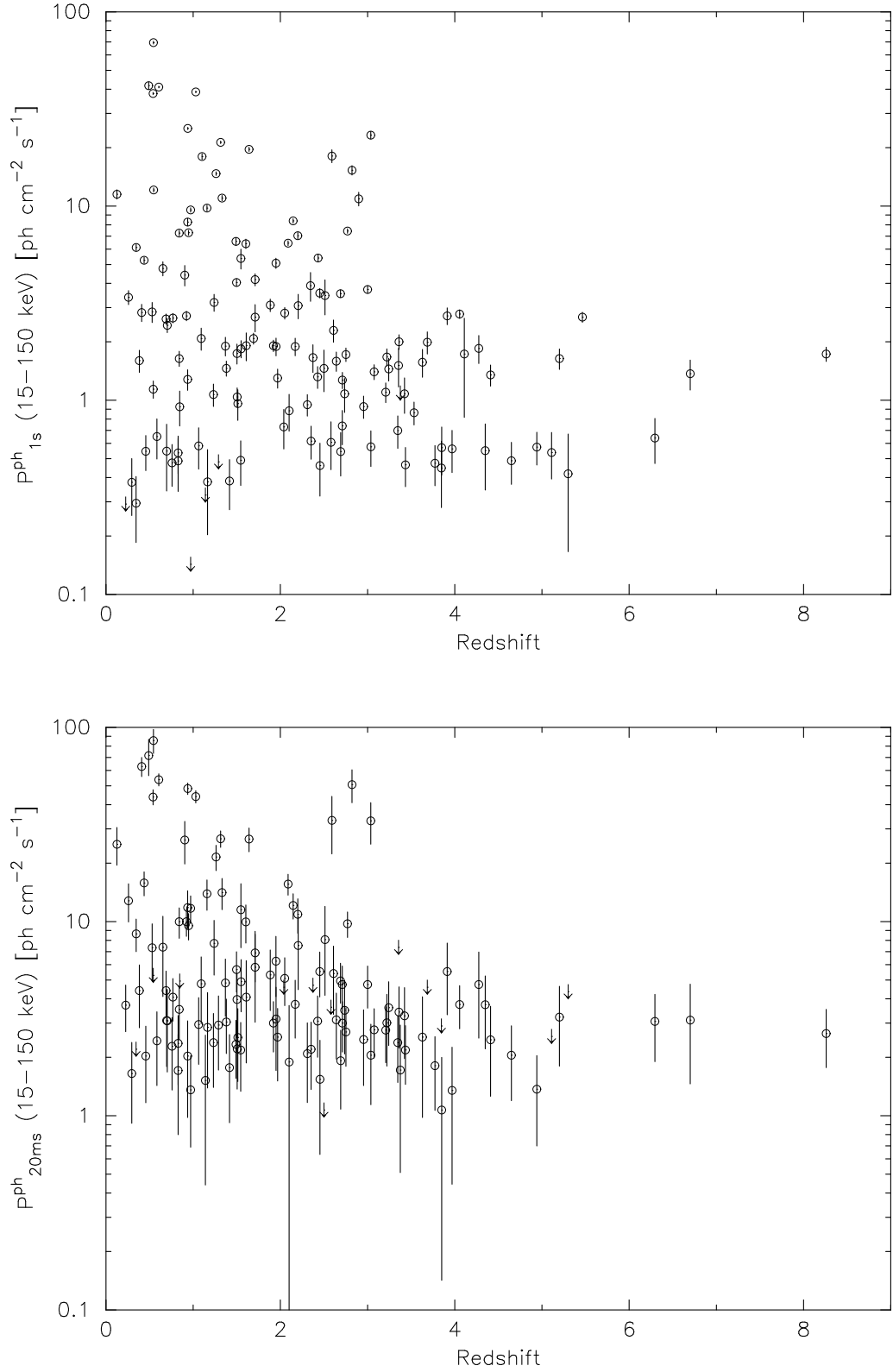

\centerline{
\includegraphics[width=10cm,angle=-90]{fig30a.eps}
}
\vspace{1cm}
\centerline{
\includegraphics[width=10cm,angle=-90]{fig30b.eps}
}
\caption{Distribution of 1 s (top) and 20 ms (bottom) observed peak photon flux 
in the 15-150 keV band vs. redshift. \label{peak_flux_z}}
\end{figure}

\begin{figure}
\centerline{
\includegraphics[width=12cm,angle=-90]{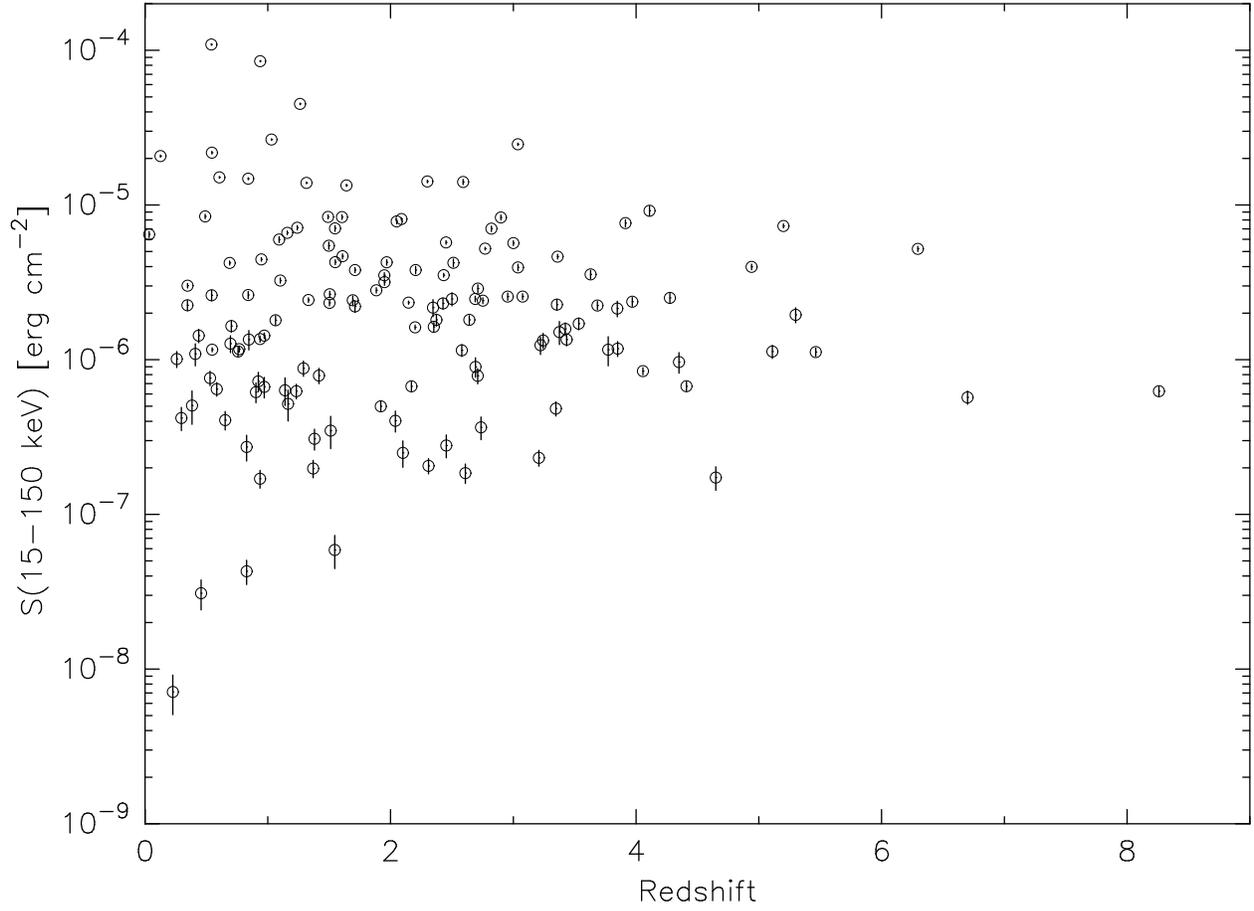}
}
\caption{Distribution of the BAT observed energy fluence in the 15-150 keV band vs. 
redshift.  \label{fluence_z}}
\end{figure}

The distributions of the peak flux and the fluence as a function of redshift are 
shown in Figures \ref{peak_flux_z} and \ref{fluence_z}.  Although the peak flux and 
the fluence measured in the narrow BAT energy band are not bolometric measurements, 
there is no obvious correlation between those parameters and the redshifts.  The fluence 
and the peak flux of the low redshift GRBs are scattered from the lowest to the 
highest values.  The high redshift GRBs are located in the small peak flux and 
the fluence region, but they are not the dimmest populations of the BAT known redshift 
GRBs.  

\subsection{Rest-frame Properties}

\begin{figure}
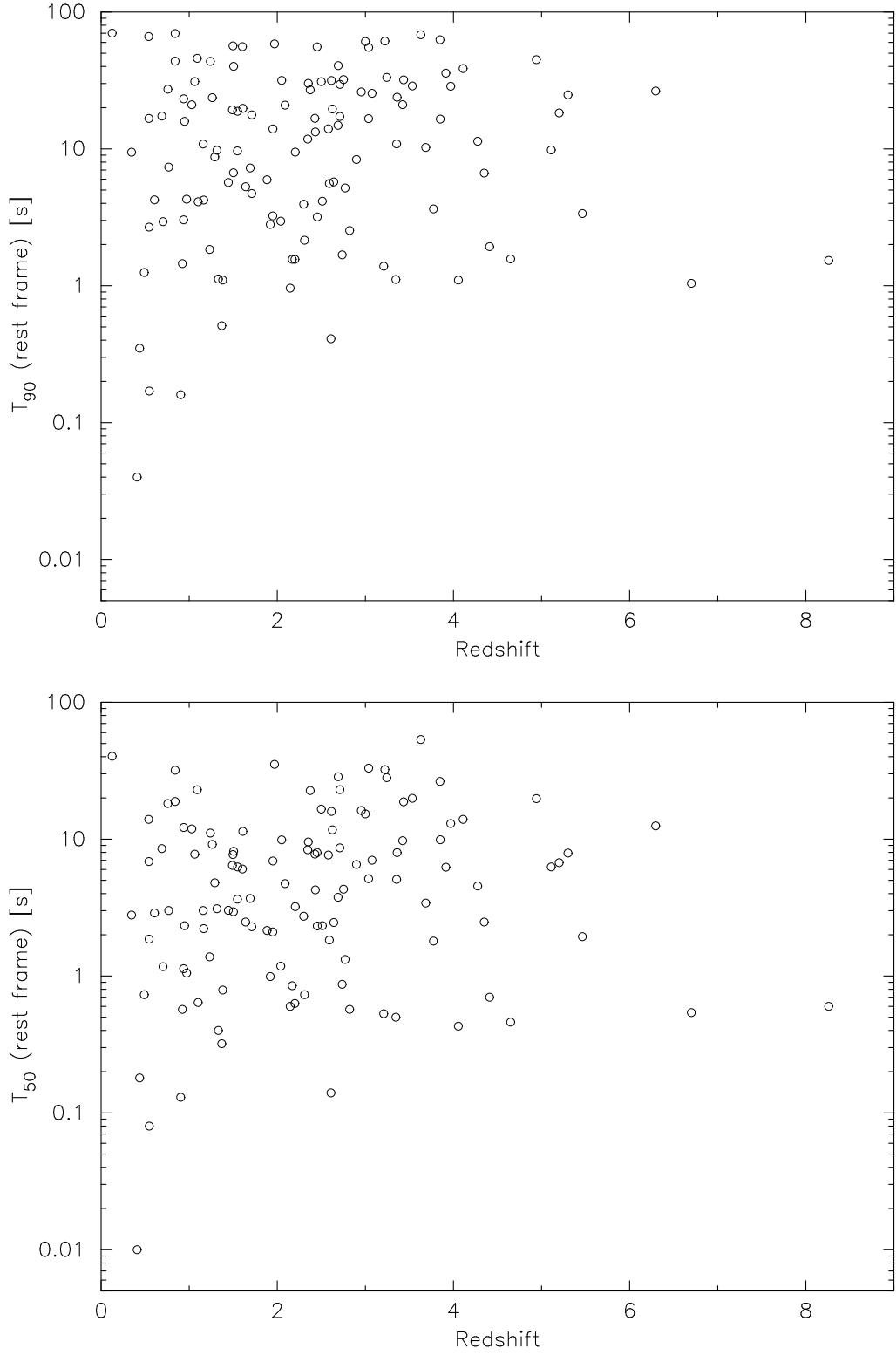

\centerline{
\includegraphics[width=10cm,angle=-90]{fig32a.eps}
}
\vspace{0.5cm}
\centerline{
\includegraphics[width=10cm,angle=-90]{fig32b.eps}
}
\caption{Distributions of $T_{90}$ (top) and $T_{50}$ (bottom) 
in the 140-220 keV band at the GRB rest frame vs. redshift. \label{t90_t50_restframe}}
\end{figure}

Figure \ref{t90_t50_restframe} shows the distribution of $T_{90}$ and $T_{50}$ 
durations calculated in the 140-220 keV band in the GRB rest frame as a function 
of redshift.  The averaged $T_{90}$ and $T_{50}$ durations in the rest frame are 
18.5 s and 8.0 s respectively.  The respective correlation coefficients 
between $T_{90}$ and $T_{50}$ and the redshifts are 0.09 in 122 samples 
(null probability of 0.3) and 0.1 in 121 samples (null probability of 0.3).  
Therefore, there is no clear trend between the duration in the 
GRB rest fame and the redshift.  

\begin{figure}
\centerline{
\includegraphics[width=12cm,angle=-90]{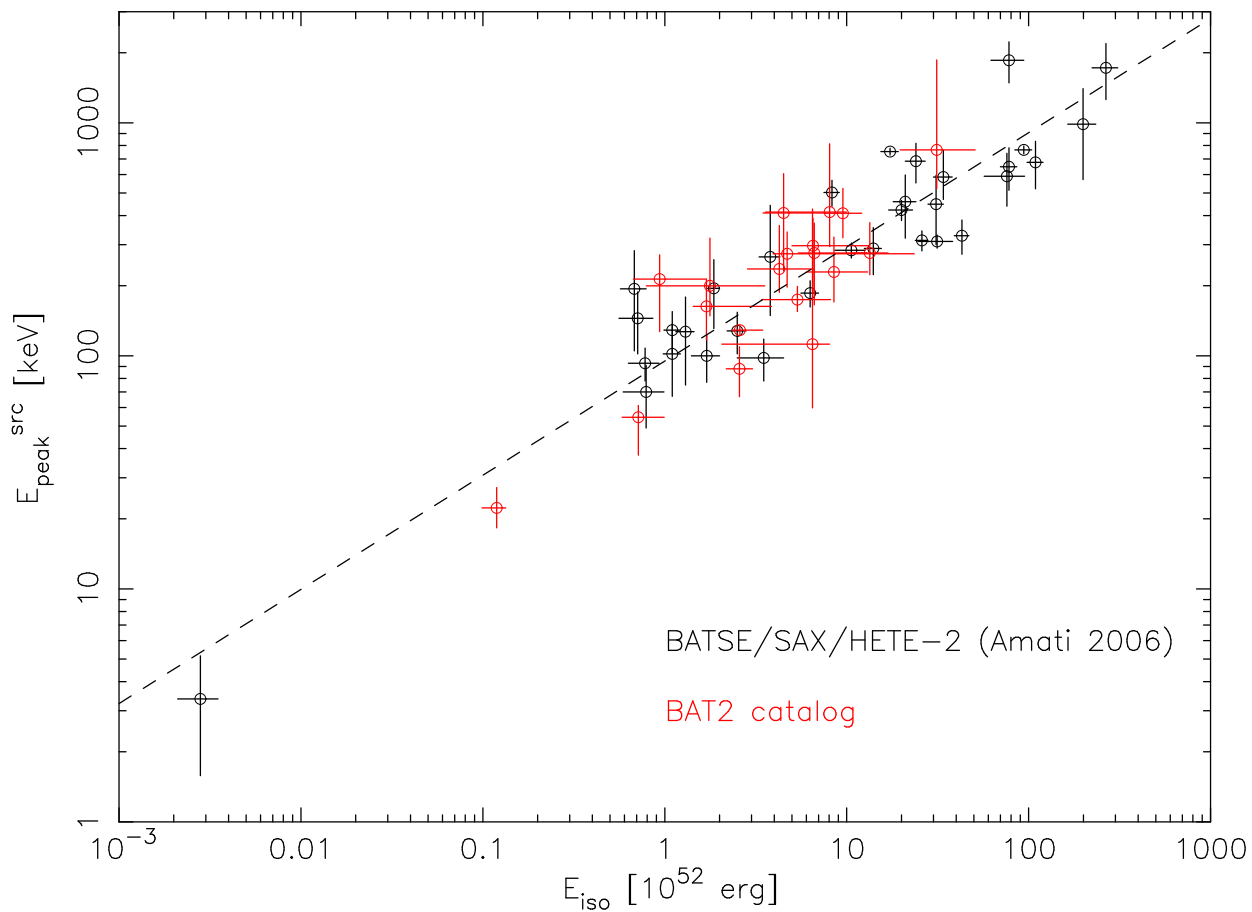}
}
\caption{The correlation between $\esp$ and $\eiso$ for the {\it Swift} GRBs (red) and 
other GRB missions (black).  The dashed line is the best fit correlation between $\esp$ 
and $\eiso$ reported by \citet{amati2006}: $\esp = 95 \times (\eiso/10^{52})^{0.49}$.  
\label{ep_iso}}
\end{figure}

Figure \ref{ep_iso} shows the correlation between $\esp$ and the isotropic-equivalent 
energy $\eiso$.  The $\esp$ and $\eiso$ values of the GRB samples from BATSE, 
{\it BeppoSAX} and {\it HETE}-2 are 
extracted from \citet{amati2006}.  We only select the {\it Swift} GRBs for which 
we reported the time-averaged CPL fits in this catalog and that also have redshift measurements.  
For those bursts, we fit the spectrum with a Band function to measure $\esp$ and 
$\eiso$.   $\eiso$ is calculated by integrating over the 1 keV - 10 MeV band in the GRB rest frame.  
We used the same cosmological parameters of \citet{amati2006} in the calculation 
of $\eiso$ for the {\it Swift} GRBs.  Table \ref{ep_eiso_swift} summarizes the values of 
$\esp$ and $\eiso$ of {\it Swift} GRBs.  As shown in Figure \ref{ep_iso}, the {\it Swift} 
GRB samples are consistent with the samples from other GRB missions, and follow 
the relation originally proposed by \citet{amati2002}.  

\section{Summary}

The BAT2 catalog includes 476 GRBs detected by BAT during 5 years of 
operation.  We present the observed temporal and spectral properties of the BAT 
GRBs mainly based on BAT event data.  In this catalog, we present not only the 
time-averaged but also the time-resolved spectral properties of the BAT GRBs.  
We also distinguish among L-GRBs, S-GRBs, and S-GRBs with E.E. 
to investigate possible distinct characteristics in the prompt emission properties.  
Comparisons of the prompt emission properties among the BATSE, the {\it BeppoSAX} and 
the {\it HETE}-2 GRB 
samples are shown.  The observed prompt emission properties for the BAT known 
redshift GRBs are also presented.  

We have shown that the BAT $T_{90}$ and $T_{50}$ durations peak around 70 s and 30 s, respectively, 
whereas the BATSE, the {\it BeppoSAX} and the {\it HETE}-2 $T_{90}$ durations peak 
around 10-30 s.  This can be understood by the differences 
in the sensitivities of the instruments.  We have confirmed that the 
spectra of the BAT S-GRBs are generally harder than those of the L-GRBs.  The overall hardness 
of the S-GRBs with E.E. is comparable to that of the L-GRBs, 
whereas, the hardness of the initial short spikes of the S-GRBs with E.E. is comparable 
to that of the S-GRBs.  
By comparing the BAT GRBs with the BATSE and the {\it HETE}-2 samples using the fluences 
in the 50-150 keV and the 15-50 keV bands, we have shown that the majority of the 
BAT GRBs are systematically softer than the bright BATSE GRBs, whereas the {\it HETE}-2   
samples overlap with the BAT GRBs in this fluence-fluence plane.  
We have confirmed that the photon indicies of 
PL fits to the BAT S-GRBs are harder than those of the L-GRBs in the time-averaged spectral 
analysis. The distribution of the time-averaged PL photon indices of the S-GRBs with E.E. 
is consistent with that of the L-GRBs.  However, the PL photon indicies of the initial 
short spikes of the S-GRBs with E.E. are much more similar to those of the S-GRBs.  
The time-averaged $\eop$ of the BAT GRBs based on a 
CPL fit shows a log-normal distribution in the peak around 80 keV which is significantly 
smaller than that of the BATSE GRBs which peak around 320 keV.  There is no significant 
difference in the $\eop$ based on a CPL fit between the time-averaged and the 
time-resolved spectra in the BAT data.  We have confirmed that only 10\% of the BAT photon indices 
in the BAT time-resolved spectra are outside the allowed range of the line of death 
(the limit from the SSM).  The intervals which violate the line of death are 
at bright peaks and/or at a rising part of a peak.  
We see no obvious observed trend in the BAT $T_{90}$ and the observed 
spectra with redshifts.  
The $T_{90}$ and $T_{50}$ durations calculated in the 140-220 keV band in the GRB rest frame for the 
BAT known redshift sample are peaked at 19 s and 8 s, respectively.  The BAT GRB samples 
are consistent with the $\esp$-$\eiso$ relation (Amati relation).  

\acknowledgements
We would like to thank C.~R. Shrader for instructing tools (CGRO ftools) to read BATSE data 
by {\tt xspec}.  We would also like to thank the anonymous referee for comments 
and suggestions that materially improved the paper.

\clearpage

\begin{deluxetable}{lccccccccccccc}
\tabletypesize{\scriptsize}
\rotate
\tablecaption{BAT GRB summary\label{tbl:bat_summary}}
\tablewidth{0pt}
\tablehead{
\colhead{GRB} &
\colhead{Trigger} &
\colhead{Trigger time} &
\colhead{R.A.} &
\colhead{Dec.} &
\colhead{R.A.(XRT)} &
\colhead{Dec.(XRT)} &
\colhead{SN$_{img}$} &
\colhead{Error} & 
\colhead{$T_{90}$} &
\colhead{$T_{50}$} &
\colhead{Start} &
\colhead{Stop} &
\colhead{Note}\\
\colhead{Name} &
\colhead{Number} &
\colhead{} &
\colhead{($^{\circ}$)} &
\colhead{($^{\circ}$)} &
\colhead{($^{\circ}$)} &
\colhead{($^{\circ}$)} &
\colhead{($\sigma$)} &
\colhead{($^{\prime}$)} &
\colhead{(s)} &
\colhead{(s)} &
\colhead{(s)} &
\colhead{(s)} &
\colhead{}}
\startdata
 041217 & 100116 & 2004-12-17 07:28:25.920 & 164.790 & -17.944 &       --- &      --- & 19.3 & 1.4 &    5.65 &    2.71 &     -2 &     18 &     \\
041219A & 100318 & 2004-12-19 01:42:18.000 &   6.154 &  62.847 &       --- &      --- &  --- & --- &     --- &     --- &    --- &    --- &  (1)\\
041219B & 100367 & 2004-12-19 15:38:48.000 & 167.674 & -33.458 &       --- &      --- &  --- & --- &     --- &     --- &    --- &    --- &  (1)\\
041219C & 100380 & 2004-12-19 20:30:34.000 & 343.882 & -76.786 &       --- &      --- & 13.2 & 1.8 &   10.00 &    4.00 &     -3 &     17 &     \\
 041220 & 100433 & 2004-12-20 22:58:26.599 & 291.301 &  60.596 &       --- &      --- & 31.9 & 1.0 &    5.58 &    2.20 &   -300 &    302 &     \\
 041223 & 100585 & 2004-12-23 14:06:17.956 & 100.186 & -37.072 &       --- &      --- & 83.7 & 1.0 &  109.08 &   29.20 &   -299 &    303 &     \\
 041224 & 100703 & 2004-12-24 20:20:57.698 &  56.192 &  -6.666 &       --- &      --- & 11.4 & 2.0 &  177.17 &   37.68 &   -299 &    303 &     \\
 041226 & 100815 & 2004-12-26 20:34:18.976 &  79.647 &  73.343 &       --- &      --- &  5.6 & 3.3 &   89.50 &   52.72 &   -299 &    303 & (11)\\
 041228 & 100970 & 2004-12-28 10:49:14.142 & 336.649 &   5.027 &       --- &      --- & 36.5 & 1.0 &   52.16 &   19.54 &   -299 &    303 &     \\
 050117 & 102861 & 2005-01-17 12:52:36.037 & 358.490 &  65.934 &       --- &      --- & 53.8 & 1.0 &  166.65 &   83.51 &   -299 &    303 &     \\
 050124 & 103647 & 2005-01-24 11:30:02.876 & 192.876 &  13.041 &  192.8773 &  13.0443 & 35.0 & 1.0 &    3.93 &    1.88 &   -299 &    303 &     \\
 050126 & 103780 & 2005-01-26 12:00:54.073 & 278.134 &  42.395 &       --- &      --- & 15.2 & 1.6 &   48.00 &   16.00 &   -299 &    303 &     \\
 050128 & 103906 & 2005-01-28 04:19:55.191 & 219.584 & -34.762 &  219.5737 & -34.7654 & 26.3 & 1.1 &   28.00 &    8.00 &   -300 &    303 &     \\
 050202 & 104298 & 2005-02-02 03:35:14.800 & 290.583 & -38.733 &       --- &      --- &  9.7 & 2.2 &    0.11 &    0.05 &   -299 &    243 & (2) \\
050215A & 106106 & 2005-02-15 02:15:28.543 & 348.411 &  49.322 &       --- &      --- &  7.8 & 2.6 &   66.41 &   49.00 &   -120 &    302 &     \\
050215B & 106107 & 2005-02-15 02:33:43.199 & 174.467 &  40.792 &  174.4473 &  40.7974 & 15.4 & 1.6 &   10.62 &    4.74 &   -240 &    229 &     \\
050219A & 106415 & 2005-02-19 12:40:01.049 & 166.413 & -40.685 &  166.4125 & -40.6841 & 34.5 & 1.0 &   23.84 &   10.06 &   -299 &    303 &     \\
\enddata
\tablenotetext{1}{The event data are not available.}
\tablenotetext{2}{Short duration GRB.}
\tablenotetext{3}{The event data of the part of the burst emission are not available.}
\tablenotetext{4}{Short duration GRB with an extended emission.}
\tablenotetext{5}{battblocks failed because of the weak nature of the burst.}
\tablenotetext{6}{GRB found by the ground process.}
\tablenotetext{7}{The detector plane histogram data are used in the fluence calculation and the spectral analysis.}
\tablenotetext{8}{Possible GRB.}
\tablenotetext{9}{GRB found by the BAT slew survey process.}
\tablenotetext{10}{T90 and T50 are lower limit.}
\tablenotetext{11}{Re-calculate the image significance using the interval determined by the flight software.}
\end{deluxetable}

\begin{deluxetable}{lcccccccccccc}
\tabletypesize{\scriptsize}
\tablecaption{BAT GRB energy fluence \label{tbl:bat_fluence}}
\tablewidth{0pt}
\tablehead{
\colhead{GRB} &
\colhead{Spectral} &
\colhead{S(15-25)} &
\colhead{S(25-50)} &
\colhead{S(50-100)} &
\colhead{S(100-150)} &
\colhead{S(15-150)} &
\colhead{Start} &
\colhead{Stop} & 
\colhead{Note}\\
\colhead{Name} &
\colhead{Model} &
\multicolumn{5}{c}{(10$^{-8}$ ergs cm$^{-2}$)} &
\colhead{(s)} &
\colhead{(s)}}
\startdata
 041217 & CPL &  $30.0 \pm 3.1$ &  $72.1 \pm 3.3$ & $106.0 \pm 5.3$  &  $61.1 \pm 8.1$ & $270.0 \pm  12.1$ &   +0.82 &   +7.89 & \\
041219A & --- &              --- &               --- &                --- &               --- &                 --- &     --- &     --- & (1)\\
041219B & --- &              --- &               --- &                --- &               --- &                 --- &     --- &     --- & (1)\\ 
041219C &  PL &  $25.5 \pm 2.0$ &  $34.5 \pm 1.6$ & $34.4 \pm 2.5$   &  $20.0 \pm 2.3$ & $114.0 \pm  5.8$  &   +0.00 &  +12.00 & \\ 
 041220 &  PL &   $5.9 \pm  0.6$ &   $9.8 \pm  0.7$ &  $12.3 \pm  1.2$ &   $8.6 \pm   1.3$ &  $36.6 \pm   2.8$ &   -0.21 &   +6.81 & \\
 041223 &  PL & $153.0 \pm   5.8$ & $347.0 \pm  7.7$ & $624.0 \pm  10.4$ & $576.0 \pm  15.3$ & $1700.0 \pm  27.2$ &  -10.54 & +145.94 & \\ 
 041224 & CPL & $133.0 \pm  11.0$ & $261.0 \pm  10.4$ & $309.0 \pm  14.2$ & $149.0 \pm  21.1$ & $853.0 \pm  34.3$ & -109.63 & +116.27 & \\ 
 041226 &  PL &   $3.8 \pm   1.5$ &   $7.4 \pm   1.8$ &  $11.0 \pm   3.4$ &   $8.9 \pm   4.3$ &  $31.0 \pm   8.2$ &   -1.34 &  +93.50 & \\ 
 041228 &  PL &  $52.8 \pm   4.2$ &  $90.2 \pm   4.2$ & $118.0 \pm   6.1$ &  $84.7 \pm   7.1$ & $345.0 \pm  14.8$ &   -0.40 &  +66.78 & \\ 
 050117 & CPL & $112.0 \pm 7.1$   & $226.0 \pm 5.9$   & $318.0 \pm 9.2$   & $211.0 \pm 15.6$ & $867.0 \pm 23.0$ & +4.28 & +215.86 &\\
 050124 & CPL & $12.7 \pm 1.7$    & $30.3 \pm 2.0$    & $46.3 \pm 3.1$    & $28.5 \pm 4.5$   & $118.0 \pm 6.9$ & -3.90 & +3.07 &\\
 050126 &  PL & $10.1 \pm 1.8$    & $20.3 \pm 2.2$    & $31.6 \pm 4.0$    & $26.1 \pm 5.0$   & $88.0 \pm 10.0$ & -11.44 & +36.56 &\\
 050128 & CPL & $51.2 \pm 6.3$    & $125.0 \pm 6.8$   & $201.0 \pm 11.3$  & $133.0 \pm 15.2$ & $511.0 \pm 23.7$ & -7.56 & +25.44 &\\
 050202 &  PL & $0.4 \pm 0.1$     & $0.7 \pm 0.1$     & $1.1 \pm 0.2$     & $0.9 \pm 0.3$    & $3.2 \pm 0.6$ & +0.00 & +0.14 &\\
050215A &  PL & $7.1 \pm 1.9$     & $14.4 \pm 2.2$    & $23.0 \pm 4.0$    & $19.4 \pm 5.5$   & $63.9 \pm 9.9$ & -0.32 & +71.07 &\\
050215B &  PL & $6.1 \pm 0.9$     & $7.4 \pm 0.8$     & $6.6 \pm 1.4$     & $3.5 \pm 1.1$    & $23.5 \pm 3.2$ & -1.98 & +10.86 &\\
050219A & CPL & $33.9 \pm 3.7$    & $102.0 \pm 4.5$   & $173.0 \pm 7.5$   & $98.4 \pm 10.6$  & $408.0 \pm 15.9$ & -5.81 & +29.98 &\\
050219B & CPL & $186.0 \pm 16.8$  & $413.0 \pm 15.5$  & $604.0 \pm 23.7$  & $381.0 \pm 29.3$ & $1580.0 \pm 49.5$ & -35.59 & +38.57 &\\
 050223 &  PL & $12.2 \pm 1.6$    & $18.4 \pm 1.6$    & $20.7 \pm 2.8$    & $13.3 \pm 2.8$   & $64.6 \pm 6.6$ & -12.01 & +16.57 &\\
 050306 & CPL & $135.0 \pm 14.6$  & $287.0 \pm 13.6$  & $426.0 \pm 21.1$  & $294.0 \pm 27.5$ & $1140.0 \pm 33.9$ & -3.42 & +187.84 &\\
 050315 &  PL & $75.9 \pm 5.4$    & $97.8 \pm 4.2$    & $92.0 \pm 6.2$    & $51.3 \pm 5.5$   & $317.0 \pm 14.3$ & -56.24 & +69.18 &\\
 050318 & --- &           ---     &            ---    &            ---    &            ---   &              --- & --- & --- & (2)\\
 050319 &  PL & $27.8 \pm 3.7$    & $39.3 \pm 3.6$    & $41.1 \pm 6.5$    & $24.9 \pm 6.0$   & $133.0 \pm 14.9$ & -134.09 & +27.76 &\\
 050326 &  PL & $90.4 \pm 3.5$    & $192.0 \pm 4.3$   & $321.0 \pm 5.9$   & $280.0 \pm 8.5$  & $884.0 \pm 15.2$ & -9.84 & +32.42 &\\
 050401 &  PL & $94.8 \pm 7.3$    & $191.0 \pm 8.7$   & $298.0 \pm 11.9$  & $247.0 \pm 15.4$ & $831.0 \pm 30.8$ & -7.68 & +29.06 &\\
 050406 & CPL & $2.2 \pm 0.5$     & $3.4 \pm 0.8$     & $1.1 \pm 0.8$     & 0.35             & $6.7 \pm 1.4$ & -2.50 & +3.90 & (3)\\
 050410 & CPL & $59.6 \pm 6.9$    & $128.0 \pm 7.3$   & $164.0 \pm 10.8$  & $82.9 \pm 15.6$  & $435.0 \pm 24.9$ & -27.02 & +36.98 &\\
 050412 &  PL & $3.3 \pm 0.7$     & $9.6 \pm 1.2$     & $23.2 \pm 2.0$    & $26.8 \pm 3.7$   & $62.9 \pm 5.7$ & -8.28 & +22.16 &\\
050416A & CPL & $19.7 \pm 2.5$    & $13.3 \pm 2.2$    & $5.9 \pm 2.0$     & $1.8 \pm 0.9$    & $40.7 \pm 5.6$ & +0.03 & +10.14 &\\
050416B & CPL & $10.3 \pm 2.3$    & $27.9 \pm 2.7$    & $45.6 \pm 4.6$    & $27.2 \pm 6.8$   & $111.0 \pm 10.1$ & +0.06 & +4.20 &\\
 050418 & CPL & $77.9 \pm 6.4$    & $143.0 \pm 5.7$   & $181.0 \pm 8.2$   & $109.0 \pm 12.8$ & $511.0 \pm 19.9$ & -8.35 & +80.50 &\\
 050421 &  PL & $2.1 \pm 0.8$     & $3.7 \pm 0.9$     & $4.9 \pm 1.7$     & $3.6 \pm 1.9$    & $14.3 \pm 4.0$ & +0.05 & +18.11 &\\
 050422 &  PL & $7.2 \pm 1.5$     & $14.1 \pm 1.8$    & $21.2 \pm 3.3$    & $17.1 \pm 4.1$   & $59.6 \pm 8.1$ & -11.30 & +68.70 &\\
050502B &  PL & $7.0 \pm 0.8$     & $12.1 \pm 0.9$    & $16.2 \pm 1.8$    & $11.8 \pm 2.0$   & $47.1 \pm 4.2$ & -17.32 & +3.18 &\\
 050505 &  PL & $30.0 \pm 4.0$    & $58.8 \pm 4.8$    & $89.3 \pm 7.1$    & $72.5 \pm 8.9$   & $251.0 \pm 18.1$ & -10.06 & +52.51 &\\
 050507 & --- &            ---    &            ---    &            ---    &            ---   &  --- & --- & --- & (2)\\
050509A &  PL & $8.3 \pm 1.0$     & $10.4 \pm 0.8$    & $9.7 \pm 1.5$     & $5.3 \pm 1.3$    & $33.6 \pm 3.3$ & -6.26 & +8.06 &\\
050509B &  PL & $0.08 \pm 0.03$   & $0.16 \pm 0.04$   & $0.26 \pm 0.08$   & $0.2 \pm 0.1$    & $0.7 \pm 0.2$ & +0.00 & +0.03 &\\
050525A & CPL & $220.0 \pm 9.8$   & $440.0 \pm 7.6$   & $555.0 \pm 8.8$   & $299.0 \pm 8.1$  & $1510.0 \pm 20.0$ & -0.12 & +12.68 &\\
 050528 &  PL & $12.1 \pm 2.2$    & $14.3 \pm 2.0$    & $12.1 \pm 3.0$    & $6.2 \pm 2.3$    & $44.7 \pm 7.0$ & -8.20 & +4.98 &\\
 050603 &  PL & $66.3 \pm 5.4$    & $147.0 \pm 7.1$   & $257.0 \pm 11.2$  & $233.0 \pm 16.1$ & $703.0 \pm 28.9$ & -17.77 & +33.23 &\\
\enddata
\tablenotetext{1}{The event data are not available.}
\tablenotetext{2}{The event data of the part of the burst emission are not available.}
\tablenotetext{3}{S(100-150) is an upper limit.}
\tablenotetext{4}{S(50-100) is an upper limit.}
\end{deluxetable}

\begin{deluxetable}{lccccccccccc}
\tabletypesize{\scriptsize}
\tablecaption{BAT GRB 1-s peak photon flux \label{tbl:bat_peak1sphflux}}
\tablewidth{0pt}
\tablehead{
\colhead{GRB} &
\colhead{Spectral} &
\colhead{F$^{\rm p}_{\rm ph}$(15-25)} &
\colhead{F$^{\rm p}_{\rm ph}$(25-50)} &
\colhead{F$^{\rm p}_{\rm ph}$(50-100)} &
\colhead{F$^{\rm p}_{\rm ph}$(100-150)} &
\colhead{F$^{\rm p}_{\rm ph}$(15-150)} &
\colhead{Start} &
\colhead{Note}\\
\colhead{Name} &
\colhead{Model} &
\multicolumn{5}{c}{(photons cm$^{-2}$ s$^{-1}$)} &
\colhead{(s)} &
}
\startdata
041217 & PL & $2.09 \pm 0.25$ & $2.24 \pm 0.17$ & $1.70 \pm 0.14$ & $0.796 \pm 0.106$ & $6.82 \pm 0.48$ & +3.26 & \\
041219A & --- & --- & --- & --- & --- & --- & --- & (1)\\
041219B & --- & --- & --- & --- & --- & --- & --- & (1)\\
041219C & PL & $0.88 \pm 0.14$ & $0.78 \pm 0.04$ & $0.48 \pm 0.03$ & $0.19 \pm 0.02$ & $2.33 \pm 0.23$ & +2.00 & \\
 041220 & PL & $0.52 \pm 0.03$ & $0.58 \pm 0.02$ & $0.46 \pm 0.02$ & $0.22 \pm 0.02$ & $1.78 \pm 0.13$ & -0.02 & \\
 041223 & PL & $1.51 \pm 0.11$ & $2.20 \pm 0.10$ & $2.38 \pm 0.10$ & $1.48 \pm 0.05$ & $7.58 \pm 0.29$ & +35.01 & \\
 041224 & PL & $0.82 \pm 0.14$ & $0.95 \pm 0.10$ & $0.79 \pm 0.04$ & $0.40 \pm 0.03$ & $2.95 \pm 0.29$ & -0.06 & \\
 041226 & PL & $0.09 \pm 0.02$ & $0.11 \pm 0.02$ & $0.09 \pm 0.02$ & $0.05 \pm 0.01$ & $0.34 \pm 0.05$ & +3.31 & \\
 041228 & PL & $0.63 \pm 0.15$ & $0.56 \pm 0.04$ & $0.34 \pm 0.03$ & $0.13 \pm 0.02$ & $1.65 \pm 0.24$ & +22.03 & \\
 050117 & PL & $0.51 \pm 0.03$ & $0.70 \pm 0.03$ & $0.70 \pm 0.03$ & $0.41 \pm 0.03$ & $2.32 \pm 0.16$ & +87.22 & \\
 050124 & CPL & $4.20 \pm 0.63$ & $11.9 \pm 0.9$ & $19.5 \pm 1.5$     & $11.1 \pm 2.2$   & $46.6 \pm 3.3$   & -0.1 &\\
 050126 &  PL & $0.64$          & $0.91$           & $1.56$           &           $2.21$ & $4.22$           & +12.1 & (3,4,5,6,7)\\
 050128 &  PL & $6.73 \pm 1.02$ & $13.9 \pm 1.3$ & $22.2 \pm 2.2$     & $18.7 \pm 2.8$   & $61.6 \pm 5.66$  & +5.4 &\\
 050202 &  PL & $0.58 \pm 0.31$ & $1.09 \pm 0.34$  & $1.57 \pm 0.55$  & $1.22 \pm 0.69$  & $4.46 \pm 1.31$  & -0.4 &\\
050215A &  PL & $0.27 \pm 0.16$ & $0.78 \pm 0.28$  & $1.86 \pm 0.46$  & $2.11 \pm 0.80$  & $5.02 \pm 1.27$  & +5.6 &\\
050215B &  PL & $0.78 \pm 0.20$ & $1.24 \pm 0.23$  & $1.47 \pm 0.42$  & $0.98 \pm 0.42$  & $4.47 \pm 1.00$  & +0.1 &\\
050219A &  PL & $2.78 \pm 0.47$ & $6.38 \pm 0.67$  & $11.6 \pm 1.2$   & $10.7 \pm 1.7$   & $31.4 \pm 3.1$   & +9.5 &\\
050219B & CPL & $23.0 \pm 2.6$  & $50.7 \pm 3.1$   & $69.6 \pm 4.9$   & $40.2 \pm 5.9$   & $184.0 \pm 10.0$ & +2.8 &\\
 050223 &  PL & $0.65 \pm 0.26$ & $1.22 \pm 0.30$  & $1.76 \pm 0.55$  & $1.37 \pm 0.67$  & $5.00 \pm 1.32$  & +1.6 &\\
 050306 &  PL & $2.88 \pm 0.71$ & $6.39 \pm 0.96$  & $11.1 \pm 1.5$   & $10.0 \pm 2.0$   & $30.4 \pm 3.9$   & +107.9 &\\
 050315 &  PL & $2.67 \pm 0.38$ & $3.47 \pm 0.36$  & $3.30 \pm 0.57$  & $1.85 \pm 0.48$  & $11.3 \pm 1.4$   & +24.6 &\\
 050318 & --- &             --- &             ---  &              --- &              --- &              --- & --- & (2)\\
 050319 &  PL & $1.99 \pm 0.35$ & $2.69 \pm 0.36$  & $2.68 \pm 0.60$  & $1.56 \pm 0.52$  & $8.92 \pm 1.43$  & +0.6 &\\
 050326 &  PL & $8.13 \pm 0.40$ & $21.0 \pm 0.6$   & $43.6 \pm 1.4$   & $44.8 \pm 2.1$   & $117.0 \pm 3.6$  & -0.1 &\\
 050401 & CPL & $6.88 \pm 1.53$ & $21.2 \pm 2.1$   & $41.7 \pm 4.1$   & $29.6 \pm 5.3$   & $99.2 \pm 7.9$   & +24.3 &\\
 050406 &  PL & $0.53 \pm 0.18$ & $0.65 \pm 0.20$  & $0.57 \pm 0.32$  & $0.30 \pm 0.25$  & $2.04 \pm 0.75$  & +0.5 &\\
 050410 &  PL & $1.91 \pm 0.62$ & $2.98 \pm 0.68$  & $3.51 \pm 1.12$  & $2.33 \pm 1.10$  & $10.7 \pm 2.7$   & -3.5 &\\
 050412 &  PL & $0.23 \pm 0.11$ & $0.74 \pm 0.22$  & $1.98 \pm 0.38$  & $2.47 \pm 0.72$  & $5.42 \pm 1.07$  & +1.1 &\\
050416A & CPL & $8.64 \pm 0.92$ & $9.44 \pm 1.13$  & $1.75 \pm 0.94$  &              --- & $19.9 \pm 1.9$   & +0.0 & (6)\\
050416B &  PL & $4.27 \pm 0.82$ & $10.6 \pm 1.3$   & $21.2 \pm 2.1$   & $21.2 \pm 3.1$   & $57.3 \pm 5.4$   & +0.1 &\\
 050418 &  PL & $2.79 \pm 0.38$ & $6.57 \pm 0.55$  & $12.3 \pm 1.0$   & $11.7 \pm 1.4$   & $33.4 \pm 2.5$   & +0.7 &\\
 050421 &  PL & $0.48 \pm 0.19$ & $0.82 \pm 0.21$  & $1.05 \pm 0.41$  & $0.75 \pm 0.45$  & $3.10 \pm 0.97$  & +0.4 &\\
 050422 &  PL & $0.36 \pm 0.19$ & $0.51 \pm 0.20$  & $0.55 \pm 0.38$  & $0.80 \pm ---$   & $1.76 \pm 0.87$  & +60.2 & (6)\\
050502B & CPL & $1.12 \pm 0.20$ & $3.10 \pm 0.32$  & $4.47 \pm 0.52$  & $2.01 \pm 0.81$  & $10.7 \pm 1.2$   & +0.2 &\\
 050505 &  PL & $1.39 \pm 0.43$ & $3.29 \pm 0.64$  & $6.17 \pm 0.95$  & $5.89 \pm 1.32$  & $16.7 \pm 2.5$   & +1.0 &\\
 050507 & --- &            ---  &              --- &              --- &              --- &              --- & --- & (2)\\
\enddata
\tablenotetext{1}{The event data are not available.}
\tablenotetext{2}{The event data of the part of the burst emission are not available.}
\tablenotetext{3}{F$^{\rm p}_{\rm ph}$(15-25) is an upper limit.}
\tablenotetext{4}{F$^{\rm p}_{\rm ph}$(25-50) is an upper limit.}
\tablenotetext{5}{F$^{\rm p}_{\rm ph}$(50-100) is an upper limit.}
\tablenotetext{6}{F$^{\rm p}_{\rm ph}$(100-150) is an upper limit.}
\tablenotetext{7}{F$^{\rm p}_{\rm ph}$(15-150) is an upper limit.}
\end{deluxetable}

\begin{deluxetable}{lccccccccc}
\tabletypesize{\scriptsize}
\tablecaption{BAT GRB 1-s peak energy flux \label{tbl:bat_peak1seneflux}}
\tablewidth{0pt}
\tablehead{
\colhead{GRB} &
\colhead{Spectral} &
\colhead{F$^{\rm p}_{\rm ene}$(15-25)} &
\colhead{F$^{\rm p}_{\rm ene}$(25-50)} &
\colhead{F$^{\rm p}_{\rm ene}$(50-100)} &
\colhead{F$^{\rm p}_{\rm ene}$(100-150)} &
\colhead{F$^{\rm p}_{\rm ene}$(15-150)} &
\colhead{Note}\\
\colhead{Name} &
\colhead{Model} &
\multicolumn{5}{c}{(10$^{-8}$ ergs cm$^{-2}$ s$^{-1}$)} &
}
\startdata
 041217 &  PL & $6.5 \pm 0.8$  & $12.7 \pm 0.9$ & $19.3 \pm 1.7$  & $15.7 \pm 2.1$ & $54.2 \pm 4.2$ &\\
041219A & --- &           ---  &            --- &            ---  &            --- &            --- & (1)\\
041219B & --- &           ---  &            --- &            ---  &            --- &            --- & (1)\\
041219C &  PL & $2.7 \pm 0.4$  & $4.4 \pm 0.4$  & $5.4 \pm 0.7$   & $3.7 \pm 0.8$  & $16.3 \pm 1.8$ &\\
 041220 &  PL & $1.6 \pm 0.2$  & $3.3 \pm 0.3$  & $5.2 \pm 0.5$   & $4.4 \pm 0.7$  & $14.5 \pm 1.3$ &\\
 041223 &  PL & $4.8 \pm 0.4$  & $12.8 \pm 0.6$ & $27.7 \pm 1.2$  & $29.4 \pm 1.9$ & $74.6 \pm 3.2$ &\\
 041224 &  PL & $2.5 \pm 0.4$  & $5.4 \pm 0.6$  & $9.0 \pm 1.0$   & $7.9 \pm 1.3$  & $24.9 \pm 2.6$ &\\
 041226 &  PL & $0.3 \pm 0.1$  & $0.6 \pm 0.2$  & $1.1 \pm 0.4$   & $1.0 \pm 0.5$  & $2.9 \pm 0.9$ &\\
 041228 &  PL & $2.0 \pm 0.5$  & $3.1 \pm 0.5$  & $3.8 \pm 0.8$   & $2.6 \pm 0.8$  & $11.4 \pm 1.8$ &\\
 050117 &  PL & $1.6 \pm 0.2$  & $4.0 \pm 0.3$  & $8.1 \pm 0.6$   & $8.2 \pm 1.0$  & $21.9 \pm 1.7$ &\\
 050124 & CPL & $4.2 \pm 0.6$  & $11.9 \pm 0.9$ & $19.5 \pm 1.5$  & $11.1 \pm 2.2$ & $46.6 \pm 3.3$ &\\
 050126 &  PL & $0.6$          &          $0.9$ & $1.6$           & $2.2$          & $4.2$ & (3,4,5,6,7)\\
 050128 &  PL & $6.7 \pm 1.0$  & $13.9 \pm 1.3$ & $22.2 \pm 2.2$  & $18.7 \pm 2.8$ & $61.6 \pm 5.7$ &\\
 050202 &  PL & $0.6 \pm 0.3$  & $1.1 \pm 0.3$  & $1.6 \pm 0.6$   & $1.2 \pm 0.7$  & $4.5 \pm 1.3$ &\\
050215A &  PL & $0.3 \pm 0.2$  & $0.8 \pm 0.3$  & $1.9 \pm 0.5$   & $2.1 \pm 0.8$  & $5.0 \pm 1.3$ &\\
050215B &  PL & $0.8 \pm 0.2$  & $1.2 \pm 0.2$  & $1.5 \pm 0.4$   & $1.0 \pm 0.4$  & $4.5 \pm 1.0$ &\\
050219A &  PL & $2.8 \pm 0.5$  & $6.4 \pm 0.7$  & $11.6 \pm 1.2$  & $10.7 \pm 1.7$ & $31.4 \pm 3.1$ &\\
050219B & CPL & $23.0 \pm 2.6$ & $50.7 \pm 3.1$ & $69.6 \pm 4.9$  & $40.2 \pm 5.9$ & $184.0 \pm 10.0$ &\\
 050223 &  PL & $0.7 \pm 0.3$  & $1.2 \pm 0.3$  & $1.8 \pm 0.6$   & $1.4 \pm 0.7$  & $5.0 \pm 1.3$ &\\
 050306 &  PL & $2.9 \pm 0.7$  & $6.4 \pm 1.0$  & $11.1 \pm 1.5$  & $10.0 \pm 2.0$ & $30.4 \pm 3.9$ &\\
 050315 &  PL & $2.7 \pm 0.4$  & $3.5 \pm 0.4$  & $3.3 \pm 0.6$   & $1.9 \pm 0.5$  & $11.3 \pm 1.4$ &\\
 050318 & --- &            --- &            --- &            ---  &            --- &             --- &(2)\\
 050319 &  PL & $2.0 \pm 0.4$  & $2.7 \pm 0.4$  & $2.7 \pm 0.6$   & $1.6 \pm 0.5$  & $8.9 \pm 1.4$ &\\
 050326 &  PL & $8.1 \pm 0.4$  & $21.0 \pm 0.6$ & $43.6 \pm 1.4$  & $44.8 \pm 2.1$ & $117.0 \pm 3.6$ &\\
 050401 & CPL & $6.9 \pm 1.5$  & $21.2 \pm 2.1$ & $41.7 \pm 4.1$  & $29.6 \pm 5.3$ & $99.2 \pm 7.9$ &\\
 050406 &  PL & $0.5 \pm 0.2$  & $0.7 \pm 0.2$  & $0.6 \pm 0.3$   & $0.3 \pm 0.3$  & $2.0 \pm 0.8$ &\\
 050410 &  PL & $1.9 \pm 0.6$  & $3.0 \pm 0.7$  & $3.5 \pm 1.1$   & $2.3 \pm 1.1$  & $10.7 \pm 2.7$ &\\
 050412 &  PL & $0.2 \pm 0.1$  & $0.7 \pm 0.2$  & $2.0 \pm 0.4$   & $2.5 \pm 0.7$  & $5.4 \pm 1.1$ &\\
050416A & CPL & $8.6 \pm 0.9$  & $9.4 \pm 1.1$  & $1.8 \pm 0.9$   & $0.1$          & $19.9 \pm 1.9$ &(6)\\
050416B &  PL & $4.3 \pm 0.8$  & $10.6 \pm 1.3$ & $21.2 \pm 2.1$  & $21.2 \pm 3.1$ & $57.3 \pm 5.4$ &\\
 050418 &  PL & $2.9 \pm 0.4$  & $6.6 \pm 0.6$  & $12.3 \pm 1.0$  & $11.7 \pm 1.4$ & $33.4 \pm 2.5$ &\\
 050421 &  PL & $0.5 \pm 0.2$  & $0.8 \pm 0.2$  & $1.1 \pm 0.4$   & $0.8 \pm 0.5$  & $3.1 \pm 1.0$ &\\
 050422 &  PL & $0.4 \pm 0.2$  & $0.5 \pm 0.2$  & $0.6 \pm 0.4$   & $0.8$          & $1.8 \pm 0.9$ &\\
050502B & CPL & $1.1 \pm 0.2$  & $3.1 \pm 0.3$  & $4.5 \pm 0.5$   & $2.0 \pm 0.8$  & $10.7 \pm 1.2$ &\\
 050505 &  PL & $1.4 \pm 0.4$  & $3.3 \pm 0.6$  & $6.2 \pm 1.0$   & $5.9 \pm 1.3$  & $16.7 \pm 2.5$ &\\
 050507 & --- &           ---  &            --- &            ---  &            --- &            --- &(2)\\
\enddata
\tablenotetext{1}{The event data are not available.}
\tablenotetext{2}{The event data of the part of the burst emission are not available.}
\tablenotetext{3}{F$^{\rm p}_{\rm ene}$(15-25) is an upper limit.}
\tablenotetext{4}{F$^{\rm p}_{\rm ene}$(25-50) is an upper limit.}
\tablenotetext{5}{F$^{\rm p}_{\rm ene}$(50-100) is an upper limit.}
\tablenotetext{6}{F$^{\rm p}_{\rm ene}$(100-150) is an upper limit.}
\tablenotetext{7}{F$^{\rm p}_{\rm ene}$(15-150) is an upper limit.}
\end{deluxetable}

\begin{deluxetable}{lccccccccc}
\tabletypesize{\scriptsize}
\tablecaption{BAT GRB 20-msec peak photon and energy flux \label{tbl:bat_peak20msflux}}
\tablewidth{0pt}
\tablehead{
\colhead{GRB} &
\colhead{Spectral} &
\colhead{F$^{\rm p}_{\rm ene}$(15-150)} &
\colhead{F$^{\rm p}_{\rm ph}$(15-150)} &
\colhead{Start} &
\colhead{Note}\\
\colhead{Name} &
\colhead{Model} &
\colhead{(10$^{-8}$ ergs cm$^{-2}$ s$^{-1}$)} &
\colhead{(photons cm$^{-2}$ s$^{-1}$)} &
\colhead{(s)} 
}
\startdata
 041217 & CPL & $75.5 \pm 31.0$  & $9.5 \pm 3.3$  &  +3.63 & \\
041219A & --- &             ---  &           ---  &    --- & (1)\\
041219B & --- &             ---  &           ---  &    --- & (1)\\
041219C &  PL & $20.8 \pm 10.8$  & $4.1 \pm 1.6$  &  +8.13 & \\
 041220 &  PL & $33.0 \pm 11.8$  & $3.1 \pm 1.0$  &  +0.25 & \\
 041223 &  PL & $124.0 \pm 27.1$ & $11.3 \pm 2.2$ & +24.30 & \\
 041224 & CPL & $44.6 \pm 18.5$  & $5.9 \pm 2.1$  &  +0.68 & \\
 041226 &  PL & $13.1 \pm 7.1$   & $1.4 \pm 0.7$  &  +2.75 & \\
 041228 & --- &              --- &           ---  &  +2.49 & (2)\\
 050117 & CPL & $41.9 \pm 16.6$   & $4.0 \pm 1.2$  & +88.02  & \\
 050124 & CPL & $69.1 \pm 20.4$   & $9.6 \pm 2.6$  & +0.36   & \\
 050126 &  PL & $28.5 \pm 13.2$   & $2.9 \pm 1.2$  & +3.44   & \\
 050128 &  PL & $154.0 \pm 54.7$  & $14.7 \pm 4.5$ & +6.18   & \\
 050202 &  PL & $41.9 \pm 15.3$   & $4.3 \pm 1.4$  & +0.08   & \\
050215A &  PL & $16.7 \pm 10.3$   & $1.9 \pm 1.0$  & +0.40   & \\
050215B &  PL & $17.0 \pm 10.8$   & $1.6 \pm 0.9$  & +1.52   & \\
050219A &  PL & $71.5 \pm 31.1$   & $6.9 \pm 2.5$  & +10.52  & \\
050219B &  PL & $344.0 \pm 103.0$ & $36.4 \pm 9.6$ & +3.06   & \\
 050223 & CPL & $13.7 \pm 7.5$    & $2.4 \pm 1.0$  & +7.74   & \\
 050306 & CPL & $70.3 \pm 35.6$   & $7.5 \pm 3.7$  & +107.96 & \\
 050315 &  PL & $22.7 \pm 11.2$   & $3.2 \pm 1.4$  & +9.86   & \\
 050318 & --- &              ---  &            --- &     --- & (3)\\
 050319 & CPL & $18.7 \pm 7.3$    & $3.6 \pm 1.3$  & +0.65   & \\
 050326 &  PL & $161.0 \pm 30.1$  & $16.0 \pm 2.4$ & +0.48   & \\
 050401 & --- &               --- &            --- & +25.17  & (2)\\
 050406 &  PL & $11.3$            & $1.3 \pm 0.6$  & +0.60   & (4)\\
 050410 & CPL & $81.0$            & $5.0 \pm 2.6$  & +7.84   & (4)\\
 050412 &  PL & $12.6 \pm 6.7$    & $1.5 \pm 0.6$  & +21.66  & \\
050416A & CPL & $25.0 \pm 12.1$   & $7.4 \pm 3.3$  & +0.72   & \\
050416B &  PL & $62.0 \pm 32.8$   & $8.6 \pm 4.2$  & +0.20   & \\
 050418 & CPL & $46.1 \pm 20.3$   & $5.7 \pm 2.0$  & +2.35   & \\
 050421 &  PL & $15.4 \pm 9.1$    & $1.4 \pm 0.8$  & +1.33   & \\
 050422 &  PL & $12.9 \pm 7.5$    & $1.6 \pm 0.7$  & +2.52   & \\
050502B &  PL & $22.2 \pm 9.0$    & $2.4 \pm 0.9$  & +0.89   & \\
 050505 &  PL & $33.8 \pm 18.3$   & $4.7 \pm 2.2$  & -8.37   & \\
 050507 & --- &               --- &            --- & --- & (3)\\
\enddata
\tablenotetext{1}{The event data are not available.}
\tablenotetext{2}{Reduced $\chi^{2}$ is greater than 2.}
\tablenotetext{3}{The event data of the part of the burst emission are not available.}
\tablenotetext{4}{F$^{\rm p}_{\rm ene}$(15-150) is an upper limit.}
\tablenotetext{5}{F$^{\rm p}_{\rm ph}$(15-150) is an upper limit.}
\tablenotetext{6}{battblocks failed because of the weak nature of the burst.}
\end{deluxetable}

\begin{deluxetable}{lccc|ccccc}
\tabletypesize{\scriptsize}
\tablecaption{BAT time-averaged spectral parameters \label{tbl:bat_timeave_spec}}
\tablewidth{0pt}
\tablehead{
\colhead{GRB} &
\colhead{$\alpha^{\rm PL}$} &
\colhead{K$^{\rm PL}_{50}$(a)} &
\colhead{$\chi^{2}_{\rm PL}$} & 
\colhead{$\alpha^{\rm CPL}$} & 
\colhead{$\eop$} &
\colhead{K$^{\rm CPL}_{50}$(b)} &
\colhead{$\chi^{2}_{\rm CPL}$} &
\colhead{Note}\\
\colhead{} &
\colhead{} &
\colhead{} &
\colhead{} &
\colhead{} &
\colhead{(keV)} &
\colhead{} &
\colhead{} &
\colhead{}
}
\startdata
041217 & $-1.452_{-0.064}^{+0.064}$ & $417.0_{-15.4}^{+15.4}$ & 73.2 & $-0.663_{-0.29}^{+0.309}$ & $91.5_{-12.0}^{+22.8}$ & $97.5_{-25.7}^{+37.3}$ & 49.2 &\\
041219A & --- & --- & --- & --- & --- & --- & --- & (1)\\
041219B & --- & --- & --- & --- & --- & --- & --- & (1)\\
041219C & $-2.007_{-0.089}^{+0.087}$ & $103.0_{-5.4}^{+5.4}$ & 44.6 & --- & --- & --- & --- &\\
041220 & $-1.672_{-0.122}^{+0.120}$ & $56.2_{-4.0}^{+3.9}$ & 30.5 & --- & --- & --- & --- &\\
041223 & $-1.153_{-0.03}^{+0.030}$ & $106.0_{-1.8}^{+1.8}$ & 33.4 & --- & --- & --- & --- &\\
041224 & $-1.731_{-0.058}^{+0.058}$ & $43.2_{-1.4}^{+1.4}$ & 62.8 & $-0.984_{-0.264}^{+0.281}$ & $68.9_{-7.0}^{+11.7}$ & $10.1_{-2.6}^{+3.7}$ & 36.9 &\\
041226 & $-1.416_{-0.417}^{+0.430}$ & $3.4_{-0.8}^{+0.8}$ & 85.1 & --- & --- & --- & --- &\\
041228 & $-1.617_{-0.077}^{+0.077}$ & $55.1_{-2.3}^{+2.3}$ & 64.6 & --- & --- & --- & --- &\\
050117 & $-1.519_{-0.042}^{+0.042}$ & $44.5_{-0.9}^{+0.9}$ & 44.4 & $-1.171_{-0.172}^{+0.182}$ & $130.3_{-26.7}^{+70.6}$ & $6.4_{-1.0}^{+1.3}$ & 32.2 &\\
050124 & $-1.412_{-0.085}^{+0.085}$ & $181.0_{-9.2}^{+9.2}$ & 55.2 & $-0.698_{-0.37}^{+0.404}$ & $100.2_{-18.3}^{+48.4}$ & $38.8_{-12.5}^{+20.3}$ & 43.6 &\\
050126 & $-1.360_{-0.168}^{+0.169}$ & $18.8_{-1.9}^{+1.9}$ & 72.3 & --- & --- & --- & --- &\\
050128 & $-1.368_{-0.071}^{+0.071}$ & $162.0_{-6.9}^{+6.9}$ & 63.1 & $-0.716_{-0.313}^{+0.333}$ & $113.7_{-20.2}^{+50.8}$ & $31.6_{-8.5}^{+12.3}$ & 49.5 &\\
050202 & $-1.357_{-0.292}^{+0.296}$ & $246.0_{-42.8}^{+42.8}$ & 44.7 & --- & --- & --- & --- &\\
050215A & $-1.326_{-0.261}^{+0.267}$ & $9.1_{-1.2}^{+1.2}$ & 43.0 & --- & --- & --- & --- &\\
050215B & $-2.173_{-0.223}^{+0.209}$ & $19.6_{-3.1}^{+3.0}$ & 57.6 & --- & --- & --- & --- &\\
050219A & $-1.305_{-0.056}^{+0.056}$ & $121.0_{-4.1}^{+4.1}$ & 100.7 & $-0.122_{-0.282}^{+0.300}$ & $90.6_{-7.9}^{+11.3}$ & $40.4_{-9.8}^{+13.8}$ & 39.0 &\\
050219B & $-1.508_{-0.05}^{+0.050}$ & $230.0_{-6.8}^{+6.8}$ & 79.4 & $-0.919_{-0.224}^{+0.236}$ & $107.9_{-15.2}^{+30.2}$ & $41.4_{-8.1}^{+10.6}$ & 57.9 &\\
050223 & $-1.826_{-0.163}^{+0.158}$ & $24.6_{-2.5}^{+2.4}$ & 49.3 & --- & --- & --- & --- &\\
050306 & $-1.467_{-0.065}^{+0.065}$ & $63.4_{-2.4}^{+2.4}$ & 59.5 & $-1.086_{-0.265}^{+0.281}$ & $140.3_{-35.8}^{+171.5}$ & $9.3_{-2.1}^{+2.9}$ & 53.3 &\\
050315 & $-2.087_{-0.085}^{+0.084}$ & $27.2_{-1.3}^{+1.3}$ & 52.3 & --- & --- & --- & --- &\\
050318 & --- & --- & --- & --- & --- & --- & --- & (2)\\
050319 & $-1.934_{-0.178}^{+0.171}$ & $8.9_{-1.0}^{+1.0}$ & 43.4 & --- & --- & --- & --- &\\
050326 & $-1.261_{-0.032}^{+0.033}$ & $210.0_{-3.6}^{+3.6}$ & 46.9 & --- & --- & --- & --- &\\
050401 & $-1.358_{-0.064}^{+0.064}$ & $232.0_{-8.6}^{+8.6}$ & 41.4 & --- & --- & --- & --- &\\
050406 & $-2.404_{-0.365}^{+0.317}$ & $12.4_{-3.5}^{+3.5}$ & 73.6 & $0.406_{-1.754}^{+2.594}$ & $29.4_{-3.2}^{+6.2}$ & $82.7_{-76.5}^{+2170.0}$ & 65.6 &\\
050410 & $-1.616_{-0.08}^{+0.080}$ & $76.3_{-3.7}^{+3.7}$ & 66.4 & $-0.812_{-0.355}^{+0.386}$ & $75.6_{-10.2}^{+21.0}$ & $19.0_{-6.2}^{+10.1}$ & 49.9 &\\
050412 & $-0.729_{-0.165}^{+0.173}$ & $17.1_{-1.6}^{+1.6}$ & 30.3 & --- & --- & --- & --- &\\
050416A & $-3.180_{-0.319}^{+0.285}$ & $30.6_{-7.7}^{+8.1}$ & 58.6 & $-0.972_{-1.012}^{+2.279}$ & $13.7_{-12.6}^{+7.9}$ & $4870$ & 52.2 & (6)\\
050416B & $-1.327_{-0.128}^{+0.128}$ & $285.0_{-21.7}^{+21.7}$ & 72.4 & $-0.391_{-0.585}^{+0.679}$ & $95.7_{-19.4}^{+65.5}$ & $76.2_{-34.4}^{+75.1}$ & 64.3 &\\
050418 & $-1.665_{-0.06}^{+0.060}$ & $63.4_{-2.1}^{+2.1}$ & 57.1 & $-1.292_{-0.246}^{+0.260}$ & $99.8_{-21.2}^{+86.7}$ & $9.6_{-2.2}^{+3.1}$ & 50.2 &\\
050421 & $-1.593_{-0.429}^{+0.412}$ & $8.5_{-2.2}^{+2.1}$ & 47.9 & --- & --- & --- & --- &\\
050422 & $-1.408_{-0.205}^{+0.204}$ & $7.7_{-0.9}^{+0.9}$ & 41.5 & --- & --- & --- & --- &\\
\enddata
\tablenotetext{a}{In units of 10$^{-4}$ ph cm$^{-2}$ s$^{-1}$ keV$^{-1}$.}
\tablenotetext{b}{In units of 10$^{-3}$ ph cm$^{-2}$ s$^{-1}$ keV$^{-1}$.}
\tablenotetext{1}{The event data are not available.}
\tablenotetext{2}{The event data of the part of the burst emission are not available.}
\tablenotetext{3}{No error because of reduced chi2 > 2.}
\tablenotetext{4}{K$^{\rm PL}_{50}=2.9_{-0.6}^{+0.5}$ for the spectrum based on the event data and K$^{\rm PL}_{50}=28.1_{-0.4}^{+0.4}$ for the 
spectrum based on the DPH data.}
\tablenotetext{5}{K$^{\rm PL}_{50}=4.4_{-0.4}^{+0.4}$ for the spectrum based on the event data and 
          K$^{\rm PL}_{50}=3.4_{0.2}^{+0.2}$ for the spectrum based on the DPH data.}
\tablenotetext{6}{K$^{\rm CPL}_{50}$ is an upper limit.}
\end{deluxetable}

\clearpage
\begin{deluxetable}{lcc|cccc|cccccc}
\tabletypesize{\scriptsize}
\rotate
\tablecaption{BAT time-resolved spectral parameters \label{tbl:bat_timeresolved_spec}}
\tablewidth{0pt}
\tablehead{
\colhead{GRB} &
\colhead{Start} &
\colhead{Stop} &
\colhead{$\alpha^{\rm PL}$} &
\colhead{K$^{\rm PL}_{50}$(a)} &
\colhead{$\chi^{2}$} & 
\colhead{F$_{PL}$(15-150)(b)} &
\colhead{$\alpha^{\rm CPL}$} & 
\colhead{$\eop$} &
\colhead{K$^{\rm CPL}_{50}$(c)} &
\colhead{$\chi^{2}$} &
\colhead{F$_{CPL}$(15-150)(b)} &
\colhead{Note}\\
\colhead{} &
\colhead{(s)} &
\colhead{(s)} &
\colhead{} &
\colhead{} &
\colhead{} &
\colhead{} &
\colhead{} &
\colhead{(keV)} &
\colhead{} &
\colhead{} &
\colhead{} &
\colhead{}
}
\startdata
041217 & 0.824 & 5.812 & $-1.385_{-0.064}^{+0.064}$ & $499.0_{-18.8}^{+18.8}$ & 78.2 & $48.4 \pm 1.9$ & $-0.517_{-0.297}^{+0.319}$ & $93.6_{-11.8}^{+21.1}$ & $126.0_{-33.3}^{+49.2}$ & 50.2 & $46.3 \pm 2.1$ &\\
041217 & 5.812 & 7.888 & $-1.763_{-0.198}^{+0.191}$ & $210.0_{-25.9}^{+25.5}$ & 54.5 & $19.3 \pm 2.5$ & --- & --- & --- & --- & --- &\\
041219C & 0.000 & 6.000 & $-1.912_{-0.096}^{+0.094}$ & $139.0_{-7.8}^{+7.8}$ & 52.7 & $12.7 \pm 0.7$ & --- & --- & --- & --- & --- &\\
041219C & 6.000 & 9.000 & $-2.168_{-0.179}^{+0.169}$ & $88.8_{-10.7}^{+10.4}$ & 54.4 & $8.3 \pm 0.9$ & --- & --- & --- & --- & --- &\\
041219C & 9.000 & 12.000 & $-2.060_{-0.332}^{+0.300}$ & $45.1_{-9.9}^{+9.5}$ & 42.3 & $4.2 \pm 0.8$ & --- & --- & --- & --- & --- &\\
041220 & -0.208 & 1.728 & $-1.479_{-0.117}^{+0.116}$ & $126.0_{-8.3}^{+8.3}$ & 52.0 & $12.0 \pm 0.9$ & --- & --- & --- & --- & --- &\\
041220 & 1.728 & 2.936 & $-1.862_{-0.25}^{+0.236}$ & $59.1_{-9.7}^{+9.4}$ & 53.9 & $5.4 \pm 0.9$ & --- & --- & --- & --- & --- &\\
041220 & 2.936 & 6.812 & $-2.029_{-0.408}^{+0.360}$ & $19.3_{-5.3}^{+5.0}$ & 40.1 & $1.8 \pm 0.4$ & --- & --- & --- & --- & --- &\\
041223 & -10.536 & -0.544 & $-1.363_{-0.381}^{+0.388}$ & $17.1_{-4.0}^{+4.0}$ & 61.5 & $1.7 \pm 0.4$ & --- & --- & --- & --- & --- &\\
041223 & -0.544 & -0.124 & $-1.529_{-0.408}^{+0.400}$ & $88.9_{-22.3}^{+21.9}$ & 48.5 & $8.4 \pm 2.3$ & --- & --- & --- & --- & --- &\\
041223 & -0.124 & 2.196 & $-0.771_{-0.081}^{+0.082}$ & $292.0_{-13.7}^{+13.6}$ & 57.6 & $34.6 \pm 1.7$ & --- & --- & --- & --- & --- &\\
041223 & 2.196 & 3.104 & $-0.977_{-0.286}^{+0.303}$ & $93.1_{-15.8}^{+15.7}$ & 49.6 & $10.2 \pm 1.8$ & --- & --- & --- & --- & --- &\\
041223 & 3.104 & 4.220 & $-1.155_{-0.146}^{+0.148}$ & $188.0_{-16.2}^{+16.2}$ & 58.1 & $19.4 \pm 1.9$ & --- & --- & --- & --- & --- &\\
041223 & 4.220 & 4.864 & $-0.896_{-0.098}^{+0.100}$ & $466.0_{-27.4}^{+27.4}$ & 53.7 & $52.3 \pm 3.4$ & --- & --- & --- & --- & --- &\\
041223 & 4.864 & 5.372 & $-1.148_{-0.152}^{+0.153}$ & $296.0_{-26.9}^{+26.8}$ & 61.6 & $30.5 \pm 3.1$ & --- & --- & --- & --- & --- &\\
041223 & 5.372 & 9.768 & $-1.159_{-0.092}^{+0.092}$ & $158.0_{-8.0}^{+8.0}$ & 48.5 & $16.2 \pm 0.9$ & --- & --- & --- & --- & --- &\\
041223 & 9.768 & 11.524 & $-1.415_{-0.365}^{+0.360}$ & $48.2_{-10.4}^{+10.3}$ & 54.7 & $4.6 \pm 1.1$ & --- & --- & --- & --- & --- &\\
041223 & 11.524 & 12.840 & $-1.105_{-0.188}^{+0.191}$ & $126.0_{-13.3}^{+13.2}$ & 66.2 & $13.1 \pm 1.5$ & --- & --- & --- & --- & --- &\\
041223 & 12.840 & 13.192 & $-0.701_{-0.162}^{+0.168}$ & $337.0_{-34.0}^{+33.6}$ & 69.8 & $41.1 \pm 4.1$ & --- & --- & --- & --- & --- &\\
041223 & 13.192 & 14.400 & $-0.937_{-0.176}^{+0.183}$ & $133.0_{-14.7}^{+14.7}$ & 77.0 & $14.7 \pm 1.7$ & --- & --- & --- & --- & --- &\\
\enddata
\tablenotetext{a}{In units of 10$^{-4}$ ph cm$^{-2}$ s$^{-1}$ keV$^{-1}$.}
\tablenotetext{b}{In units of 10$^{-8}$ ergs cm$^{-2}$ s$^{-1}$.}
\tablenotetext{c}{In units of 10$^{-3}$ ph cm$^{-2}$ s$^{-1}$ keV$^{-1}$.}
\tablenotetext{1}{Not enough statistics to perform the spectral fit.}
\tablenotetext{2}{No error because reduced chi2 is greater than 2.}
\tablenotetext{3}{K50PL is an upper limit.}
\tablenotetext{4}{F$_{PL}$(15-150) is an upper limit.}
\tablenotetext{5}{Bad quality of the spectrum.}
\tablenotetext{6}{Spectral paramters are not constrained by a CPL fit.}
\end{deluxetable}

\begin{table}
\caption{BAT GRB $T_{90}$ and $T_{50}$ durations in the 140-220 keV band at the GRB rest frame.\label{tbl:bat_t90_t50_src}}
\begin{center}
\begin{tabular}{lccc}\hline
GRB & $T_{90}^{src}$ & $T_{50}^{src}$ & Note\\
    & (s)           & (s)          &     \\\hline
050126 & 8.73 & 4.80 &\\
050223 & --- & --- & (1)\\
050315 & 13.99 & 6.93 &\\
050318 & 5.68 & 3.02 &\\
050319 & 33.29 & 28.15 &\\
050401 & 8.37 & 6.53 &\\
050416A & --- & --- & (1)\\
050505 & 11.37 & 4.55 &\\
050509B & --- & --- & (1)\\
050525A & 4.24 & 2.89 &\\
050603 & 2.53 & 0.57 &\\
050724 & --- & --- & (1)\\
050730 & 28.67 & 13.00 &\\
050801 & 1.10 & 0.79 &\\
050802 & 4.72 & 2.29 &\\
050814 & 24.81 & 7.91 &\\
050820A & --- & --- & (2)\\
050904 & 26.48 & 12.50 &\\
050908 & 1.11 & 0.50 &\\
050922C & 1.56 & 0.63 &\\
051016B & --- & --- & (1)\\
051109A & 11.78 & 8.36 &\\
051111 & 18.82 & 6.27 &\\
051221A & 0.17 & 0.08 &\\
060115 & 28.80 & 19.86 &\\
060124 & 3.94 & 2.73 &\\
060206 & 1.10 & 0.43 &\\
060210 & 35.74 & 6.24 &\\
060218 & --- & --- & (2)\\
060223A & 1.93 & 0.70 &\\
060418 & 19.28 & 6.43 &\\
060502A & 6.69 & 2.95 &\\
060505 & --- & --- & (1)\\
060510B & 44.87 & 19.78 &\\
\end{tabular}
\tablenotetext{1}{battblocks failed because of the weak signal in the light curve.}
\tablenotetext{2}{The event data of the part of the burst emission are not available.}
\end{center}
\end{table}

\begin{table}
\caption{Redshifts of Swift GRBs\label{tbl:redshift}}
\begin{center}
\begin{tabular}{lcc}\hline
GRB & Redshift & Reference\\\hline
050126 & 1.290 & (2)\\
050223 & 0.584 & (3)\\
050315 & 1.950 & (4)\\
050318 & 1.444 & (4)\\
050319 & 3.2425 & (1)\\
050401 & 2.8983 & (1)\\
050416A & 0.6528 & (5)\\
050505 & 4.275 & (6)\\
050509B & 0.226 & (7)\\
050525A &  0.606 & (8)\\
050603 & 2.821 & (9)\\
050724 & 0.258 & (10)\\
050730 & 3.9693 & (10)\\
050801 & 1.38 & (1)\\
050802 & 1.7102 & (1)\\
050814 & 5.3 & (11)\\
050820A &  2.6147 & (1)\\
050824 & 0.8278 & (1) \\
050826 & 0.296 & (12)\\
050904 & 6.295 & (13)\\
050908 & 3.3467 & (1)\\
050922C & 2.1995 & (1)\\\hline
\end{tabular}
\tablenotetext{1}{Fynbo et al.  ApJS, 185, 526 (2009)}
\tablenotetext{2}{Berger E. et al. ApJ, 629, 328 (2005)}
\tablenotetext{3}{Pellizza, L.J. et al. A\&A, 459, L5 (2006)}
\tablenotetext{4}{Berger, E. et al. ApJ, 634, 501 (2005)}
\tablenotetext{5}{Soderberg, A.M. et al. ApJ, 661, 982 (2007)}
\tablenotetext{6}{Berger, E. et al. ApJ, 642, 979 (2006)}
\tablenotetext{7}{Gehrels, N. et al. Nature, 437, 851 (2005)}
\tablenotetext{8}{Della Valle, M. et al. ApJ, 642, L103 (2006)}
\tablenotetext{9}{Berger, E. et al. GCN Circ. 3520}
\tablenotetext{10}{Berger, E. et al. Nature, 438, 988 (2005)}
\tablenotetext{11}{Jakobsson, P. et al. A\&A, 447, 897 (2006)}
\tablenotetext{12}{Mirabal, N. et al. ApJ, 661, L127 (2007)}
\tablenotetext{13}{Kawai, N. et al. Nature, 440, 184 (2006)}
\end{center}
\end{table}

\begin{table}
\caption{Statistics of the BAT2 GRB catalog\label{statistics}}
\begin{center}
\begin{tabular}{lccc}\hline
Class        & Number of GRBs & Fraction & Classification\\\hline
L-GRB        & 424    & 89\%     & $T_{90} \geq 2$ s\\
S-GRB        & 38     & 8\%      & $T_{90} < 2$ s\\
S-GRB w/E.E. & 10     & 2\%      & \citet{norris2010}\\ 
Unknown     & 4      & 1\%      & Incomplete/lost data\\\hline
\end{tabular}
\end{center}
\end{table}

\begin{deluxetable}{lcccccc}
\tabletypesize{\small}
\tablecaption{Spectral parameters of the initial short spikes of the short GRBs with E.E.\label{tbl:shortEE_spec}}
\tablewidth{0pt}
\tablehead{
\colhead{GRB} &
\colhead{start} & 
\colhead{stop} &
\colhead{$\alpha^{\rm PL}$} &
\colhead{K$^{\rm PL}_{50}$(a)} &
\colhead{$\chi^{2}$} \\
\colhead{} &
\colhead{(s)} &
\colhead{(s)} &
\colhead{} &
\colhead{} &
\colhead{}
}
\startdata

050724 & $-0.024$ & 0.416 & $-1.50_{-0.14}^{+0.14}$ & $515.2_{-41.1}^{+40.9}$ & 50.9\\
051227 & $-0.848$ & 0.828 & $-0.94_{-0.25}^{+0.23}$ & $61.0_{-8.0}^{+7.9}$ & 65.6\\
061006 & $-23.24$ & $-22.20$ & $-0.86_{-0.07}^{+0.07}$ & $445.3_{-17.8}^{+17.7}$ & 50.8\\
061210 & $-0.004$ & 0.080 & $-0.69_{-0.12}^{+0.12}$ & $2755.1_{-209.8}^{+209.0}$ & 58.2\\
070714B & $-1$ & $-2$ & $-0.98_{-0.08}^{+0.08}$ & $158.2_{-7.2}^{+7.2}$ & 43.0\\
071227 & $-0.144$ & 1.872 & $-0.90_{-0.24}^{+0.22}$ & $98.6_{-12.3}^{+12.1}$ & 51.0\\
080503 & $-0.048$ & 0.436 & $-1.62_{-0.23}^{+0.23}$ & $130.6_{-19.6}^{+19.3}$ & 66.0\\
090531B   & 0.252 & 1.300 & $-0.99_{-0.16}^{+0.16}$ & $148.4_{-12.1}^{+12.0}$ & 63.3\\
090715A & $-0.12$ & 0.84 & $-0.99_{-0.20}^{+0.19}$ & $348.9_{-38.8}^{+38.3}$ & 55.0\\
090916 & 0.0 & 0.35 & $-1.38_{-0.30}^{+0.30}$ & $236.5_{-44.5}^{+44.5}$ & 72.6\\
\enddata
\tablenotetext{a}{In units of 10$^{-4}$ ph cm$^{-2}$ s$^{-1}$ keV$^{-1}$.}
\end{deluxetable}

\begin{deluxetable}{lcccccc}
\tabletypesize{\small}
\tablecaption{BAT GRB energy fluence of the initial short spkes of the short GRB with E.E.\label{tbl:shortEE_fluence}}
\tablewidth{0pt}
\tablehead{
\colhead{GRB} &
\colhead{Spectral} &
\colhead{S(15-25)} &
\colhead{S(25-50)} &
\colhead{S(50-100)} &
\colhead{S(100-150)} &
\colhead{S(15-150)} \\
\colhead{Name} &
\colhead{Model} &
\multicolumn{5}{c}{(10$^{-8}$ ergs cm$^{-2}$)}
}
\startdata
050724 &  PL & 2.9 $\pm$ 0.4 & 5.3 $\pm$ 0.4 & 7.5 $\pm$ 0.8 & 5.7 $\pm$ 0.9 & 21.5 $\pm$ 1.9\\
051227 &  PL & 0.8 $\pm$ 0.2 & 2.0 $\pm$ 0.3 & 4.2 $\pm$ 0.6 & 4.3 $\pm$ 1.0 & 11.3 $\pm$ 1.6\\
061006 &  PL & 3.3 $\pm$ 0.3 & 8.9 $\pm$ 0.4 & 19.6 $\pm$ 0.8 & 21.0 $\pm$ 1.4 & 52.7 $\pm$ 2.3\\
061210 &  PL & 1.4 $\pm$ 0.2 & 4.2 $\pm$ 0.4 & 10.5 $\pm$ 0.8 & 12.4 $\pm$ 1.5 & 28.4 $\pm$ 2.4\\
070714B &  PL & 3.7 $\pm$ 0.4 & 9.4 $\pm$ 0.5 & 19.2 $\pm$ 1.0 & 19.5 $\pm$ 1.6 & 51.8 $\pm$ 2.6\\
071227 &  PL & 1.4 $\pm$ 0.4 & 3.9 $\pm$ 0.6 & 8.3 $\pm$ 1.1 & 8.8 $\pm$ 1.8 & 22.4 $\pm$ 2.9\\
080503 &  PL & 0.9 $\pm$ 0.2 & 1.5 $\pm$ 0.2 & 2.0 $\pm$ 0.4 & 1.5 $\pm$ 0.4 & 5.9 $\pm$ 0.9\\
090531B &  PL & 1.2 $\pm$ 0.2 & 3.1 $\pm$ 0.3 & 6.3 $\pm$ 0.6 & 6.3 $\pm$ 1.0 & 16.8 $\pm$ 1.6\\
090715A &  PL & 2.7 $\pm$ 0.6 & 6.7 $\pm$ 0.9 & 13.4 $\pm$ 1.5 & 13.5 $\pm$ 2.4 & 36.3 $\pm$ 4.0\\
090916 & PL & 0.9 $\pm$ 0.3 & 1.9 $\pm$ 0.4 & 2.9 $\pm$ 0.6 & 2.3 $\pm$ 0.8 & 8.0 $\pm$ 1.6\\
\enddata
\end{deluxetable}

\begin{deluxetable}{llll}
\tabletypesize{\small}
\tablecaption{$\esp$ and $\eiso$ values for {\it Swift} GRBs. \label{ep_eiso_swift}}
\tablewidth{0pt}
\tablehead{
\colhead{GRB} &
\colhead{z} &
\colhead{$\esp$} &
\colhead{$\eiso$} \\
\colhead{} &
\colhead{} &
\colhead{(keV)} &
\colhead{(10$^{52}$ erg)}
}
\startdata
050416A & 0.6528 & $22_{-4}^{+5}$ & $0.12_{-0.02}^{+0.01}$\\
050525A & 0.606  & $129_{-7}^{+6}$      & $2.6_{-0.2}^{+0.9}$\\
060115  & 3.5328 & $297_{-111}^{+92}$   & $6.5_{-1.5}^{+6.7}$\\
060206  & 4.0559 & $410_{-179}^{+195}$  & $4.5_{-1.0}^{+3.2}$\\
060707  & 3.4240 & $274_{-77}^{+66}$    & $4.7_{-0.8}^{+18.8}$\\
060908  & 1.8836 & $414_{-120}^{+399}$  & $8.1_{-4.5}^{+1.9}$\\
060927  & 5.4636 & $276_{-54}^{+97}$    & $13_{-3}^{+3}$\\
071010B & 0.947  & $88_{-21}^{+21}$     & $2.6_{-0.4}^{+0.5}$\\
071117  & 1.3308 & $112_{-52}^{+315}$   & $6.5_{-4.4}^{+1.6}$\\
080413B & 1.1014 & $163_{-46}^{+51}$    & $1.7_{-0.3}^{+2.2}$\\
080603B & 2.6892 & $277_{-111}^{+95}$   & $6.6_{-1.2}^{+6.2}$ \\
080605  & 1.6403 & $766_{-243}^{+1100}$  &  $31_{-12}^{+20}$\\
080916A & 0.6887 & $200_{-51}^{+120}$   & $1.8_{-1.0}^{+1.8}$ \\
090205  & 4.6497 & $214_{-87}^{+58}$  & $0.9_{-0.3}^{+0.7}$\\
090423  & 8.26   & $410_{-88}^{+115}$ & $9.5_{-1.9}^{+2.6}$ \\
090424  & 0.544  & $236_{-49}^{+127}$ & $4.3_{-1.4}^{+2.4}$\\
090926B & 1.24   & $175_{-20}^{+24}$ & $5.4_{-2.0}^{+2.8}$ \\
091018  & 0.971  & $55_{-17}^{+6.8}$ & $0.7_{-0.1}^{+0.3}$\\
091029  & 2.752  & $229_{-59}^{+94}$ & $8.5_{-2.5}^{+4.5}$ \\
\enddata
\end{deluxetable}

\clearpage

\clearpage

\clearpage

\clearpage

\clearpage

\clearpage

\clearpage

\clearpage

\clearpage

\clearpage

\clearpage

\clearpage
\begin{figure}
\centerline{
\includegraphics[width=12cm,angle=-90]{fig18.eps}
}
\caption{Histograms of the BAT $\ep$ in a CPL fit for the 
time-resolved (solid) and the time-averaged (dashed) spectra.  The 
histograms are normalized by a total number of spectra. 
\label{time_resolved_cpl_ep_hist}}
\end{figure}

\clearpage

\clearpage

\clearpage

\clearpage

\clearpage

\clearpage

\clearpage

\clearpage

\clearpage

\clearpage

\clearpage

\clearpage

\end{document}